\documentclass[preprint,showpacs,preprintnumbers,amsmath,amssymb]{revtex4}

\usepackage{graphicx}
\usepackage{dcolumn}
\usepackage{bm,morefloats,color}

\usepackage{amssymb,bm}

\def\Eqref#1{Eq.~(\ref{#1})}

\def\Eq#1{\begin{equation} #1 \end{equation}}
\def\Eqr#1{\begin{eqnarray} #1 \end{eqnarray}}
\def\Eqrsubl#1#2{\begin{subequations}\label{#1}\Eqr{#2}\end{subequations}}

\newcommand{\nn}{\nonumber}
\newcommand{\pd}{\partial}

\newcommand{\bea}{\begin{eqnarray}}
\newcommand{\eea}{\end{eqnarray}}

\def\Xsp{{\rm X}}

\def\Ysp{{\rm Y}}
\def\Zsp{{\rm Z}}
\def\X5sp{{\rm X}_5}
\def\Y3sp{{\rm Y}_3}
\def\Z3sp{{\rm Z}_3}

\def\Msp{{\rm M}}
\def\Nsp{{\rm N}}
\def\lap{{\triangle}}
\def\e{{\rm e}}

\begin{document}

\preprint{YITP-11-100}

\title{
Dynamics of partially localized brane systems
}

\author{Masato Minamitsuji}
\affiliation{
Yukawa Institute for Theoretical Physics
Kyoto University, Kyoto 606-8502, Japan.
}%

\author{Kunihito Uzawa}
\affiliation{%
Department of Physics, Kinki University,
Higashi-Osaka, Osaka 577-8502, Japan
}%

\date{\today}

\begin{abstract}
We study dynamical partially localized brane solutions in 
higher dimensions. 
We give new descriptions of 
the relevant solutions of dynamical branes which are localized
along both the overall and relative transverse directions. 
The starting point is a system of $p_r$-branes ending on
a $p_s$-brane with a time-dependent warp factor. 
This system can be related to D$p_r$-D$p_s$ brane system in string theory,
 where one brane is localized
at the delocalized other brane. 
We then show that these 
give Friedmann-Lemaitre-Robertson-Walker cosmological solutions. 
Our approach leads to a new and manifest 
description of the brane configurations near the delocalized branes, 
and new solutions in the wave or KK-monopole background  
in terms of certain partial differential equations in $D$ dimensions
including ten and eleven dimensions. 

\end{abstract}

\maketitle


\section{Introduction}
\label{sec:introduction}

In recent years 
much effort has been devoted 
to the construction of 
cosmological models in string theory produced by enlarging 
static $p$-brane solutions \cite{Binetruy:2007tu, Gibbons:2009dr, 
Maeda:2009zi, Minamitsuji:2010kb, Minamitsuji:2010uz}. 
Although these calculations are complicated by 
the occurrence of time-dependences, there has been active development 
in constructing time-dependent supergravity
solutions of $p$-branes and other solitons in string theory 
\cite{Gibbons:2005rt, Maeda:2009tq, Maeda:2009ds, Maeda:2010yk, 
Minamitsuji:2010fp, Maeda:2010ja, Nozawa:2010zg, Maeda:2011sh}. 
These classical solutions in string theory made it possible to 
discuss the dynamics such as cosmological evolution of our Universe 
and brane collision within the framework of string theory. 
In these studies, brane world models were 
obtained 
by wrapping or intersecting higher-dimensional $p$-branes around compact 
manifolds. In the course of compactifying $p$-branes, the dynamical 
solutions become smeared or delocalized along the compactified directions, 
which include possibly some of overall transverse directions and 
relative transverse directions that corresponds to the transverse directions 
which are longitudinal to some of other constituent branes. 
Such intersecting $p$-brane solutions in higher dimensions 
thus become localized only along the relative or 
overall transverse directions. 
The dynamical intersecting brane solutions which we have mainly constructed 
are such delocalized type \cite{Binetruy:2007tu, Maeda:2009zi, 
Minamitsuji:2010kb, Minamitsuji:2010uz}. 
There are several works to construct the static localized  
intersecting brane solutions with the restricted ansatz of fields which has 
the same form as the corresponding delocalized intersecting brane solutions 
\cite{Tseytlin:1996as, Tseytlin:1997cs, Youm:1997hw, 
Yang:1999ze}. 
The equations of motion along with such simplified assumption 
for fields require that one of the branes has to be delocalized
on the relative transverse directions. However, it is difficult 
to obtain the exact localized solutions even if we use such simplified 
ansatz because harmonic functions that specify branes satisfy coupled 
partial differential equations. The solutions of these differential 
equations in general have a complicated form. 
On the other hand, the dynamical localized intersecting brane solution  
is not well-known, and nobody mentions the explicit expressions for harmonic 
functions. This article will describe a 
method of dealing with the extension of the time-dependent solutions 
in the partially localized intersecting brane system,  
where branes are localized along the relative transverse directions 
but delocalized along the overall transverse directions \cite{Youm:1999zs}. 
For the purposes of construction of cosmological model, 
we employ the same ansatz of fields as the static 
$p$-brane solutions. 
It is, in general, possible to derive intersecting brane solutions
in terms of applying duality transformations in string theory. 
For instance, we compactify the direction which becomes
delocalized through smearing or uniform array of branes along it   
to apply T-duality transformations in the transverse directions. Then, 
the power of the radial coordinate in the harmonic function changes.
Hence, we will construct such localized 
intersecting brane solutions case by case. 

It is the purpose 
of this paper to construct various explicit partially localized 
intersecting dynamical brane solutions in various dimensions. 
We give classification of these dynamical intersecting 
brane solutions involving two branes, and discuss the 
application of these solutions to cosmology. 
We also study the arbitrary single brane on the KK-monopole 
and wave background. 
It is possible to derive the time-dependent solution if the form
of the static solution is explicitly known,
so calculations have generally relied on an assumption of fields and 
even strictly metric form. Also, even where a coupling between scalar 
field and gauge field strength in the action is known, 
the intersection rule of the brane can be obtained explicitly. 
Since a warp factor arises from a field strength, the dynamics of 
a system composed of two branes can be characterized by two warp
factors arising from two field strengths. For M-branes and D-branes, 
among these warp factors only one function can depend on time.

The procedure is described here in generally higher-dimensional 
gravity model as well as the supergravity, the solutions of a D-brane 
or M-brane in a wave or KK-monopole background. This is simple
enough to illustrate the use of the method without the general idea being
lost in the complications of higher-dimensional gravity theory, 
and yet sufficiently general so that we can see how to deal 
with an arbitrary expansion. As we will see, these methods yield a 
prescription for intersecting brane solutions that depend 
not only on the overall transverse directions, but world-volume and 
the relative transverse directions.

The paper is organized as follows. In Sec.~\ref{sec:two}, we show that 
the partially localized dynamical intersecting brane solutions of 
two $p$-branes exist as an almost immediate generalization of 
the static brane solution where one of branes is delocalized. 
We will also study explicit partially localized $p$-brane solutions 
in KK-monopole or wave background. 
We will apply these solutions to construct various explicit 
partially localized intersecting M-brane solutions in 
Sec.~\ref{sec:eleven}, and various partially localized intersecting brane 
solutions in ten dimensions in Sec.~\ref{sec:ten}.
We then go on in Sec.~\ref{sec:co} to apply these solutions 
to cosmology. Sec.~\ref{sec:discussions} is devoted to discussions.


\section{The intersection of two branes in $D$-dimensional theory}
  \label{sec:two}

We study the dynamical brane in $D$-dimensional theory. We describe  
the relation of the partially localized static brane solution to 
time-dependent solutions in $D$-dimensions. 
We also study solutions in the wave and KK-monopole background. 

\subsection{The intersection of dynamical 
$p_r-p_s$-branes in $D$-dimensional theory}
\label{sec:rs}

In this section, we consider a $D$-dimensional theory 
composed of the metric $g_{MN}$, scalar field $\phi$, and 
antisymmetric tensor field strengths of rank $(p_r+2)$ and 
$(p_s+2)$: 
\Eqr{
S&=&\frac{1}{2\kappa^2}\int \left[R\ast{\bf 1}
 -\frac{1}{2}\ast d\phi \wedge d\phi
 -\frac{1}{2}\frac{1}{\left(p_r+2\right)!}
 \e^{\epsilon_rc_r\phi}\ast F_{(p_r+2)}\wedge F_{(p_r+2)}\right.\nn\\
 &&\left. -\frac{1}{2}\frac{1}{\left(p_s+2\right)!}
 \e^{\epsilon_sc_s\phi}\ast F_{(p_s+2)}\wedge F_{(p_s+2)}
 \right],
\label{rs:action:Eq}
}
where $\kappa^2$ is the $D$-dimensional gravitational constant,  
$\ast$ is the Hodge operator in the $D$-dimensional space-time, 
$F_{\left(p_r+2\right)}$ and $F_{\left(p_s+2\right)}$ 
denote $\left(p_r+2\right)$-form, 
$\left(p_s+2\right)$-form field strengths, and 
$c_I$, $\epsilon_I~(I=r,~s)$ are constants given by 
\Eqrsubl{rs:parameters:Eq}{
c_I^2&=&N_I-\frac{2(p_I+1)(D-p_I-3)}{D-2},
   \label{rs:coupling:Eq}\\
\epsilon_I&=&\left\{
\begin{array}{cc}
 +&~{\rm if}~~p_I-{\rm brane~is~electric}\\
 -&~~~{\rm if}~~p_I-{\rm brane~is~magnetic}\,,
\end{array} \right.
 \label{rs:sig:Eq}
   }
where $N_I$ is constant. 
The field strength $F_{(p_r+2)}$, $F_{(p_s+2)}$ are given by the 
$(p_r+1)$-form, $(p_s+1)$-form gauge potentials $A_{(p_r+1)}$, 
$A_{(p_s+1)}$, respectively
\Eq{
F_{(p_r+2)}=dA_{(p_r+1)}\,,~~~~~F_{(p_s+2)}=dA_{(p_s+1)}\,.
}
 
After variations of the action
with respect to the metric, the scalar field, 
and the $\left(p_r+1\right)$-form and $\left(p_s+1\right)$-form 
gauge fields, we obtain the field equations,
\Eqrsubl{rs:equations:Eq}{
&&R_{MN}=\frac{1}{2}\pd_M\phi \pd_N \phi
+\frac{1}{2}\frac{\e^{\epsilon_rc_r\phi}}
{\left(p_r+2\right)!}
\left[\left(p_r+2\right)
F_{MA_2\cdots A_{\left(p_r+2\right)}} 
{F_N}^{A_2\cdots A_{\left(p_r+2\right)}}
-\frac{p_r+1}{D-2} g_{MN} F^2_{\left(p_r+2\right)}\right]\nn\\
&&~~~~~~+\frac{1}{2}\frac{\e^{\epsilon_sc_s\phi}}
   {\left(p_s+2\right)!}
\left[\left(p_s+2\right)
F_{MA_2\cdots A_{\left(p_s+2\right)}} 
{F_N}^{A_2\cdots A_{\left(p_s+2\right)}}
-\frac{p_s+1}{D-2} g_{MN} F_{\left(p_s+2\right)}^2\right],
   \label{rs:Einstein:Eq}\\
&&d\ast d\phi-\frac{1}{2}\frac{\epsilon_rc_r}
{\left(p_r+2\right)!}
\e^{\epsilon_rc_r\phi}\ast F_{\left(p_r+2\right)}\wedge F_{\left(p_r+2\right)}
-\frac{1}{2}\frac{\epsilon_sc_s}{\left(p_s+2\right)!}
\e^{\epsilon_sc_s\phi}\ast F_{\left(p_s+2\right)}\wedge 
F_{\left(p_s+2\right)}=0,
   \label{rs:scalar:Eq}\\
&&d\left[\e^{\epsilon_rc_r\phi}\ast F_{\left(p_r+2\right)}\right]=0,
   \label{rs:gauge-r:Eq}\\
&&d\left[\e^{\epsilon_sc_s\phi}\ast F_{\left(p_s+2\right)}\right]=0.
   \label{rs:gauge-s:Eq}
}

We look for solutions whose spacetime
metric has the form 
\Eqr{
ds^2&=&h^{a_r}_r(x, y, z)h_s^{a_s}(x, v, z)q_{\mu\nu}
(\Xsp)dx^{\mu}dx^{\nu}+h^{b_r}_r(x, y, z)h_s^{a_s}(x, v, z)\gamma_{ij}
(\Ysp_1)dy^idy^j\nn\\
&&+h^{a_r}_r(x, y, z)h_s^{b_s}(x, v, z)w_{mn}(\Ysp_2)dv^{m}dv^{n}
+h^{b_r}_r(x, y, z)h_s^{b_s}(x, v, z)u_{ab}(\Zsp)dz^adz^b, 
 \label{rs:metric:Eq}
}
where $q_{\mu\nu}(\Xsp)$ is the $(p+1)$-dimensional metric which
depends only on the $(p+1)$-dimensional coordinates $x^{\mu}$, 
$\gamma_{ij}(\Ysp_1)$ is the $(p_s-p)$-dimensional metric which
depends only on the $(p_s-p)$-dimensional coordinates $y^i$, 
$w_{mn}(\Ysp_2)$ is the $(p_r-p)$-dimensional metric which
depends only on the $(p_r-p)$-dimensional coordinates $v^m$
and finally $u_{ab}(\Zsp)$ is the $(D+p-p_r-p_s-1)$-dimensional metric which
depends only on the $(D+p-p_r-p_s-1)$-dimensional coordinates $z^a$. 
The constants $a_I~(I=r,~s)$ and $b_I~(I=r,~s)$ in the metric 
(\ref{rs:metric:Eq}) are written by 
\Eq{
a_I=-\frac{4(D-p_I-3)}{N_I(D-2)},~~~~~b_I=\frac{4(p_I+1)}{N_I(D-2)}.
 \label{rs:paremeter:Eq}
}

The brane configuration is given as follows (See Table \ref{rs}.): 
\begin{table}[h]
\caption{\baselineskip 14pt
Intersections of two $p$-branes in the metric \eqref{rs:metric:Eq}.
}
{\scriptsize
\begin{center}
\begin{tabular}{|c||c|c|c|c|c|c|c|c|c|c|c|c|c|c|}
\hline
Case&&0&1& $\cdots$ & $p$ & $p+1$ & $\cdots$ & $p_s$ & $p_s+1$ & 
$\cdots$ & $p_s+p_r-p$ & $p_s+p_r-p+1$ & $\cdots$ & $D-1$
\\
\hline
 &$p_r$ & $\circ$ & $\circ$ & $\circ$ & $\circ$ &&&& $\circ$ & $\circ$ &
  $\circ$ &&& 
\\
\cline{3-15}
$p_r$-$p_s$ & $p_s$ & $\circ$ & $\circ$ & $\circ$ & $\circ$ & $\circ$ &
 $\circ$ & $\circ$ && & & & &
\\
\cline{3-15}
&$x^N$ & $t$ & $x^1$ & $\cdots$ & $x^p$ & $y^1$ & $\cdots$ & $y^{p_s-p}$ 
& $v^1$ & $\cdots$ & $v^{p_r-p}$ & $z^1$ & $\cdots$ & $z^{D+p-p_r-p_s-1}$
\\
\hline
\end{tabular}
\end{center}
}
\label{rs}
\end{table}

The metric of $D$-dimensional spacetime (\ref{rs:metric:Eq}) implies that 
the solutions are expressed by functions, $h_r$ and $h_s$, 
which are the functions of the coordinates transverse to the brane 
as well as the world-volume coordinate.
The powers of harmonic functions for the configurations of 
$p_r-p_s$-branes have to satisfy the intersection rule, 
and split the coordinates in three parts. 
One is the overall world-volume coordinates $x^{\mu}$, which are common 
to the $p_r-p_s$-branes. The others are the coordinates of 
overall transverse space $z^a$, and the relative transverse coordinates 
$y^i$ and $v^m$, which are transverse to only one 
of the $p_r-p_s$-branes.  
Each of $h_r$ and $h_s$ depends not only on overall transverse coordinates 
but also on the corresponding relative coordinates. 

We also assume that the scalar field $\phi$ and field 
strengths $F_{\left(p_r+2\right)}$, $F_{\left(p_s+2\right)}$ are written by
\Eqrsubl{rs:ansatz:Eq}{
\e^{\phi}&=&h_r^{2\epsilon_rc_r/N_r}\,
h_s^{2\epsilon_sc_s/N_s},
  \label{rs:scalar-a:Eq}\\
F_{\left(p_r+2\right)}&=&\frac{2}{\sqrt{N_r}}
d\left[h^{-1}_r(x, y, z)\right]\wedge\Omega(\Xsp)\wedge\Omega(\Ysp_2),
  \label{rs:ra:Eq}\\
F_{\left(p_s+2\right)}&=&\frac{2}{\sqrt{N_s}}
d\left[h^{-1}_s(x, v, z)\right]\wedge\Omega(\Xsp)\wedge\Omega(\Ysp_1),
  \label{rs:sa:Eq}
}
where $\Omega(\Xsp)$, $\Omega(\Ysp_1)$, and $\Omega(\Ysp_2)$ 
are the volume $(p+1)$-form, $(p_s-p)$-form, and $(p_r-p)$-form, 
respectively: 
\Eqrsubl{rs:volume:Eq}{
\Omega(\Xsp)&=&\sqrt{-q}\,dx^0\wedge dx^1\wedge \cdots \wedge 
dx^p,\\
\Omega(\Ysp_1)&=&\sqrt{\gamma}\,dy^1\wedge dy^2\wedge \cdots \wedge 
dy^{p_s-p},\\
\Omega(\Ysp_2)&=&\sqrt{w}\,dv^1\wedge dv^2\wedge \cdots \wedge 
dv^{p_r-p}.
}
Here, $q$, $\gamma$, and $w$ are the determinants of the metrics $q_{\mu\nu}$, 
$\gamma_{ij}$, and $w_{mn}$, respectively.

First we consider the Einstein Eq.~(\ref{rs:Einstein:Eq}). 
Using the ansatz (\ref{rs:metric:Eq}), (\ref{rs:ansatz:Eq}), 
the Einstein equations are written by
\Eqrsubl{rs:cEinstein:Eq}{
&&\hspace{-0.5cm}R_{\mu\nu}(\Xsp)
-\frac{4}{N_r}h_r^{-1}D_{\mu}D_{\nu}h_r
-\frac{4}{N_s}h_s^{-1}D_{\mu}D_{\nu}h_s
+\frac{2}{N_r}\pd_{\mu}\ln h_r\left[\left(1-\frac{4}{N_r}\right)
\pd_{\nu}\ln h_r-\frac{4}{N_s}\pd_{\nu}\ln h_s\right]\nn\\
&&~~~~+\frac{2}{N_s}\pd_{\mu}\ln h_s\left[\left(1-\frac{4}{N_s}\right)
\pd_{\nu}\ln h_s-\frac{4}{N_r}\pd_{\nu}\ln h_r\right]\nn\\
&&~~~~-\frac{1}{2}q_{\mu\nu}h_r^{-4/N_r}h_s^{-4/N_s}
\left[a_rh_r^{-1}\left(h_s^{4/N_s}\lap_{\Ysp_1}h_r+\lap_{\Zsp}h_r\right)
+a_sh_s^{-1}\left(h_r^{4/N_r}\lap_{\Ysp_2}h_s+\lap_{\Zsp}h_s\right)\right]\nn\\
&&~~~~-\frac{1}{2}q_{\mu\nu}\left[a_r
h_r^{-1}\lap_{\Xsp}h_r-a_rq^{\rho\sigma}\pd_{\rho}\ln h_r
\left\{\left(1-\frac{4}{N_r}\right)
\pd_{\sigma}\ln h_r-\frac{4}{N_s}
\pd_{\sigma}\ln h_s\right\}\right.\nn\\
&&\left. ~~~~+a_sh_s^{-1}\lap_{\Xsp}h_s
-a_sq^{\rho\sigma}\pd_{\rho}\ln h_s\left\{\left(1-\frac{4}{N_s}\right)
\pd_{\sigma}\ln h_s-\frac{4}{N_r}\pd_{\sigma}\ln h_r\right\}\right]=0,
 \label{rs:cEinstein-mu:Eq}\\
&&\frac{2}{N_r}h_r^{-1}\left(\pd_{\mu}\pd_i h_r
+\frac{4}{N_s}\pd_{\mu}\ln h_s\pd_ih_r\right)=0,
 \label{rs:cEinstein-mi:Eq}\\
&&\frac{2}{N_s}h_s^{-1}\left(\pd_{\mu}\pd_m h_s
+\frac{4}{N_r}\pd_{\mu}\ln h_r\pd_mh_s\right)=0,
 \label{rs:cEinstein-mm:Eq}\\
&&\frac{2}{N_r}h_r^{-1}\pd_{\mu}\pd_a h_r
+\frac{2}{N_s}h_s^{-1}\pd_{\mu}\pd_a h_s=0,
 \label{rs:cEinstein-ma:Eq}\\
&&R_{ij}(\Ysp_1)-\frac{1}{2}h_r^{4/N_r}\gamma_{ij}\left[b_r
h_r^{-1}\lap_{\Xsp}h_r-b_rq^{\rho\sigma}\pd_{\rho}\ln h_r
\left\{\left(1-\frac{4}{N_r}\right)
\pd_{\sigma}\ln h_r-\frac{4}{N_s}
\pd_{\sigma}\ln h_s\right\}\right.\nn\\
&&\left. ~~~~+a_sh_s^{-1}\lap_{\Xsp}h_s
-a_sq^{\rho\sigma}\pd_{\rho}\ln h_s\left\{\left(1-\frac{4}{N_s}\right)
\pd_{\sigma}\ln h_s-\frac{4}{N_r}\pd_{\sigma}\ln h_r\right\}\right]\nn\\
&&~~~~-\frac{1}{2}\gamma_{ij}h_s^{-4/N_s}\left[b_rh_r^{-1}
\left(h_s^{4/N_s}\lap_{\Ysp_1}h_r+\lap_{\Zsp}h_r\right)
+a_sh_s^{-1}\left(h_r^{4/N_r}\lap_{\Ysp_2}h_s+\lap_{\Zsp}h_s\right)\right]=0,
 \label{rs:cEinstein-ij:Eq}\\
&&\frac{8}{N_rN_s(D-2)^2}\left[(p_r+1)(p_s+1)-(D-2)(p_r+p_s+2)
\right]\pd_i\ln h_r\pd_m\ln h_s=0\,,
 \label{rs:cEinstein-im:Eq}\\
&&R_{mn}(\Ysp_2)-\frac{1}{2}h_s^{4/N_s}w_{mn}\left[a_r
h_r^{-1}\lap_{\Xsp}h_r-a_rq^{\rho\sigma}\pd_{\rho}\ln h_r
\left\{\left(1-\frac{4}{N_r}\right)
\pd_{\sigma}\ln h_r-\frac{4}{N_s}
\pd_{\sigma}\ln h_s\right\}\right.\nn\\
&&\left. ~~~~+b_sh_s^{-1}\lap_{\Xsp}h_s
-b_sq^{\rho\sigma}\pd_{\rho}\ln h_s\left\{\left(1-\frac{4}{N_s}\right)
\pd_{\sigma}\ln h_s-\frac{4}{N_r}\pd_{\sigma}\ln h_r\right\}\right]\nn\\
&&~~~~-\frac{1}{2}w_{mn}h_r^{-4/N_r}\left[a_rh_r^{-1}
\left(h_s^{4/N_s}\lap_{\Ysp_1}h_r+\lap_{\Zsp}h_r\right)
+b_sh_s^{-1}\left(h_r^{4/N_r}\lap_{\Ysp_2}h_s+\lap_{\Zsp}h_s\right)
\right]=0,
 \label{rs:cEinstein-mn:Eq}\\
&&
R_{ab}(\Zsp)-\frac{1}{2}h_r^{4/N_r}h_s^{4/N_s}u_{ab}\left[b_r
h_r^{-1}\lap_{\Xsp}h_r-b_rq^{\rho\sigma}\pd_{\rho}\ln h_r
\left\{\left(1-\frac{4}{N_r}\right)
\pd_{\sigma}\ln h_r-\frac{4}{N_s}\pd_{\sigma}\ln h_s\right\}\right.\nn\\
&&\left. ~~~~+b_sh_s^{-1}\lap_{\Xsp}h_s
-b_sq^{\rho\sigma}\pd_{\rho}\ln h_s\left\{\left(1-\frac{4}{N_s}\right)
\pd_{\sigma}\ln h_s-\frac{4}{N_r}\pd_{\sigma}\ln h_r\right\}\right]\nn\\
&&~~~~-\frac{1}{2}u_{ab}\left[b_rh_r^{-1}
\left(h_s^{4/N_s}\lap_{\Ysp_1}h_r+\lap_{\Zsp}h_r\right)
+b_sh_s^{-1}\left(h_r^{4/N_r}\lap_{\Ysp_2}h_s+\lap_{\Zsp}h_s\right)
\right]=0,
  \label{rs:cEinstein-ab:Eq}
}
where $D_{\mu}$ is the covariant derivative with respect to
the metric $q_{\mu\nu}$, and $\triangle_{\Xsp}$, $\triangle_{\Ysp_1}$, 
$\triangle_{\Ysp_2}$, 
$\triangle_{\Zsp}$ the Laplace operators on 
$\Xsp$, $\Ysp_1$, $\Ysp_2$, $\Zsp$ spaces, 
$R_{\mu\nu}(\Xsp)$, $R_{ij}(\Ysp_1)$, $R_{mn}(\Ysp_2)$,
and $R_{ab}(\Zsp)$ are the Ricci tensors
constructed from the metrics $q_{\mu\nu}(\Xsp)$, $\gamma_{ij}(\Ysp_1)$,
$w_{mn}(\Ysp_2)$ and $u_{ab}(\Zsp)$, respectively, 
and we have used the intersection 
rule $\chi=0$. Here $\chi$ is defined by
\Eq{
\chi=p+1-\frac{\left(p_r+1\right)\left(p_s+1\right)}{D-2}
+\frac{1}{2}\epsilon_r\epsilon_sc_rc_s.
   \label{rs:ir:Eq}
}
The relation $\chi=0$ is consistent with the intersection rule 
\cite{Tseytlin:1996bh, Argurio:1997gt, Argurio:1998cp, Ohta:1997gw, 
Binetruy:2007tu, Maeda:2009zi, Minamitsuji:2010kb, Minamitsuji:2010uz}.

From Eqs.~(\ref{rs:cEinstein-mi:Eq}), (\ref{rs:cEinstein-mm:Eq}) 
and (\ref{rs:cEinstein-ma:Eq}), 
the functions $h_r$ and $h_s$ have to be of the form
\Eqrsubl{rs:warp:Eq}{
&&h_r= h_0(x)+h_1(y, z),~~~~h_s=h_s(v, z)\,,~~~~~~{\rm For}~~
\pd_{\mu}h_s=0\,,
  \label{rs:warp1:Eq}\\
&&h_r= h_r(y, z),~~~~h_s= k_0(x)+k_1(v, z)\,,~~~~~~{\rm For}~~
\pd_{\mu}h_r=0.
  \label{rs:warp2:Eq}  
}
Let us consider the case $\pd_{\mu}h_s=0$. 
The components of the Einstein Eqs.~(\ref{rs:cEinstein:Eq}) are rewritten as
\Eqrsubl{rs:c2Einstein:Eq}{
&&\hspace{-0.3cm}R_{\mu\nu}(\Xsp)
-\frac{2}{N_r}\left[2h_r^{-1}D_{\mu}D_{\nu}h_0
-\left(1-\frac{4}{N_r}\right)\pd_{\mu}\ln h_r
\pd_{\nu}\ln h_r\right]\nn\\
&&-\frac{1}{2}q_{\mu\nu}h_r^{-4/N_r}h_s^{-4/N_s}
\left[a_rh_r^{-1}\left(h_s^{4/N_s}\lap_{\Ysp_1}h_1+\lap_{\Zsp}h_1\right)
+a_sh_s^{-1}\left(h_r^{4/N_r}\lap_{\Ysp_2}h_s+\lap_{\Zsp}h_s\right)\right]\nn\\
&&-\frac{1}{2}a_rq_{\mu\nu}\left[
h_r^{-1}\lap_{\Xsp}h_0-\left(1-\frac{4}{N_r}\right)
q^{\rho\sigma}\pd_{\rho}\ln h_r\pd_{\sigma}\ln h_r\right]=0,
 \label{rs:c2Einstein-mu:Eq}\\
&&\hspace{-0.3cm}R_{ij}(\Ysp_1)-\frac{1}{2}b_rh_r^{4/N_r}\gamma_{ij}\left[
h_r^{-1}\lap_{\Xsp}h_0-\left(1-\frac{4}{N_r}\right)
q^{\rho\sigma}\pd_{\rho}\ln h_r\pd_{\sigma}\ln h_r\right]\nn\\
&&-\frac{1}{2}\gamma_{ij}h_s^{-4/N_s}\left[b_rh_r^{-1}
\left(h_s^{4/N_s}\lap_{\Ysp_1}h_1+\lap_{\Zsp}h_1\right)
+a_sh_s^{-1}\left(h_r^{4/N_r}\lap_{\Ysp_2}h_s+\lap_{\Zsp}h_s\right)\right]=0,
 \label{rs:c2Einstein-ij:Eq}\\
&&\hspace{-0.3cm}\frac{8}{N_rN_s(D-2)^2}\left[(p_r+1)(p_s+1)
-(D-2)(p_r+p_s+2)\right]\pd_i\ln h_r\pd_m\ln h_s=0\,,
 \label{rs:c2Einstein-im:Eq}\\
&&\hspace{-0.3cm}R_{mn}(\Ysp_2)-\frac{1}{2}a_rh_s^{4/N_s}w_{mn}\left[
h_r^{-1}\lap_{\Xsp}h_0-\left(1-\frac{4}{N_r}\right)
q^{\rho\sigma}\pd_{\rho}\ln h_r\pd_{\sigma}\ln h_r\right]\nn\\
&&-\frac{1}{2}w_{mn}h_r^{-4/N_r}\left[a_rh_r^{-1}
\left(h_s^{4/N_s}\lap_{\Ysp_1}h_1+\lap_{\Zsp}h_1\right)
+b_sh_s^{-1}\left(h_r^{4/N_r}\lap_{\Ysp_2}h_s+\lap_{\Zsp}h_s\right)\right]=0,
 \label{rs:c2Einstein-mn:Eq}\\
&&\hspace{-0.3cm}R_{ab}(\Zsp)-\frac{1}{2}b_rh_r^{4/N_r}h_s^{4/N_s}u_{ab}\left[
h_r^{-1}\lap_{\Xsp}h_0-\left(1-\frac{4}{N_r}\right)
q^{\rho\sigma}\pd_{\rho}\ln h_r\pd_{\sigma}\ln h_r\right]\nn\\
&&-\frac{1}{2}u_{ab}\left[b_rh_r^{-1}
\left(h_s^{4/N_s}\lap_{\Ysp_1}h_1+\lap_{\Zsp}h_1\right)
+b_sh_s^{-1}\left(h_r^{4/N_r}\lap_{\Ysp_2}h_s+\lap_{\Zsp}h_s\right)\right]=0.
  \label{rs:c2Einstein-ab:Eq}
}

Let us next consider the gauge field Eqs.~(\ref{rs:gauge-r:Eq}), 
(\ref{rs:gauge-s:Eq}).
Under the assumption (\ref{rs:ra:Eq}) and  
(\ref{rs:sa:Eq}), the field equations are written by 
\Eqrsubl{rs:gauge2:Eq}{
&&d\left[h_s^{4(\chi+1)/N_s}\pd_i h_r\left(\ast_{\Ysp_1}dy^i\right)
\wedge\Omega(\Zsp)+h_s^{4\chi/N_s}\pd_a h_r\left(\ast_{\Zsp}dz^a\right)
\wedge\Omega(\Ysp_1)\right]=0,
  \label{rs:gauge2-r:Eq}\\
&&d\left[h_r^{4(\chi+1)/N_r}\pd_m h_s\left(\ast_{\Ysp_2}dv^m\right)
\wedge\Omega(\Zsp)+h_r^{4\chi/N_r}\pd_a h_s\left(\ast_{\Zsp}dz^a\right)
\wedge\Omega(\Ysp_2)\right]=0,
  \label{rs:gauge2-s:Eq}
 }
where $\ast_{\Ysp_1}$, $\ast_{\Ysp_2}$, and $\ast_{\Zsp}$ 
denote the Hodge operator on $\Ysp_1$, $\Ysp_2$, and $\Zsp$, 
respectively, and $\chi$ is given by
\eqref{rs:ir:Eq}\,.
For $\chi=0$, the Eq.~(\ref{rs:gauge2-r:Eq}) thus reduces to
\Eq{
h_s\lap_{\Ysp_1}h_r+\lap_{\Zsp}h_r=0,
~~~\pd_{\mu}\pd_i h_r+\frac{4}{N_s}\pd_{\mu}\ln h_s\pd_ih_r=0,
~~~\pd_{\mu}\pd_a h_r=0,
  \label{rs:gauge3:Eq}
}
where $\triangle_{\Ysp_1}$, and $\triangle_{\Zsp}$ are 
the Laplace operators on the space of $\Ysp_1$, and 
$\Zsp$, respectively. On the other hand, 
it follows from (\ref{rs:gauge2-s:Eq}) that 
\Eq{
h_r\lap_{\Ysp_2}h_s+\lap_{\Zsp}h_s=0,
~~~\pd_{\mu}\pd_m h_s+\frac{4}{N_r}\pd_{\mu}\ln h_r\pd_mh_s=0, 
~~~\pd_{\mu}\pd_a h_s=0\,,
   \label{rs:gauge4:Eq}
}
where $\triangle_{\Ysp_2}$ is 
the Laplace operator on the space of $\Ysp_2$.

Finally we consider the scalar field equation. 
Substituting Eq.~(\ref{rs:warp:Eq}), the ansatz (\ref{rs:ansatz:Eq}), and 
the intersection rule $\chi=0$ into Eq.~(\ref{rs:scalar:Eq}), we have
\Eqr{
&&\frac{\epsilon_rc_r}{N_r}h_r^{4/N_r}h_s^{4/N_s}\left[
h_r^{-1}\lap_{\Xsp}h_r-\left(1-\frac{4}{N_r}\right)
q^{\rho\sigma}\pd_{\rho}\ln h_r\pd_{\sigma}\ln h_r
+\frac{4}{N_s}q^{\rho\sigma}\pd_{\rho}\ln h_r\pd_{\sigma}\ln h_s\right]\nn\\
&&+\frac{\epsilon_sc_s}{N_s}h_r^{4/N_r}h_s^{4/N_s}\left[
h_s^{-1}\lap_{\Xsp}h_s-\left(1-\frac{4}{N_s}\right)
q^{\rho\sigma}\pd_{\rho}\ln h_s\pd_{\sigma}\ln h_s
+\frac{4}{N_r}q^{\rho\sigma}\pd_{\rho}\ln h_r\pd_{\sigma}\ln h_s\right]\nn\\
&&~~~~~+\frac{\epsilon_rc_r}{N_r}h_r^{-1}\left(h_s^{4/N_s}\lap_{\Ysp_1}h_r
+\lap_{\Zsp}h_r\right)
+\frac{\epsilon_sc_s}{N_s}h_s^{-1}\left(h_r^{4/N_r}\lap_{\Ysp_2}h_s
+\lap_{\Zsp}h_s\right)=0.
  \label{rs:scalar-e:Eq}
}
If we set $\pd_{\mu}h_s=0$, 
the functions $h_r$ and $h_s$ satisfy the equations
\Eqrsubl{rs:scalar-s:Eq}{
&&h_r\triangle_{\Xsp}h_0
-\left(1-\frac{4}{N_r}\right)q^{\rho\sigma}\pd_{\rho}h_0\pd_{\sigma}h_0=0,
~~~h_s^{4/N_s}\lap_{\Ysp_1}h_1+\lap_{\Zsp}h_1=0,
   \label{rs:scalar-s1:Eq}\\
&&h_r^{4/N_r}\lap_{\Ysp_2}h_s+\lap_{\Zsp}h_s=0.
   \label{rs:scalar-s2:Eq}
}

Combining these, the field equations reduce to
\Eqrsubl{rs:solution1:Eq}{
&&R_{\mu\nu}(\Xsp)=0,~~~~R_{ij}(\Ysp_1)=0,~~~~
R_{mn}(\Ysp_2)=0,~~~~R_{ab}(\Zsp)=0,
   \label{rs:Ricci:Eq}\\
&&h_r=h_0(x)+h_1(y, z),~~~~h_s=h_s(v, z),~~~~\pd_ih_r\pd_mh_s=0\,,
   \label{rs:h:Eq}\\
&&D_{\mu}D_{\nu}h_0=0, ~~~\left(1-\frac{4}{N_r}\right)\pd_{\mu}h_0
\pd_{\nu}h_0=0,
~~~h_s^{4/N_s}\lap_{\Ysp_1}h_1+\triangle_{\Zsp}h_1=0,
   \label{rs:warp1-1:Eq}\\
&&h_r^{4/N_r}\lap_{\Ysp_2}h_s+\triangle_{\Zsp}h_s=0.
   \label{rs:warp1-2:Eq}
 }
The function $h_r$ can depend on the 
coordinate $x^{\mu}$ only if $N_r=4$\,.
We can also choose the solution in which the
$p_s$-brane part depends on $x^{\mu}$. Then, we have 
\Eqrsubl{rs:solution2:Eq}{
&&R_{\mu\nu}(\Xsp)=0,~~~~R_{ij}(\Ysp_1)=0,~~~~
R_{mn}(\Ysp_2)=0,~~~~R_{ab}(\Zsp)=0,
   \label{rs:Ricci2:Eq}\\
&&h_r=h_r(y, z),~~~~h_s=k_0(x)+k_1(v, z),~~~~\pd_ih_r\pd_mh_s=0\,,
   \label{rs:h2:Eq}\\
&&D_{\mu}D_{\nu}k_0=0, ~~~\left(1-\frac{4}{N_s}\right)\pd_{\mu}k_0
\pd_{\nu}k_0=0,
~~~h_r^{4/N_r}\lap_{\Ysp_2}k_1+\triangle_{\Zsp}k_1=0,
   \label{rs:warp2-1:Eq}\\
&&h_s^{4/N_s}\lap_{\Ysp_1}h_r+\triangle_{\Zsp}h_r=0.
   \label{rs:warp2-2:Eq}
 }
It is clear that there is no solution for $k_0(x)$
such as $\partial_{\mu}h_s\ne 0$ unless $N_s=4$.
If $F_{\left(p_r+2\right)}=0$ and $F_{\left(p_s+2\right)}=0$,
the functions $h_1$ and $k_1$ become trivial, 
and the $D$-dimensional spacetime is no longer warped~
\cite{Kodama:2005fz, Kodama:2005cz}. 
Moreover, the \Eqref{rs:h:Eq} 
$\pd_ih_r\pd_mh_s=0$ implies the following two cases :\\
(i) Two branes are delocalized, which are localized only 
along the overall transverse directions. \\ 
(ii) One brane is completely localized on the other brane 
which is localized only along the overall transverse directions.

As a special example, we consider the case
\Eqrsubl{rs:flat:Eq}{
&&q_{\mu\nu}=\eta_{\mu\nu}\,,~~~\gamma_{ij}=\delta_{ij}\,,~~~
w_{mn}=\delta_{mn}\,,~~~u_{ab}=\delta_{ab}\,,\\
&&N_r=N_s=4,~~~h_s=h_s(z)\,,
 }
where $\eta_{\mu\nu}$ is the $(p+1)$-dimensional
Minkowski metric and $\delta_{ij}$, $\delta_{mn}$, $\delta_{ab}$ are
the $(p_s-p)$-, $(p_r-p)$- and $(D+p-p_r-p_s-1)$-dimensional 
Euclidean metrics, respectively. This means that both branes have 
physically the same total amount of charge. 
Since the function $h_0$ obeys the equation $\pd_{\mu}\pd_{\nu}h_0=0$, 
we can easily get the solution
\Eqr{
h_0(x)=A_{\mu}x^{\mu}+B\,,
} 
where $A_{\mu}$ and $B$ are constants.
On the other hand, the functions $h_1$ and $h_s$ satisfy 
the coupled partial differential equations 
\Eq{
h_s\lap_{\Ysp_1}h_1+\lap_{\Zsp}h_1=0,~~~~\lap_{\Zsp}h_s=0\,.
    \label{rs:hrhs:Eq}
} 
The harmonic function $h_s$ that satisfies the second differential 
equation in (\ref{rs:warp1-2:Eq}) has the form 
\Eqr{
h_s=1+\sum_l\frac{M_l}{|\bm z-\bm z_\ell|^{d_z-2}}, 
}
where $d_z\equiv D+p-p_r-p_s-1$, and 
$z_l^a$ are locations of the $l$-th $p_s$-brane
with charge $M_l$.
We will mainly discuss the case in which the $p_s$-branes 
coincide at the same location in the overall transverse directions.
Now we choose the following form of the harmonic function $h_s$:
\Eq{
h_s(z)=\frac{M}{|\bm z-\bm z_0|^{d_z-2}}, 
}
where $M$ is constant, 
$\bm z_0$ is the location of the stack of $p_s$-branes. 
It is not so easy to find solutions for the harmonic function 
$h_1$ in the case where each of the $p_s$-branes are located
at different points along the $z$-directions. 

If the dimensionality of the overall transverse space is  
$d_z\neq 2$ and $d_z\neq 4$, 
the equation (\ref{rs:hrhs:Eq}) can be solved as \cite{Youm:1999zs}
\Eqr{
h_1(y, z)=1+\frac{M_\ell}{\left[|\bm y-\bm y_\ell|^2
+\frac{4M}{\left(4-d_z\right)^2}
|\bm z-\bm z_0|^{4-d_z}\right]^{\frac{1}{2}
(p_s-p-1+\frac{d_z}{4-d_z})}}\,,
   \label{rs:h_r solution2:Eq}
}
where $M_\ell$ is constant.
Hence, the functions $h_r$ and $h_s$ can be written explicitly as 
\Eqrsubl{rs:solutions1:Eq}{
h_r(x, y, z)&=&A_{\mu}x^{\mu}+B
+\sum_{\ell}\frac{M_\ell}{\left[|\bm y-\bm y_\ell|^2
+\frac{4M}{(4-d_z)^2}|\bm z-\bm z_0|^{4-d_z}\right]^
{\frac{1}{2}(p_s-p-1+\frac{d_z}{4-d_z})}},
 \label{rs:solution-r:Eq}\\
h_s(z)&=&\frac{M}{|\bm z-\bm z_0|^{d_z-2}},
 \label{rs:solution-s:Eq}
}
where $A_{\mu}$, $B$, $M_\ell$ and $M$ are constant parameters,
and $\bm y_\ell$ and $\bm z_0$ are
constant vectors representing the positions of the branes.
Since the functions coincide, the locations of the 
branes will also coincide. 
There are curvature singularities at $h_r=0$ or $h_s=0$ 
in the $D$-dimensional metric (\ref{rs:metric:Eq}).
Moreover, we have a singularity at $\bm z=\bm z_0$ unless the 
scalar field is trivial.

In the case of $d_z=2$, we have 
\Eqrsubl{rs:solutions1-2:Eq}{
h_r(x, y, z)&=&A_{\mu}x^{\mu}+B
+\sum_{\ell}\frac{M_\ell}{\left[|\bm y-\bm y_\ell|^2
+M|\bm z-\bm z_0|^2\right]^
{\frac{1}{2}(p_s-p)}},
 \label{rs:hr-2:Eq}\\
h_s(z)&=&M\ln|\bm z-\bm z_0|\,.
 \label{rs:hs-2:Eq}
}
For $d_z=4$, the solution of \Eqref{rs:hrhs:Eq} can be written by 
\Eqrsubl{rs:solutions1-3:Eq}{
h_r(x, y, z)&=&A_{\mu}x^{\mu}+B
+\sum_{\ell}M_\ell\left[|\bm y-\bm y_\ell|^2
-(p_s-p)M\ln |\bm z-\bm z_0|\right]\,,
 \label{rs:hr-3:Eq}\\
h_s(z)&=&\frac{M}{|\bm z-\bm z_0|^2}\,.
 \label{rs:hs-3:Eq}
}
We note that the solutions \eqref{rs:solutions1-2:Eq} and 
\eqref{rs:solutions1-3:Eq} have curvature singularities not only  
at $h_r=0$ but also at the infinity due to the logarithmic spatial 
dependence of the metric.
There is also a singularity at $\bm z=\bm z_0$ if the scalar field is 
nontrivial. 

One can easily get the solution for 
$\pd_{\mu}h_r=0$ and $\pd_{\mu}h_s\ne 0$ if the roles of $\Ysp_1$ and 
$\Ysp_2$ are exchanged. The solution of field equations for 
$d_z\neq2$ and $d_z\neq4$ are written as
\Eqrsubl{rs:solutions2:Eq}{
h_s(x, v, z)&=&A_{\mu}x^{\mu}+B
+\sum_{\ell}\frac{M_\ell}{\left[|\bm v-\bm v_\ell|^2
+\frac{4M}{(4-d_z)^2}|\bm z-\bm z_0|^{4-d_z}\right]^
{\frac{1}{2}(p_r-p-1+\frac{d_z}{4-d_z})}},
 \label{rs:solution2-s:Eq}\\
h_r(z)&=&\frac{M}{|\bm z-\bm z_0|^{d_z-2}}\,,
 \label{rs:solution2-r:Eq}
}
where $\bm v_\ell$ is constant vector representing the position of 
the $p_s$-branes.
 
For $d_z=2$ and $d_z=4$, the harmonic functions have 
logarithmic spatial dependence like \eqref{rs:solutions1-2:Eq} and 
\eqref{rs:solutions1-3:Eq}.

Let us briefly summarize the intersecting rules 
in eleven-dimensional supergravity and in ten-dimensional string theory.
For the M-branes in eleven-dimensional supergravity, there
is 4-form field strength without scalar field, 
the intersection rule $\chi=0$ gives 
\Eq{
p=\frac{(p_r+1)(p_s+1)}{9}-1\,,
}
where $p$ is the number of overlapping dimensions of the $p_r$ and 
$p_s$ branes. Hence, we find the intersections involving the M2 and M5-branes
\cite{Argurio:1997gt,Ohta:1997gw, Maeda:2009zi, Minamitsuji:2010kb}
\Eq{
{\rm M2}\cap {\rm M2}=0,~~~~{\rm M2}\cap {\rm M5}=1,~~~~
{\rm M5}\cap {\rm M5}=3.~~~~
   \label{rs:int M:Eq}
}

On the other hand, for the ten-dimensional string theory, 
the couplings to the scalar field for the RR-charged D-branes are 
expressed as
\Eq{
\epsilon_rc_r=\frac{1}{2}\left(3-p_r\right),~~~~
\epsilon_sc_s=\frac{1}{2}\left(3-p_s\right).
}
Then, the condition $\chi=0$ gives
\Eq{
p=\frac{1}{2}\left(p_r+p_s-4\right).
}
The intersection rules for the D-branes are thus given 
by~\cite{Argurio:1997gt,Ohta:1997gw, Minamitsuji:2010kb}
\Eq{
{\rm D}p_r\cap {\rm D}p_s = \frac{1}{2}(p_r+p_s)-2.
   \label{rs:int D:Eq}
}
Let us finally consider the intersections for NS-branes. 
The constants $c_r$ for fundamental string (F1) and solitonic 
5-brane are $\epsilon_1c_1=-1$
(for F1) and $\epsilon_5c_5=1$ (for NS5), respectively.
The intersection rules involving the
F1 and NS5-branes are \cite{Argurio:1997gt,Ohta:1997gw, Minamitsuji:2010kb}
\Eqrsubl{rs:int DNS:Eq}{
&&
{\rm F1}\cap {\rm NS5}=1,~~~~
{\rm NS5}\cap {\rm NS5}=3,
   \label{rs:int NS:Eq}\\
&&{\rm F1}\cap {\rm D}\bar{p}=0,\\
&&{\rm D}\bar{p}\cap {\rm NS}5=\bar{p}-1,~~~~1\le\bar{p}\le 6.
}
We cannot construct solutions for the F1-F1 and D0-NS5 intersecting 
brane systems because the numbers of space dimensions for
each pairwise overlap are negative by the intersection rule.

\subsection{The intersection of $p$-brane and KK monopole system}
Now we discuss the dynamical intersecting brane solutions 
including KK-monopoles in $D$ dimensions. 
The $p$-branes we have described above carry a charge
in $D$-dimensions. The Kaluza-Klein (KK) charged objects are also in general 
branes living in the compactified space-time and carrying a electric or 
magnetic charge with respect to the 2-form field strength generated by 
dimensional reduction in $D$-dimensional theory. 
If the space-time dimension is $D$ after compactification on one direction, 
we can obtain an electric KK 0-brane and a magnetic KK $(D-4)$-brane. 
In the $(D+1)$-dimensional uncompactified space-time, 
these two objects correspond to configurations where the only 
non-trivial field is the metric and which are identified, 
to a KK-wave and a KK-monopole, respectively. 
The metric in the uncompactified space has nontrivial 
off-diagonal terms necessarily. In this section, we discuss 
the KK-monopole and summarize those objects. 
We extend our brane solutions to the cases with waves next section.

We will start from the $D$-dimensional theory, 
for which the action in the Einstein frame contains the metric $g_{MN}$,
the dilaton $\phi$, 
and the antisymmetric tensor field of rank $(p+2)$, $F_{(p+2)}$
\Eq{
S=\frac{1}{2\kappa^2}\int \left[R\ast{\bf 1}
 -\frac{1}{2}\ast d\phi \wedge d\phi
 -\frac{1}{2\cdot (p+2)!}\e^{\epsilon c\phi}
 \ast F_{(p+2)}\wedge F_{(p+2)}\right],
\label{k:action:Eq}
}
where $\kappa^2$ is the $D$-dimensional gravitational constant,
$\ast$ is the Hodge operator in the $D$-dimensional space-time,
$F_{(p+2)}$ is the $(p+2)$-form field strength,
and $c$, $\epsilon$ are constants given by
\Eqrsubl{k:parameters:Eq}{
c^2&=&N-\frac{2(p+1)(D-p-3)}{D-2},
   \label{k:c:Eq}\\
\epsilon&=&\left\{
\begin{array}{cc}
 +&~{\rm if}~~p-{\rm brane~is~electric}\\
 -&~~~{\rm if}~~p-{\rm brane~is~magnetic}\,.
\end{array} \right.
 \label{k:epsilon:Eq}
   }
Here $N$ is a constant. 
The field strength $F_{(p+2)}$ is given by the
$(p+1)$-form gauge potential $A_{(p+1)}$
\Eq{F_{(p+2)}=dA_{(p+1)}.}

The field equations are given by
\Eqrsubl{k:field equations:Eq}{
&&\hspace{-1cm}R_{MN}=\frac{1}{2}\pd_M\phi \pd_N \phi\nn\\
&&\hspace{-1cm}~~~~+\frac{1}{2\cdot (p+2)!}\e^{\epsilon c\phi}
\left[(p+2)F_{MA_2\cdots A_{p+2}} {F_N}^{A_2\cdots A_{p+2}}
-\frac{p+1}{D-2}g_{MN} F^2_{(p+2)}\right],
   \label{k:Einstein:Eq}\\
&&\hspace{-1cm}d\ast d\phi-\frac{\epsilon c}{2\cdot (p+2)!}
\e^{\epsilon c\phi}\ast F_{(p+2)}\wedge F_{(p+2)}=0,
   \label{k:scalar:Eq}\\
&&\hspace{-1cm}d\left[\e^{\epsilon c\phi}\ast F_{(p+2)}\right]=0.
   \label{k:gauge:Eq}
}

We assume that the $D$-dimensional metric
takes the form
\Eqr{
ds^2&=&h^{a}(x, y, z)q_{\mu\nu}(\Xsp)dx^{\mu}dx^{\nu}
+h^{b}(x, y, z)\left[\gamma_{ij}
(\Ysp)dy^idy^j\right.\nn\\
&&\left.+h_k(x, z)u_{ab}(\Zsp)dz^adz^b
+h_k^{-1}(x, z)\left(dv+A_adz^a\right)^2\right], 
 \label{k:metric:Eq}
}
where $q_{\mu\nu}$ is the $(p+1)$-dimensional metric which
depends only on the $(p+1)$-dimensional coordinates $x^{\mu}$, 
$\gamma_{ij}$ is the $(D-p-d_z-2)$-dimensional metric which
depends only on the $(D-p-d_z-2)$-dimensional coordinates $y^i$, 
and finally $u_{ab}$ is the $d_z$-dimensional metric which
depends only on the $d_z$-dimensional coordinates $z^a$. 
The parameters $a$ and $b$ in the metric 
(\ref{k:metric:Eq}) are given by 
\Eq{
a=-\frac{4(D-p-3)}{N(D-2)},~~~~~b=\frac{4(p+1)}{N(D-2)}.
 \label{k:paremeter:Eq}
}
The brane configuration is given in Table \ref{table_B}\,. 
\begin{table}[h]
\caption{\baselineskip 14pt
Intersections of $p$-brane and KK-monopole in the metric
\eqref{k:metric:Eq}. }
\label{KKDp}
{\scriptsize
\begin{center}
\begin{tabular}{|c||c|c|c|c|c|c|c|c|c|c|c|c|}
\hline
Case&&0&1& $\cdots$ & $p$ & $p+1$ & $\cdots$ & $D-d_z-2$ & $D-d_z-1$ 
& $D-d_z$ & $\cdots$ & $D-1$
\\
\hline
&$p$ & $\circ$ & $\circ$ & $\circ$ & $\circ$ & & &&
 &&& 
\\
\cline{3-13}
$p$-KK
& KK & $\circ$ & $\circ$ & $\circ$ & $\circ$ & $\circ$ &
 $\circ$ & $\circ$ & & $A_1$ & $\cdots$ & $A_{d_z}$ 
\\
\cline{3-13}
&$x^N$ & $t$ & $x^1$ & $\cdots$ & $x^p$ & $y^1$ & $\cdots$ & $y^{D-p-d_z-2}$ 
& $v$ & $z^1$ & $\cdots$ & $z^{d_z}$
\\
\hline
\end{tabular}
\end{center}
}
\label{table_B}
\end{table} 
The $D$-dimensional metric (\ref{k:metric:Eq}) implies that 
the solutions are characterized by 
two functions, $h$ and $h_k$,
which depend 
on the coordinates transverse to the brane as well as 
the world volume coordinate.

We also assume that the scalar field $\phi$ and the gauge field 
strength $F_{\left(p+2\right)}$ are given by
\Eqrsubl{k:ansatz:Eq}{
\e^{\phi}&=&h^{2\epsilon c/N}\,,
  \label{k:ansatz for scalar:Eq}\\
F_{\left(p+2\right)}&=&\frac{2}{\sqrt{N}}
d\left[h^{-1}(x, y, z)\right]\wedge\Omega(\Xsp),
  \label{k:ansatz for gauge:Eq}
}
where $\Omega(\Xsp)$ 
denotes the volume $(p+1)$-form 
\Eqr{
\Omega(\Xsp)=\sqrt{-q}\,dx^0\wedge dx^1\wedge \cdots \wedge 
dx^p\,.
   \label{k:volume:Eq}
}
Here, $q$ is the determinant of the metric $q_{\mu\nu}$.

First we consider the Einstein Eq.~(\ref{k:Einstein:Eq}). 
Using the assumptions (\ref{k:metric:Eq}), (\ref{k:ansatz:Eq}) and 
setting $d_z=3$, 
the Einstein equations are given by
\Eqrsubl{k:cEinstein:Eq}{
&&\hspace{-0.3cm}
R_{\mu\nu}(\Xsp)-\frac{4}{N}h^{-1}D_{\mu}D_{\nu}h
-h^{-1}_kD_{\mu}D_{\nu}h_k
+\frac{2}{N}\left(1-\frac{4}{N}\right)h^{-1}\pd_{\mu}\ln h\pd_{\nu}\ln h\nn\\
&&~~~~~-\frac{a}{2}q_{\mu\nu}\left[h^{-1}\lap_{\Xsp}h
+q^{\rho\sigma}\pd_{\rho}\ln h\left\{
\left(\frac{4}{N}-1\right)\pd_{\sigma}\ln h
+\pd_{\sigma}\ln h_k\right\}\right]\nn\\
&&~~~~~-\frac{a}{2}q_{\mu\nu}h^{-1-\frac{4}{N}}\left(\lap_{\Ysp}h
+h_k^{-1}\lap_{\Zsp}h\right)
-\frac{2}{N}\left(\pd_{\mu}\ln h\pd_{\nu}\ln h_k
+\pd_{\mu}\ln h_k\pd_{\nu}\ln h\right)=0\,,
 \label{k:cEinstein-mn:Eq}\\
&&\hspace{-0.3cm}
\frac{2}{N}\left(h^{-1}\pd_{\mu}\pd_i h+\pd_{\mu}\ln h_k
\pd_i\ln h\right)=0\,,
 \label{k:cEinstein-mi:Eq}\\
&&\hspace{-0.3cm}
\frac{2}{N}h^{-1}\pd_{\mu}\pd_a h+\frac{1}{2}h_k^{-1}\pd_{\mu}\pd_a h_k=0\,,
 \label{k:cEinstein-ma:Eq}\\
&&\hspace{-0.3cm}
R_{ij}(\Ysp)-\frac{b}{2}h^{4/N}\gamma_{ij}\left[h^{-1}\lap_{\Xsp}h
+q^{\rho\sigma}\pd_{\rho}\ln h\left\{\left(\frac{4}{N}-1\right)
\pd_{\sigma}\ln h+\pd_{\sigma}\ln h_k\right\}\right]\nn\\
&&~~~~~-\frac{b}{2}h^{-1}\gamma_{ij}\left(\lap_{\Ysp}h
+h_k^{-1}\lap_{\Zsp}h\right)=0\,,
 \label{k:cEinstein-ij:Eq}\\
&&\hspace{-0.3cm}
R_{ab}(\Zsp)-\frac{b}{2}h^{4/N}h_k\left(u_{ab}+h_k^{-2}A_aA_b\right)
\left[h^{-1}\lap_{\Xsp}h
+q^{\rho\sigma}\pd_{\rho}\ln h\left\{\left(\frac{4}{N}-1\right)
\pd_{\sigma}\ln h+\pd_{\sigma}\ln h_k\right\}\right]\nn\\
&&~~-\frac{1}{2}h^{4/N}h_k\left(u_{ab}-h_k^{-2}A_aA_b\right)
\left(h_k^{-1}\lap_{\Xsp}h_k
+\frac{8}{N}q^{\rho\sigma}\pd_{\rho}\ln h_k\pd_{\sigma}\ln h\right)\nn\\
&&~~-\frac{b}{2}h^{-1}h_k\left(u_{ab}+h_k^{-2}A_aA_b\right)
\left(\lap_{\Ysp}h+h_k^{-1}\lap_{\Zsp}h\right)
-\frac{1}{2}h_k^{-1}\left(u_{ab}-h_k^{-2}A_aA_b\right)\lap_{\Zsp}h_k=0\,,
  \label{k:cEinstein-ab:Eq}  \\
&&\hspace{-0.3cm}-bh^{4/N}h_k^{-1}\left[h^{-1}\lap_{\Xsp}h
+q^{\rho\sigma}\pd_{\rho}\ln h\left\{\left(\frac{4}{N}-1\right)
\pd_{\sigma}\ln h+\pd_{\sigma}\ln h_k\right\}\right]\nn\\
&&~~~~~+h^{4/N}h_k^{-1}\left(h_k^{-1}\lap_{\Xsp}h_k
+\frac{8}{N}q^{\rho\sigma}\pd_{\rho}\ln h_k\pd_{\sigma}\ln h
\right)\nn\\
&&~~~~~-b\left(hh_k\right)^{-1}
\left(\lap_{\Ysp}h+h_k^{-1}\lap_{\Zsp}h\right)
+h_k^{-3}\lap_{\Zsp}h_k=0,
 \label{k:cEinstein-mm:Eq}
}
where we assumed $dh_k=\ast_{\Zsp}dA_{(1)}$, 
$D_{\mu}$ is the covariant derivative with respect to the metric $q_{\mu\nu}$, 
and $\triangle_{\Xsp}$, $\triangle_{\Ysp}$, $\lap_{\Zsp}$ are
the Laplace operators on 
$\Xsp$, $\Ysp$, $\Zsp$ space, and $R_{\mu\nu}(\Xsp)$, 
$R_{ij}(\Ysp)$, $R_{ab}(\Zsp)$ are the Ricci tensors  
associated with the metrics $q_{\mu\nu}(\Xsp)$, 
$\gamma_{ij}(\Ysp)$, $u_{ab}(\Zsp)$, 
respectively. Here $\ast_{\Zsp}$ denotes the Hodge operator on $\Zsp$. 
We see from Eqs.~(\ref{k:cEinstein-mi:Eq}), and (\ref{k:cEinstein-ma:Eq})
that the warp factors $h$ must 
take the form
\Eqr{
&&h= h_0(x)+h_1(y, z),~~~~h_k=h_k(z)\,,~~~~~~{\rm For}~~
\pd_{\mu}h_k=0\,.
  \label{k:warp:Eq}
}
If we set $\pd_{\mu}h_k=0$ and $N=4$, 
the components of the Einstein Eqs.~(\ref{k:cEinstein:Eq}) are rewritten as
\Eqrsubl{k:c2Einstein:Eq}{
&&\hspace{-1cm}R_{\mu\nu}(\Xsp)-h^{-1}D_{\mu}D_{\nu}h_0
-\frac{a}{2}q_{\mu\nu}h^{-1}\left[\lap_{\Xsp}h_0
+h^{-1}\left(\lap_{\Ysp}h_1+h_k^{-1}\lap_{\Zsp}h_1\right)\right]=0\,,
 \label{k:c2Einstein-mn:Eq}\\
&&\hspace{-1cm}R_{ij}(\Ysp)-\frac{b}{2}\gamma_{ij}\left[\lap_{\Xsp}h_0
+h^{-1}\left(\lap_{\Ysp}h_1+h_k^{-1}\lap_{\Zsp}h_1\right)\right]=0\,,
 \label{k:c2Einstein-ij:Eq}\\
&&\hspace{-1cm}
R_{ab}(\Zsp)-\frac{b}{2}h_k\left(u_{ab}+h_k^{-2}A_aA_b\right)
\left[\lap_{\Xsp}h_0+h^{-1}\left(\lap_{\Ysp}h_1+h_k^{-1}\lap_{\Zsp}h_1\right)
\right]\nn\\
&&-\frac{1}{2}h_k^{-1}\left(u_{ab}-h_k^{-2}A_aA_b\right)\lap_{\Zsp}h_k=0\,,
  \label{k:c2Einstein-ab:Eq}\\
&&\hspace{-1cm}
b\left(hh_k\right)^{-1}
\left(h\lap_{\Xsp}h_0+\lap_{\Ysp}h_1+h_k^{-1}\lap_{\Zsp}h_1\right)
-h_k^{-3}\lap_{\Zsp}h_k=0\,.
 \label{k:c2Einstein-mm:Eq}
}

Next we consider the gauge field Eqs.~(\ref{k:gauge:Eq}).
Under the assumption (\ref{k:ansatz for gauge:Eq}) and 
setting $d_z=3$, we find 
\Eq{
d\left[h_k\pd_i h\left(\ast_{\Ysp}dy^i\right)
\wedge\Omega(\Zsp)+\pd_a h\left(\ast_{\Zsp}dz^a\right)
\wedge\Omega(\Ysp)\right]\wedge dv=0,
  \label{k:gauge2:Eq}
 }
where $\ast_{\Ysp}$, $\ast_{\Zsp}$ 
denote the Hodge operator on 
$\Ysp$, $\Zsp$, respectively, 
and $\Omega(\Ysp)$, $\Omega(\Zsp)$ denote the volume $(D-p-5)$-, 
3-form respectively:
\Eqrsubl{k:volume z:Eq}{
\Omega(\Ysp)&=&\sqrt{\gamma}\,dy^1\wedge dy^2\wedge \cdots \wedge 
dy^{D-p-5}\,,\\
\Omega(\Zsp)&=&\sqrt{u}\,dz^1\wedge dz^2\wedge dz^3\,.
}
Then, the Eq.~(\ref{k:gauge2:Eq}) gives
\Eq{
h_k\lap_{\Ysp}h+\lap_{\Zsp}h=0,
~~~\pd_{\mu}\pd_i h+\pd_{\mu}h_k\pd_ih=0,
~~~\pd_{\mu}\pd_a h=0,
   \label{k:gauge3:Eq}
}
where $\triangle_{\Ysp}$, $\triangle_{\Zsp}$ are 
the Laplace operators on the space of $\Ysp$, 
$\Zsp$, respectively.

Finally we should consider the scalar field equation.
Substituting Eqs.~(\ref{k:ansatz:Eq}) and (\ref{k:warp:Eq}) 
into Eq.~(\ref{k:scalar:Eq}), we obtain
\Eqr{
&&\frac{2\epsilon c}{N}h^{-b+4/N}\left[
h^{-1}\lap_{\Xsp}h_0+q^{\rho\sigma}\pd_{\rho}\ln h\pd_{\sigma}\ln h_k
-\left(1-\frac{4}{N}\right)
q^{\rho\sigma}\pd_{\rho}\ln h\pd_{\sigma}\ln h\right.\nn\\
&&~~~~~\left.
+h^{-1-4/N}\left(\lap_{\Ysp}h_1+h_k^{-1}\lap_{\Zsp}h_1\right)\right]=0.
  \label{k:scalar equation:Eq}
}
Thus, for $N=4$, the warp factor $h$ should satisfy the equations
\Eq{
\lap_{\Xsp}h_0=0,~~~~\pd_{\mu}h_0\pd_{\nu}h_k=0,~~~~
\lap_{\Ysp}h_1+h_k^{-1}\lap_{\Zsp}h_1=0\,.
   \label{k:scalar solution:Eq}
}
Combining these, we find that these field equations lead to
\Eqrsubl{k:solution1:Eq}{
&&R_{\mu\nu}(\Xsp)=0,~~~~R_{ij}(\Ysp)=0,~~~~R_{ab}(\Zsp)=0,~~~~
d_z=3,
   \label{k:Ricci:Eq}\\
&&h=h_0(x)+h_1(z),
   \label{k:h:Eq}\\
&&D_{\mu}D_{\nu}h_0=0, ~~\pd_{\mu}h_0\pd_{\nu}h_k=0,~~~
\pd_{\mu}h_k\pd_ih_1=0,~~~\left(1-\frac{4}{N}\right)
q^{\rho\sigma}\pd_{\rho}\ln h\pd_{\sigma}\ln h=0,
   \label{k:warp1-1:Eq}\\
&&h_k^{4/N}\lap_{\Ysp}h_1+\triangle_{\Zsp}h_1=0\,,
~~~\triangle_{\Zsp}h_1=0\,.
   \label{k:warp1-2:Eq}
 }
The function $h$ can depend on the 
coordinate $x$ only if $N=4$\,. 
If $F_{\left(p+2\right)}=0$, 
the function $h_1$ becomes trivial.

As a special example, let us consider the case
\Eqrsubl{k:flat metric:Eq}{
&&q_{\mu\nu}=\eta_{\mu\nu}\,,~~~\gamma_{ij}=\delta_{ij}\,,~~~
u_{ab}=\delta_{ab}\,,\\ 
&&N=4,~~~h_k=h_k(z)\,,
 }
where $\eta_{\mu\nu}$ is the $(p+1)$-dimensional
Minkowski metric and $\delta_{ij}$, $\delta_{ab}$ are
the $(D-5-p)$-, three-dimensional Euclidean metrics,
respectively. 
The solution for $h$ and $h_k$ can be
obtained explicitly as
\Eqrsubl{k:solutions1:Eq}{
h(x, y, z)&=&A_{\mu}x^{\mu}+\tilde{c}
+\sum_{\ell}\frac{M_\ell}{\left[|\bm y-\bm y_\ell|^2
+4M|\bm z-\bm z_0|\right]^
{\frac{1}{2}(D-p-3)}},
 \label{k:solution-r:Eq}\\
h_k(z)&=&\frac{M}{|\bm z-\bm z_0|},
 \label{k:solution-s:Eq}
}
where $A_{\mu}$, $\tilde{c}$, $M_\ell$ and $M$ are constant parameters,
and $\bm y_\ell$ and $\bm z_0$ are
constant vectors representing the positions of the branes.
Since the functions coincide, the locations of the 
branes will also coincide. 
The $D$-dimensional metric \eqref{k:metric:Eq} exists for $h>0$ 
and has curvature singularities at $h=0$.


\subsection{The intersection of $p$-brane and plane wave system}
Let us next consider the solutions with the plane wave.  
One can obtain the electric 0-brane and the magnetic 
$(D-5)$-brane solutions in $(D-1)$-dimensional spacetime because 
the dimensional reduction generates the Kaluza-Klein charge 
in the 2-form field strengths. 
After we lift up those solutions by one dimension, 
we obtain the plane wave solutions in $D$-dimensions.  
We briefly discuss the plane wave solution in this section.
 
Now we look for solution whose spacetime metric has the form
\Eqr{
ds^2&=&h^{a_w}(t, z)\left[-dt^2+dx^2+\left\{h_w(t, y, z)-1\right\}
\left(dt-dx\right)^2+\gamma_{ij}(\Ysp)dy^idy^j\right]\nn\\
&&+h^{b_w}(t, z)u_{ab}(\Zsp)dz^adz^b, 
 \label{p:metric:Eq}
}
where $\gamma_{ij}$ is the $(p-1)$-dimensional metric which
depends only on the $(p-1)$-dimensional coordinates $y^i$, 
and finally $u_{ab}$ is the $(D-p-1)$-dimensional metric which
depends only on the $(D-p-1)$-dimensional coordinates $z^a$. 
The parameters $a$ and $b$ in the metric 
(\ref{p:metric:Eq}) are given by 
\Eq{
a_w=-\frac{4(D-p-3)}{N(D-2)},~~~~~b_w=\frac{4(p+1)}{N(D-2)}.
 \label{p:paremeter:Eq}
}
We show the brane configuration in Table \ref{table_C}\,.
\begin{table}[h]
\caption{\baselineskip 14pt
Intersections of $p$-brane and plane wave in the metric
\eqref{p:metric:Eq}. }
\label{WDp}
{\scriptsize
\begin{center}
\begin{tabular}{|c||c|c|c|c|c|c|c|c|c|}
\hline
Case&&0& $1$ & 2& $\cdots$ & $p$ & $p+1$ & $\cdots$ & $D-1$
\\
\hline
 &$p$ & $\circ$ & $\circ$ & $\circ$ & $\circ$ & $\circ$ 
 &&& 
\\
\cline{3-10}
$p$-W & W & $\circ$ &  &  & & && & 
\\
\cline{3-10}
&$x^N$ & $t$ & $x$ & $y^1$ & $\cdots$ & $y^{p-1}$ &
$z^1$ & $\cdots$ & $z^{D-p-1}$
\\
\hline
\end{tabular}
\end{center}
}
\label{table_C}
\end{table} 

We also assume that the 
scalar field $\phi$ and the gauge field strength $F_{\left(p+2\right)}$ 
are given by
\Eqrsubl{p:ansatz:Eq}{
\e^{\phi}&=&h^{2\epsilon c/N}\,
   \label{p:ansatz for scalar:Eq}\\
F_{\left(p+2\right)}&=&\frac{2}{\sqrt{N}}
d\left[h^{-1}(t, z)\right]\wedge dt\wedge dx\wedge\Omega(\Ysp)\,,
  \label{p:ansatz for gauge:Eq}
}
where $c$, $\epsilon$ are constants given by
(\ref{k:parameters:Eq}), and 
$\Omega(\Ysp)$ denotes the volume $(p-1)$-form :
\Eq{
\Omega(\Ysp)=\sqrt{\gamma}\,dy^1\wedge dy^2\wedge \cdots \wedge 
dy^{p-1}\,.
   \label{p:volume y:Eq}
}
Here, $\gamma$ is the determinant of the metric $\gamma_{ij}$. 

First, we consider the Einstein Eqs.~(\ref{k:Einstein:Eq}). 
Using the assumptions (\ref{p:metric:Eq}) and (\ref{p:ansatz:Eq}), 
the Einstein equations are given by
\Eqrsubl{p:cEinstein:Eq}{
&&\hspace{-0.5cm}
\left[a_w(2-h_w)h_w+\frac{8}{N}\right]h^{-1}\pd_t^2h
+(2-h_w)\pd_t^2h_w+\lap_{\Ysp}h_w+h^{-\frac{4}{N}}\lap_{\Zsp}h_w
+\frac{1}{2}(2-h_w)h^{-1-\frac{4}{N}}\lap_{\Zsp}h\nn\\
&&~~~~+\frac{4}{N}\left(\frac{4}{N}-1\right)\left(\pd_t\ln h\right)^2
+a_wh_w(2-h_w)\left[\left(\frac{4}{N}-1\right)\pd_t\ln h+\pd_t\ln h_w\right]
\pd_t\ln h\nn\\
&&~~~~+\frac{4}{N}(2-h_w)\pd_t\ln h\pd_th_w=0\,,
 \label{p:cEinstein-tt:Eq}\\
&&\pd_t\pd_i h_w+\frac{4}{N}\pd_t\ln h\pd_ih_w=0\,,
 \label{p:cEinstein-ti:Eq}\\
&&\pd_t\pd_a h_w+\frac{4}{N}h^{-1}\pd_t\pd_a h=0\,,
 \label{p:cEinstein-ta:Eq}\\
&&a_wh_w^2h^{-1}\pd_t^2h
+h_w\pd_t^2h_w-\lap_{\Ysp}h_w-h^{-4/N}\lap_{\Zsp}h_w
-a_wh_wh^{-1-\frac{4}{N}}\lap_{\Zsp}h\nn\\
&&~~~~-a_wh_w^2\left[\left(\frac{4}{N}-1\right)\pd_t\ln h
+\pd_t\ln h_w\right]\pd_t\ln h=0\,,
 \label{p:cEinstein-xx:Eq}\\
&&R_{ij}(\Ysp)+\frac{a_w}{2}h_w\gamma_{ij}\left[h^{-1}\pd_t^2h
+\left\{\left(\frac{4}{N}-1\right)\pd_t\ln h
+\pd_t\ln h_w\right\}\pd_t\ln h\right]\nn\\
&&~~~~-\frac{a_w}{2}h^{-1-\frac{4}{N}}\gamma_{ij}\lap_{\Zsp}h=0\,,
 \label{p:cEinstein-ij:Eq}\\
&&R_{ab}(\Zsp)+\frac{a_w}{2}h^{4/N}h_wu_{ab}\left[h^{-1}
\pd_t^2h+\left\{\left(\frac{4}{N}-1\right)\pd_t\ln h
+\pd_t\ln h_w\right\}\pd_t\ln h\right]\nn\\
&&~~~~-\frac{b_w}{2}h^{-1}u_{ab}\lap_{\Zsp}h=0\,,
  \label{p:cEinstein-ab:Eq}
}
where $\triangle_{\Ysp}$, $\lap_{\Zsp}$ are
the Laplace operators on $\Ysp$, $\Zsp$ space, and
$R_{ij}(\Ysp)$, $R_{ab}(\Zsp)$ are the Ricci tensors 
associated with the metrics $\gamma_{ij}(\Ysp)$, $u_{ab}(\Zsp)$, 
respectively.

We see from Eqs.~(\ref{p:cEinstein-ti:Eq}), and (\ref{p:cEinstein-ta:Eq})
that the warp factors $h$ must take the form
\Eqrsubl{p:warp:Eq}{
&&h= h_0(t)+h_1(z),~~~~h_w=h_w(y, z)\,,~~~~~~{\rm For}~~
\pd_{t}h_w=0\,,
  \label{p:warp1:Eq}\\
&&h= h(z),~~~~h_w= k_0(t)+k_1(y, z)\,,~~~~~~{\rm For}~~
\pd_th=0.
  \label{p:warp2:Eq}  
}
If we set $\pd_th=0$, 
the components of the Einstein Eqs.~(\ref{p:cEinstein:Eq}) are rewritten as
\Eqrsubl{p:c2Einstein:Eq}{
&&(2-h_w)\pd_t^2k_0+\lap_{\Ysp}k_1+h^{-1}\lap_{\Zsp}k_1
+\frac{1}{2}(2-h_w)h^{-2}\lap_{\Zsp}h=0\,,
 \label{p:c2Einstein-tt:Eq}\\
&&h_w\pd_t^2k_0-\lap_{\Ysp}k_1-h^{-1}\lap_{\Zsp}k_1
-a_wh_wh^{-2}\lap_{\Zsp}h=0\,,
 \label{p:c2Einstein-xx:Eq}\\
&&R_{ij}(\Ysp)-\frac{a_w}{2}\gamma_{ij}h^{-2}\lap_{\Zsp}h=0\,,
 \label{p:c2Einstein-ij:Eq}\\
&&R_{ab}(\Zsp)-\frac{b_w}{2}u_{ab}h^{-1}\lap_{\Zsp}h=0\,.
  \label{p:c2Einstein-ab:Eq}
}

Now we consider the gauge field Eqs.~(\ref{k:gauge:Eq}).
Under the assumption (\ref{p:ansatz for gauge:Eq}), we find 
\Eq{
d\left[\pd_a h\left(\ast_{\Zsp}dz^a\right)\right]=0\,,
  \label{p:gauge2:Eq}
 }
where $\ast_{\Zsp}$ 
denote the Hodge operator on $\Zsp$\,.

Then, the Eq.~(\ref{p:gauge2:Eq}) gives
\Eq{
\lap_{\Zsp}h=0,
~~~\pd_t\pd_a h=0\,.
   \label{p:gauge3:Eq}
}

Finally we should consider the scalar field equation.
Substituting Eqs.~(\ref{p:ansatz:Eq}) and (\ref{p:warp:Eq}) 
into Eq.~(\ref{k:scalar:Eq}), we obtain
\Eqr{
&&\hspace{-0.8cm}\frac{2\epsilon c}{N}h^{-b+4/N-1}h_w\left[
\pd_t^2h_0
+\left\{\left(\frac{4}{N}-1\right)\pd_t\ln h+\pd_t \ln h_w\right\}
\pd_th_0-h^{-4/N}h_w^{-1}\lap_{\Zsp}h_1\right]=0\,,
  \label{p:scalar equation:Eq}
}
where we used \eqref{p:warp1:Eq}\,.
Thus, the warp factor $h$ should satisfy the equations
\Eq{
\pd_t^2h_0=0,~~~~\left(\frac{4}{N}-1\right)\pd_t\ln h+\pd_t \ln h_w=0\,,
~~~~\lap_{\Zsp}h_1=0\,.
   \label{p:scalar solution:Eq}
}
Combining these, we find that these field equations lead to 
\Eqrsubl{p:solution1:Eq}{
&&R_{ij}(\Ysp)=0,~~~~R_{ab}(\Zsp)=0,
   \label{p:Ricci:Eq}\\
&&h=h_0(t)+h_1(z),
   \label{p:h:Eq}\\
&&\pd_t^2h_0=0, ~~~\pd_th_0\pd_th_w=0,
~~~\left(\frac{4}{N}-1\right)\pd_t\ln h+\pd_t\ln h_w=0,
~~~\triangle_{\Zsp}h_1=0,
   \label{p:warp1-1:Eq}\\
&&h^{4/N}\lap_{\Ysp}h_w+\triangle_{\Zsp}h_w=0.
   \label{p:warp1-2:Eq}
 }
The function $h$ can depend on the 
coordinate $t$ only if $N=4$\,.
We can also choose the solution in which the
function $h_w$ depends on $t$. Then, we have 
\Eqrsubl{p:solution2:Eq}{
&&R_{ij}(\Ysp)=0,~~~~R_{ab}(\Zsp)=0,
   \label{p:Ricci2:Eq}\\
&&h=h(z),~~~~h_w=k_0(t)+k_1(y, z),
   \label{p:h2:Eq}\\
&&\pd_t^2k_0=0, 
~~~h^{4/N}\lap_{\Ysp}k_1+\triangle_{\Zsp}k_1=0\,.
   \label{p:warp2-1:Eq}
 }
If $F_{\left(p+2\right)}=0$, 
the functions $h_1$ become trivial. 

As a special example, let us consider the case
\Eqrsubl{p:flat metric:Eq}{
&&\gamma_{ij}=\delta_{ij}\,,~~~u_{ab}=\delta_{ab}\,,\\
&&N=4,~~~h=h(z)\,,
 }
where $\delta_{ij}$, $\delta_{ab}$ are
the $(p-1)$-, $(D-p-1)$-dimensional Euclidean metrics,
respectively. This physically means that both branes have the same 
total amount of charge. 
The solution for $h$ and $h_w$ can be
obtained explicitly as
\Eqrsubl{p:solutions1:Eq}{
h_w(t, z)&=&A_{\mu}x^{\mu}+B
+\sum_{\ell}\frac{M_\ell}{\left[|\bm y-\bm y_\ell|^2
+\frac{4M}{(D-p-5)^2}|\bm z-\bm z_0|^{-D+p+5}\right]^
{\frac{1}{2}(p-2-\frac{D-p-1}{D-p-5})}},
 \label{p:solution-r:Eq}\\
h(z)&=&\frac{M}{|\bm z-\bm z_0|^{D-p-3}},
 \label{p:solution-s:Eq}
}
where $A_{\mu}$, $B$, $M_\ell$ and $M$ are constant parameters,
and $\bm y_\ell$ and $\bm z_0$ are
constant vectors representing the positions of the branes.
Since the functions coincide, the locations of the 
branes will also coincide. 
Even if the near-brane structure is regular, 
we expect another type of singularity may appear at $h_w=0$ due to 
the presence of the time dependence. 
For $c\ne 0$, the $D$-dimensional spacetime has curvature singularities
where $\bm z=\bm z_0$ since the scalar field diverges there.
%

\section{The intersection of dynamical branes in eleven-dimensional theory}
  \label{sec:eleven}
In this section, we apply the above 
solutions to eleven-dimensional theory. 
In this theory, we have a 4-form field strength and no dilaton. 
The 4-form gives rise to 2- and 5-branes, called, respectively, M2 and M5. 
We also obtain the KK-wave and KK-monopole in eleven dimensions. 
In particular, KK-wave is called "M-wave" in eleven-dimensional theory
\cite{FigueroaO'Farrill:1999tx, FigueroaO'Farrill:2002tc, Maeda:2009zi}.
The 11-dimensional action which contains 
the metric $g_{MN}$, and 4-form field strength $F_{(4)}$ is given by
\Eq{
S=\frac{1}{2\kappa^2}\int \left[R\ast{\bf 1}
 -\frac{1}{2\cdot 4!}\ast F_{(4)}\wedge F_{(4)}\right],
\label{m:action:Eq}
}
where $\kappa^2$ is the eleven-dimensional gravitational constant,
$\ast$ is the Hodge operator in the eleven-dimensional space-time. 
The field strengths $F_{(4)}$ is given by the 3-form gauge potential 
\Eq{
F_{(4)}=dC_{(3)}\,.}

The field equations are given by
\Eqrsubl{m:field equations:Eq}{
&&\hspace{-1cm}R_{MN}=\frac{1}{2\cdot 4!}\left[4F_{MABC} {F_N}^{ABC}
-\frac{1}{2}g_{MN} F^2_{(4)}\right],
   \label{m:Einstein:Eq}\\
&&\hspace{-1cm}d\left[\ast F_{(4)}\right]=0\,,~~~~~dF_{(4)}=0\,.
   \label{m:gauge:Eq}
}

In the following, we discuss the dynamical brane solution 
for all the possible combinations of intersecting brane pairs 
in the eleven-dimensional theory. 

\subsection{The intersection of two M2-branes}

Let us consider the solution of two M2-branes. 
We assume that the eleven-dimensional metric is written by
\Eqr{
ds^2&=&h^{-2/3}_2(t, y, z)\bar{h}_2^{-2/3}(z)
\left[-dt^2+h_2(t, y, z)\gamma_{ij}(\Ysp_1)dy^idy^j\right.\nn\\
&&\left.+\bar{h}_2(z)w_{mn}(\Ysp_2)dv^mdv^n
+h_2(t, y, z)\bar{h}_2(z)u_{ab}(\Zsp)dz^adz^b\right], 
 \label{m2:metric:Eq}
}
where $\gamma_{ij}$ is the two-dimensional metric which
depends only on the two-dimensional coordinates $y^i$, 
$w_{mn}$ is the two-dimensional metric which
depends only on the two-dimensional coordinates $v^m$, 
and finally $u_{ab}$ is the six-dimensional metric which
depends only on the six-dimensional coordinates $z^a$. 

We also assume that the gauge field 
strength $F_{\left(4\right)}$ is given by
\Eqr{
F_{\left(4\right)}= d\left[h_2^{-1}(t, y, z)
\,dt\wedge \Omega(\Ysp_2)+\bar{h}_2^{-1}(z)
\,dt\wedge \Omega(\Ysp_1)\right],
  \label{m2:ansatz:Eq}
}
where $\Omega(\Ysp_1)$ and $\Omega(\Ysp_2)$ 
denote the volume two-form and two-form, respectively 
\Eqrsubl{m2:volume:Eq}{
\Omega(\Ysp_1)&=&\sqrt{\gamma}\,dy^1\wedge dy^2\,,
   \label{m2:volume y1:Eq}\\
\Omega(\Ysp_2)&=&\sqrt{w}\,dv^1\wedge dv^2\,.
   \label{m2:volume y2:Eq}
}
Here, $\gamma$ and $w$ are the determinant of the metric 
$\gamma_{ij}$, and $w_{mn}$, respectively.

In terms of ansatz for fields \eqref{m2:metric:Eq} and 
\eqref{m2:ansatz:Eq}, the field equations lead to
\Eqrsubl{m2:solution1:Eq}{
&&R_{ij}(\Ysp_1)=0,~~~~R_{mn}(\Ysp_2)=0,~~~~
R_{ab}(\Zsp)=0,
   \label{m2:Ricci:Eq}\\
&&h_2=h_0(t)+h_1(y, z)\,,~~~
\pd_t^2h_0=0, ~~~\bar{h}_2\lap_{\Ysp_1}h_1+\triangle_{\Zsp}h_1=0\,,
~~~\triangle_{\Zsp}\bar{h}_2=0\,,
   \label{m2:warp:Eq}
 }
where $\triangle_{\Ysp_1}$, $\lap_{\Zsp}$ are
the Laplace operators on $\Ysp_1$, $\Zsp$ space, and  
$R_{ij}(\Ysp_1)$, $R_{mn}(\Ysp_2)$, $R_{ab}(\Zsp)$ are the Ricci tensors
associated with the metrics 
$\gamma_{ij}(\Ysp_1)$, $w_{mn}(\Ysp_2)$, $u_{ab}(\Zsp)$, 
respectively. 
As a special example, let us consider the case
\Eq{
\gamma_{ij}=\delta_{ij}\,,~~~w_{mn}=\delta_{mn}\,,~~~
u_{ab}=\delta_{ab}\,,
 \label{m2:flat metric:Eq}
 }
where $\delta_{ij}$, $\delta_{mn}$, $\delta_{ab}$ are
the two-, two-, six-dimensional Euclidean metrics,
respectively. 
The solution for $h_2$ and $\bar{h}_2$ can be
obtained explicitly as
\Eqrsubl{m2:solutions1:Eq}{
h_2(t, y, z)&=&\bar{c}t+\tilde{c}
+\sum_{\ell}M_\ell\left[|\bm y-\bm y_\ell|^2
+M|\bm z-\bm z_0|^{-2}\right],
 \label{m2:solution-r:Eq}\\
\bar{h}_2(z)&=&\frac{M}{|\bm z-\bm z_0|^4},
 \label{m2:solution-s:Eq}
}
where $\bar{c}$, $\tilde{c}$, $M_\ell$ and $M$ are constant parameters,
and $\bm y_\ell$ and $\bm z_0$ are
constant vectors representing the positions of the branes.
Since the functions coincide, the locations of the 
branes will also coincide. 
If we delocalize along $n$ of the overall transverse directions, 
the harmonic functions take the following form:
\Eqrsubl{m2:solutions2:Eq}{
h_2(t, y, z)&=&\bar{c}t+\tilde{c}
+\sum_{\ell}\frac{M_\ell}{\left[|\bm y-\bm y_\ell|^2
+\frac{4M}{(n-2)^2}|\bm z-\bm z_0|^{n-2}\right]^{2/(n-2)}},
 \label{m2:solution-r2:Eq}\\
\bar{h}_2(z)&=&\frac{M}{|\bm z-\bm z_0|^{4-n}}\,.
 \label{m2:solution-s2:Eq}
}

\subsection{Intersecting M2- and M5-branes}
Next we consider the solution of M2-M5 branes. 
We assume that the eleven-dimensional metric is written by
\Eqr{
ds^2&=&h^{-2/3}_2(x, y, z)h_5^{-1/3}(z)
\left[q_{\mu\nu}(\Xsp)
dx^{\mu}dx^{\nu}+h_2(x, y, z)\gamma_{ij}(\Ysp)dy^idy^j\right.\nn\\
&&\left.+h_5(z)dv^2+h_2(x, y, z)h_5(z)u_{ab}(\Zsp)dz^adz^b\right], 
 \label{m25:metric:Eq}
}
where $q_{\mu\nu}$ is the two-dimensional metric which
depends only on the two-dimensional coordinates $x^{\mu}$, 
$\gamma_{ij}$ is the four-dimensional metric which
depends only on the four-dimensional coordinates $y^i$, 
and finally $u_{ab}$ is the four-dimensional metric which
depends only on the four-dimensional coordinates $z^a$. 

We also assume that the gauge field 
strength $F_{\left(4\right)}$ is given by
\Eqr{
F_{\left(4\right)}&=& d\left[h_2^{-1}(x, y, z)\right]\wedge
\,\Omega(\Xsp)\wedge dv+\ast \left[dh_5^{-1}(z)
\,\Omega(\Xsp)\wedge \Omega(\Ysp)\right],
  \label{m25:ansatz:Eq}
}
where $\Omega(\Xsp)$ and $\Omega(\Ysp)$ 
denote the volume two-form and four-form, respectively 
\Eqrsubl{m25:volume:Eq}{
\Omega(\Xsp)&=&\sqrt{-q}\,dx^0\wedge dx^1 \,,
   \label{m25:volume x:Eq}\\
\Omega(\Ysp)&=&\sqrt{\gamma}\,dy^1\wedge dy^2\wedge dy^3\wedge dy^4\,.
   \label{m25:volume y:Eq}
}
Here, $q$ and $\gamma$ are the determinant of the metric $q_{\mu\nu}$ and 
$\gamma_{ij}$, respectively.

In terms of ansatz for fields \eqref{m25:metric:Eq} and 
\eqref{m25:ansatz:Eq}, the field equations lead to
\Eqrsubl{m25:solution1:Eq}{
&&R_{\mu\nu}(\Xsp)=0,~~~~R_{ij}(\Ysp)=0,~~~~R_{ab}(\Zsp)=0,
   \label{m25:Ricci:Eq}\\
&&h_2=h_0(x)+h_1(y, z),
   \label{m25:h:Eq}\\
&&D_{\mu}D_{\nu}h_0=0, 
~~~h_5\lap_{\Ysp}h_1+\triangle_{\Zsp}h_1=0\,,
~~~\triangle_{\Zsp}h_5=0\,,
   \label{m25:warp:Eq}
 }
where $D_{\mu}$ is the covariant derivative constructed by the metric 
$q_{\mu\nu}$, and $\triangle_{\Ysp}$, $\lap_{\Zsp}$ are
the Laplace operators on $\Ysp$, $\Zsp$ space, and $R_{\mu\nu}(\Xsp)$, 
$R_{ij}(\Ysp)$, $R_{ab}(\Zsp)$ are the Ricci tensors
associated with the metrics $q_{\mu\nu}(\Xsp)$, 
$\gamma_{ij}(\Ysp)$, $u_{ab}(\Zsp)$, 
respectively. 
As a special example, let us consider the case
\Eq{
q_{\mu\nu}=\eta_{\mu\nu}\,,~~~
\gamma_{ij}=\delta_{ij}\,,~~~u_{ab}=\delta_{ab}\,,
 \label{m25:flat metric:Eq}
 }
where $\eta_{\mu\nu}$ is the two-dimensional
Minkowski metric and $\delta_{ij}$, $\delta_{ab}$ are
the four-, four-dimensional Euclidean metrics,
respectively. 
The solution for $h_2$ and $h_5$ can be
obtained explicitly as
\Eqrsubl{m25:solutions1:Eq}{
h_2(x, y, z)&=&c_{\mu}x^{\mu}+\tilde{c}
+\sum_{\ell}M_\ell\left[|\bm y-\bm y_\ell|^2
-4M\ln|\bm z-\bm z_0|\right],
 \label{m25:solution-r:Eq}\\
h_5(z)&=&\frac{M}{|\bm z-\bm z_0|^2},
 \label{m25:solution-s:Eq}
}
where $c_{\mu}$, $\tilde{c}$, $M_\ell$ and $M$ are constant parameters,
and $\bm y_\ell$ and $\bm z_0$ are
constant vectors representing the positions of the branes.

\subsection{The intersection of two M5-branes}

Let us consider the solution of two M5-branes. 
We assume that the eleven-dimensional metric is written by
\Eqr{
ds^2&=&h^{-1/3}_5(x, y, z)\bar{h}_5^{-1/3}(z)
\left[q_{\mu\nu}(\Xsp)
dx^{\mu}dx^{\nu}+h_5(x, y, z)\gamma_{ij}(\Ysp_1)dy^idy^j\right.\nn\\
&&\left.+\bar{h}_5(z)w_{mn}(\Ysp_2)dv^mdv^n
+h_5(x, y, z)\bar{h}_5(z)u_{ab}(\Zsp)dz^adz^b\right], 
 \label{m5:metric:Eq}
}
where $q_{\mu\nu}$ is the four-dimensional metric which
depends only on the four-dimensional coordinates $x^{\mu}$, 
$\gamma_{ij}$ is the two-dimensional metric which
depends only on the two-dimensional coordinates $y^i$, 
$w_{mn}$ is the two-dimensional metric which
depends only on the two-dimensional coordinates $v^m$, 
and finally $u_{ab}$ is the three-dimensional metric which
depends only on the three-dimensional coordinates $z^a$. 

We also assume that the gauge field 
strength $F_{\left(4\right)}$ is given by
\Eqr{
F_{\left(4\right)}&=& \ast d\left[h_5^{-1}(x, y, z)
\,\Omega(\Xsp)\wedge \Omega(\Ysp_2)+\bar{h}_5^{-1}(z)
\,\Omega(\Xsp)\wedge \Omega(\Ysp_1)\right],
  \label{m5:ansatz:Eq}
}
where $\Omega(\Xsp)$, $\Omega(\Ysp_1)$ and $\Omega(\Ysp_2)$ 
denote the volume four-form, two-form and two-form, respectively 
\Eqrsubl{m5:volume:Eq}{
\Omega(\Xsp)&=&\sqrt{-q}\,dx^0\wedge dx^1 \wedge dx^2\wedge dx^{3}\,,
   \label{m5:volume x:Eq}\\
\Omega(\Ysp_1)&=&\sqrt{\gamma}\,dy^1\wedge dy^2\,,
   \label{m5:volume y1:Eq}\\
\Omega(\Ysp_2)&=&\sqrt{w}\,dv^1\wedge dv^2\,.
   \label{m5:volume y2:Eq}
}
Here, $q$, $\gamma$ and $w$ are the determinant of the metric $q_{\mu\nu}$, 
$\gamma_{ij}$, and $w_{mn}$, respectively.

In terms of ansatz for fields \eqref{m5:metric:Eq} and 
\eqref{m5:ansatz:Eq}, the field equations lead to
\Eqrsubl{m5:solution1:Eq}{
&&R_{\mu\nu}(\Xsp)=0\,,~~~~R_{ij}(\Ysp_1)=0\,,~~~~R_{mn}(\Ysp_2)=0\,,~~~~
R_{ab}(\Zsp)=0,
   \label{m5:Ricci:Eq}\\
&&h_5=h_0(x)+h_1(y, z)\,,
   \label{m5:h:Eq}\\
&&D_{\mu}D_{\nu}h_0=0\,, 
~~~\bar{h}_5\lap_{\Ysp_1}h_1+\triangle_{\Zsp}h_1=0\,,
~~~\triangle_{\Zsp}\bar{h}_5=0\,,
   \label{m5:warp:Eq}
 }
where $D_{\mu}$ is the covariant derivative constructed by the metric 
$q_{\mu\nu}$, and $\triangle_{\Ysp_1}$, $\lap_{\Zsp}$ are
the Laplace operators on 
$\Ysp_1$, $\Zsp$ space, and $R_{\mu\nu}(\Xsp)$, $R_{ij}(\Ysp_1)$, 
$R_{mn}(\Ysp_2)$, $R_{ab}(\Zsp)$ are the Ricci tensors
associated with the metrics $q_{\mu\nu}(\Xsp)$, 
$\gamma_{ij}(\Ysp_1)$, $w_{mn}(\Ysp_2)$, $u_{ab}(\Zsp)$, 
respectively. 
As a special example, let us consider the case
\Eq{
q_{\mu\nu}=\eta_{\mu\nu}\,,~~~\gamma_{ij}=\delta_{ij}\,,~~~
w_{mn}=\delta_{mn}\,,~~~u_{ab}=\delta_{ab}\,,
 \label{m5:flat metric:Eq}
 }
where $\eta_{\mu\nu}$ is the four-dimensional
Minkowski metric and $\delta_{ij}$, $\delta_{mn}$, $\delta_{ab}$ are
the two-, two-, three-dimensional Euclidean metrics,
respectively. 
The solution for $h_5$ and $\bar{h}_5$ can be
obtained explicitly as
\Eqrsubl{m5:solutions1:Eq}{
h_2(x, y, z)&=&c_{\mu}x^{\mu}+\tilde{c}
+\sum_{\ell}\frac{M_\ell}{\left[|\bm y-\bm y_\ell|^2
+4M|\bm z-\bm z_0|\right]^2},
 \label{m5:solution-r:Eq}\\
h_5(z)&=&\frac{M}{|\bm z-\bm z_0|},
 \label{m5:solution-s:Eq}
}
where $c_{\mu}$, $\tilde{c}$, $M_\ell$ and $M$ are constant parameters,
and $\bm y_\ell$ and $\bm z_0$ are
constant vectors representing the positions of the branes.

\subsection{The intersection of M2-brane and one Kaluza-Klein monopole}
Now we discuss the KK-monopole in the transverse space of M2-brane. 
We assume that the eleven-dimensional metric takes the form
\Eqr{
ds^2&=&h_2^{-2/3}(x, y, z)q_{\mu\nu}(\Xsp)dx^{\mu}dx^{\nu}
+h_2^{1/3}(x, y, z)\left[\gamma_{ij}(\Ysp)dy^idy^j\right.\nn\\
&&\left.+h_k(z)u_{ab}(\Zsp)dz^adz^b
+h_k^{-1}(z)\left(dv+A_adz^a\right)^2\right], 
 \label{kkM2:metric:Eq}
}
where $q_{\mu\nu}$ is the three-dimensional metric which
depends only on the three-dimensional coordinates $x^{\mu}$, 
$\gamma_{ij}$ is the four-dimensional metric which
depends only on the four-dimensional coordinates $y^i$, 
and finally $u_{ab}$ is the three-dimensional metric which
depends only on the three-dimensional coordinates $z^a$. 

We also assume that the gauge field 
strength $F_{\left(4\right)}$ is given by
\Eqr{
F_{\left(4\right)}&=&d\left[h_2^{-1}(x, y, z)\right]\wedge\Omega(\Xsp),
  \label{kkM2:ansatz:Eq}
}
where $\Omega(\Xsp)$ denotes the volume three-form 
\Eqr{
\Omega(\Xsp)=\sqrt{-q}\,dx^0\wedge dx^1\wedge dx^2\,.
   \label{kKM2:volume:Eq}
}
Here, $q$ is the determinant of the metric $q_{\mu\nu}$.

In terms of ansatz for fields \eqref{kkM2:metric:Eq} and 
\eqref{kkM2:ansatz:Eq}, the field equations lead to
\Eqrsubl{kkM2:solution1:Eq}{
&&R_{\mu\nu}(\Xsp)=0,~~~~R_{ij}(\Ysp)=0,~~~~R_{ab}(\Zsp)=0,
   \label{kkM2:Ricci:Eq}\\
&&h_2=h_0(x)+h_1(y, z),~~~~dh_k=\ast_{\Zsp}dA\,,
   \label{kkM2:h:Eq}\\
&&D_{\mu}D_{\nu}h_0=0, ~~~~h_k\lap_{\Ysp}h_1+\triangle_{\Zsp}h_1=0\,,~~~
\triangle_{\Zsp}h_k=0\,,
   \label{kkM2:warp1-2:Eq}
 }
where $D_{\mu}$ is the covariant derivative with respect to
the metric $q_{\mu\nu}$, and 
$\ast_{\Zsp}$ denotes the Hodge operator on $\Zsp$, 
and $\triangle_{\Ysp}$, $\lap_{\Zsp}$ are
the Laplace operators on 
$\Xsp$, $\Ysp$, $\Zsp$ space, and $R_{\mu\nu}(\Xsp)$, 
$R_{ij}(\Ysp)$, $R_{ab}(\Zsp)$ are the Ricci tensors  
associated with the metrics $q_{\mu\nu}(\Xsp)$, 
$\gamma_{ij}(\Ysp)$, $u_{ab}(\Zsp)$, respectively. 

As a special example, let us consider the case
\Eq{
q_{\mu\nu}=\eta_{\mu\nu}\,,~~~\gamma_{ij}=\delta_{ij}\,,
~~~u_{ab}=\delta_{ab}\,,
 \label{kkM2:flat metric:Eq}
 }
where $\eta_{\mu\nu}$ is the three-dimensional
Minkowski metric and $\delta_{ij}$, $\delta_{ab}$ are
the four-, three-dimensional Euclidean metrics,
respectively. 
The solution for $h_2$ and $h_k$ can be
obtained explicitly as
\Eqrsubl{kkM2:solutions:Eq}{
h_2(x, y, z)&=&c_{\mu}x^{\mu}+\tilde{c}
+\sum_{\ell}\frac{M_\ell}{\left[|\bm y-\bm y_\ell|^2
+4M|\bm z-\bm z_0|\right]^3},
 \label{kkM2:solution-r:Eq}\\
h_k(z)&=&\frac{M}{|\bm z-\bm z_0|}\,,
 \label{kkM2:solution-s:Eq}
}
where $c_{\mu}$, $\tilde{c}$, $M_\ell$ and $M$ are constant parameters,
and $\bm y_\ell$ and $\bm z_0$ are
constant vectors representing the positions of the branes.
Since the functions coincide, the locations of the 
branes will also coincide.

\subsection{The intersection of M5-brane and one Kaluza-Klein monopole}
In this subsection, 
we discuss the KK-monopole in the transverse space of the M5-brane. 
We assume that the eleven-dimensional metric takes the form
\Eqr{
ds^2&=&h_5^{-1/3}(x, y, z)q_{\mu\nu}(\Xsp)dx^{\mu}dx^{\nu}
+h_5^{2/3}(x, y, z)\left[dy^2\right.\nn\\
&&\left.+h_k(z)u_{ab}(\Zsp)dz^adz^b
+h_k^{-1}(z)\left(dv+A_adz^a\right)^2\right], 
 \label{kkM5:metric:Eq}
}
where $q_{\mu\nu}$ is the six-dimensional metric which
depends only on the six-dimensional coordinates $x^{\mu}$, 
and finally $u_{ab}$ is the three-dimensional metric which
depends only on the three-dimensional coordinates $z^a$. 

We also assume that the gauge field 
strength $F_{\left(4\right)}$ is given by
\Eqr{
F_{\left(4\right)}&=&\ast d\left[h_5^{-1}(x, y, z)
\wedge\Omega(\Xsp)\right],
  \label{kkM5:ansatz:Eq}
}
where $\Omega(\Xsp)$ 
denotes the volume six-form 
\Eqr{
\Omega(\Xsp)=\sqrt{-q}\,dx^0\wedge dx^1\wedge \cdots \wedge dx^5\,.
   \label{kkM5:volume:Eq}
}
Here, $q$ is the determinant of the metric $q_{\mu\nu}$.

In terms of ansatz for fields \eqref{kkM5:metric:Eq} and 
\eqref{kkM5:ansatz:Eq}, the field equations lead to
\Eqrsubl{kkM5:solution1:Eq}{
&&R_{\mu\nu}(\Xsp)=0,~~~~R_{ab}(\Zsp)=0,
   \label{kkM5:Ricci:Eq}\\
&&h_5=h_0(x)+h_1(y, z),~~~~dh_k=\ast_{\Zsp}dA\,,
   \label{kkM5:h:Eq}\\
&&D_{\mu}D_{\nu}h_0=0, ~~~
h_k\pd^2_yh_1+\triangle_{\Zsp}h_1=0\,,~~~\triangle_{\Zsp}h_k=0\,,
   \label{kkM5:warp1-2:Eq}
 }
where $D_{\mu}$ is the covariant derivative with respect to
the metric $q_{\mu\nu}$, and 
$\ast_{\Zsp}$ denotes the Hodge operator on $\Zsp$, 
$\triangle_{\Xsp}$, $\lap_{\Zsp}$ are
the Laplace operators on $\Xsp$, $\Zsp$ space, and $R_{\mu\nu}(\Xsp)$, 
 $R_{ab}(\Zsp)$ are the Ricci tensors 
associated with the metrics $q_{\mu\nu}(\Xsp)$, 
 $u_{ab}(\Zsp)$, respectively. 

As a special example, let us consider the case
\Eq{
q_{\mu\nu}=\eta_{\mu\nu}\,,
~~~u_{ab}=\delta_{ab}\,,
 \label{kkM5:flat metric:Eq}
 }
where $\eta_{\mu\nu}$ is the six-dimensional
Minkowski metric and $\delta_{ab}$ is
the three-dimensional Euclidean metric. 
The solution for $h_5$ and $h_k$ can be
obtained explicitly as
\Eqrsubl{kkM5:solutions1:Eq}{
h_5(x, y, z)&=&c_{\mu}x^{\mu}+\tilde{c}
+\sum_{\ell}\frac{M_\ell}{\left[|\bm y-\bm y_\ell|^2
+4M|\bm z-\bm z_0|\right]^{3/2}},
 \label{kkM5:solution-r:Eq}\\
h_k(z)&=&\frac{M}{|\bm z-\bm z_0|},
 \label{kkM5:solution-s:Eq}
}
where $c_{\mu}$, $\tilde{c}$, $M_\ell$ and $M$ are constant parameters,
and $\bm y_\ell$ and $\bm z_0$ are
constant vectors representing the positions of the branes.
Since the functions coincide, the locations of the 
branes will also coincide.

\subsection{The intersection involving plane wave and M2-brane}

We present the M2-brane with the plane wave propagating along
its longitudinal direction. 
We assume that the eleven-dimensional metric takes the form
\Eqr{
ds^2&=&h_2^{-2/3}(z)\left[-dt^2+dx^2+dy^2
+\left\{h_w(t, y, z)-1\right\}\left(dt-dx\right)^2\right.\nn\\
&&\left.+h_2(z)u_{ab}(\Zsp)dz^adz^b\right], 
 \label{wm2:metric:Eq}
}
where $u_{ab}$ is the eight-dimensional metric which
depends only on the eight-dimensional coordinates $z^a$. 

We also assume that the gauge field 
strength $F_{\left(4\right)}$ is given by
\Eqr{
F_{\left(4\right)}&=& d\left[h_2^{-1}(z)\right]\wedge dt
\wedge dx\wedge dy\,.
  \label{wm2:ansatz:Eq}
}

In terms of ansatz for fields \eqref{wm2:metric:Eq} and 
\eqref{wm2:ansatz:Eq}, the field equations lead to
\Eqrsubl{wm2:solution1:Eq}{
&&R_{ab}(\Zsp)=0,
   \label{wm2:Ricci:Eq}\\
&&h_w=h_0(t)+h_1(y, z)\,,~~~
\pd_t^2h_0=0,~~~h_2\pd_y^2h_1+\triangle_{\Zsp}h_1=0\,,~~~
\triangle_{\Zsp}h_2=0\,,
   \label{wm2:warp1:Eq}
 }
where $\lap_{\Zsp}$ is
the Laplace operator on $\Zsp$ space, and 
$R_{ab}(\Zsp)$ are the Ricci tensor
associated with the metric $u_{ab}(\Zsp)$. 
As a special example, let us consider the case
\Eq{
u_{ab}=\delta_{ab}\,,
 \label{wm2:flat metric:Eq}
 }
where $\delta_{ab}$ is 
the eight-dimensional Euclidean metrics, respectively. 
The solution for $h_2$ and $h_w$ can be
obtained explicitly as
\Eqrsubl{wm2:solutions1:Eq}{
h_w(t, y, z)&=&\bar{c}t+\tilde{c}
+\sum_{\ell}\frac{M_\ell}{\left[|y-y_\ell|^2
+\frac{M}{4}|\bm z-\bm z_0|^{-4}\right]^{-1}},
 \label{wm2:solution-r:Eq}\\
h_2(z)&=&\frac{M}{|\bm z-\bm z_0|^6},
 \label{wm2:solution-s:Eq}
}
where $\bar{c}$, $\tilde{c}$, $M_\ell$ and $M$ are constant parameters,
and $y_\ell$ and $\bm z_0$ are
constant vectors representing the positions of the branes.
If we delocalize along $n$ of the overall transverse directions, 
the harmonic functions take the following form:
\Eqrsubl{wm2:solutions2:Eq}{
h_w(t, y, z)&=&\bar{c}t+\tilde{c}
+\sum_{\ell}\frac{M_\ell}{\left[|y-y_\ell|^2
+\frac{4M}{(n-4)^2}|\bm z-\bm z_0|^{n-4}\right]^{\frac{8-n}{2(n-4)}}},
 \label{wm2:solution-r2:Eq}\\
h_2(z)&=&\frac{M}{|\bm z-\bm z_0|^{6-n}}\,.
 \label{wm2:solution-s2:Eq}
}

\subsection{The intersection involving wave and M5-brane}

We present the M5-brane with the plane wave propagating along
its longitudinal direction. 
We assume that the eleven-dimensional metric takes the form
\Eqr{
ds^2&=&h_5^{-1/3}(z)\left[-dt^2+dx^2
+\left\{h_w(t, y, z)-1\right\}\left(dt-dx\right)^2\right.\nn\\
&&\left.+\gamma_{ij}(\Ysp)dy^idy^j+h_5(z)u_{ab}(\Zsp)dz^adz^b\right], 
 \label{wm5:metric:Eq}
}
where $\gamma_{ij}$ is the four-dimensional metric which
depends only on the four-dimensional coordinates $y^i$, 
and finally $u_{ab}$ is the five-dimensional metric which
depends only on the five-dimensional coordinates $z^a$. 

We also assume that the gauge field 
strength $F_{\left(4\right)}$ is given by
\Eqr{
F_{\left(4\right)}&=&\ast d\left[h_5^{-1}(z)\wedge dt
\wedge dx\wedge\Omega(\Ysp)\right],
  \label{wm5:ansatz:Eq}
}
where $\Omega(\Ysp)$ 
denotes the volume four-form 
\Eqr{
\Omega(\Ysp)=\sqrt{\gamma}\,dy^1\wedge dy^2\wedge dy^3\wedge dy^4\,.
   \label{wm5:volume:Eq}
}
Here, $\gamma$ is the determinant of the metric $\gamma_{ij}$.

In terms of ansatz for fields \eqref{wm5:metric:Eq} and 
\eqref{wm5:ansatz:Eq}, the field equations lead to
\Eqrsubl{wm5:solution1:Eq}{
&&R_{ij}(\Ysp)=0,~~~~R_{ab}(\Zsp)=0,
   \label{wm5:Ricci:Eq}\\
&&h_w=h_0(t)+h_1(y, z)\,,~~~
\pd_t^2h_0=0,~~~h_5\lap_{\Ysp}h_1+\triangle_{\Zsp}h_1=0\,,~~~
\triangle_{\Zsp}h_5=0\,,
   \label{wm5:warp1:Eq}
 }
where $\triangle_{\Ysp}$, $\lap_{\Zsp}$ are
the Laplace operators on 
$\Ysp$, $\Zsp$ space, and 
$R_{ij}(\Ysp)$, $R_{ab}(\Zsp)$ are the Ricci tensors
associated with the metrics $\gamma_{ij}(\Ysp)$, $u_{ab}(\Zsp)$, 
respectively. 
As a special example, let us consider the case
\Eq{
\gamma_{ij}=\delta_{ij}\,,
~~~u_{ab}=\delta_{ab}\,,
 \label{wm5:flat metric:Eq}
 }
where $\delta_{ij}$, $\delta_{ab}$ are
the four-, five-dimensional Euclidean metrics, respectively. 
The solution for $h_5$ and $h_w$ can be
obtained explicitly as
\Eqrsubl{wm5:solutions1:Eq}{
h_w(t, y, z)&=&\bar{c}t+\tilde{c}
+\sum_{\ell}\frac{M_\ell}{\left[|\bm y-\bm y_\ell|^2
+4M|\bm z-\bm z_0|^{-1}\right]^{-1}},
 \label{wm5:solution-r:Eq}\\
h_5(z)&=&\frac{M}{|\bm z-\bm z_0|^3},
 \label{wm5:solution-s:Eq}
}
where $\bar{c}$, $\tilde{c}$, $M_\ell$ and $M$ are constant parameters,
and $\bm y_\ell$ and $\bm z_0$ are
constant vectors representing the positions of the branes. 
If we delocalize along $n$ of the overall transverse directions, 
the harmonic functions take the following form:
\Eqrsubl{wm5:solutions2:Eq}{
h_w(t, y, z)&=&\bar{c}t+\tilde{c}
+\sum_{\ell}\frac{M_\ell}{\left[|\bm y-\bm y_\ell|^2
+\frac{4M}{(n-1)^2}|\bm z-\bm z_0|^{n-1}\right]^{(n+1)/(n-1)}},
 \label{wm5:solution-r2:Eq}\\
h_5(z)&=&\frac{M}{|\bm z-\bm z_0|^{3-n}}\,.
 \label{wm5:solution-s2:Eq}
}

\subsection{The plane wave in the KK-monopole background}

We consider the plane wave propagating in the background of the KK-monopole. 
The solution of ten-dimensional metric is given by 
\Eqr{
ds^2&=&-dt^2+dx^2
+\left\{h_w(t, y, z)-1\right\}\left(dt-dx\right)^2
+\gamma_{ij}(\Ysp)dy^idy^j\nn\\
&&+h_k(z)u_{ab}(\Zsp)dz^adz^b
+h_k^{-1}(z)\left(dv+A_adz^a\right)^2, 
 \label{mkkw:metric:Eq}
}
where $\gamma_{ij}$ is the five-dimensional metric which
depends only on the five-dimensional coordinates $y^i$, 
and finally $u_{ab}$ is the three-dimensional metric which
depends only on the three-dimensional coordinates $z^a$. 

The ten-dimensional metric and the function $h_k$ obey 
\Eqrsubl{mkkw:solution1:Eq}{
&&R_{ij}(\Ysp)=0,~~~~R_{ab}(\Zsp)=0,
   \label{mkkw:Ricci:Eq}\\
&&h_w=h_0(t)+h_1(y, z),~~~~\pd_t^2h_0=0,~~~~
h_k\lap_{\Ysp}h_1+\triangle_{\Zsp}h_1=0\,,~~~\triangle_{\Zsp}h_k=0\,,
   \label{mkkw:h:Eq}\\
&&dh_k=\ast_{\Zsp}dA\,,
   \label{mkkw:warp:Eq}
 }
where $\ast_{\Zsp}$ is the Hodge operator in the Z space, and 
 $\triangle_{\Ysp}$, $\lap_{\Zsp}$ are
the Laplace operators on 
$\Ysp$, $\Zsp$ space, and 
$R_{ij}(\Ysp)$, $R_{ab}(\Zsp)$ are the Ricci tensors
associated with the metrics $q_{\mu\nu}(\Xsp)$, 
$\gamma_{ij}(\Ysp)$, $u_{ab}(\Zsp)$, 
respectively. 
As a special example, let us consider the case
\Eq{
\gamma_{ij}=\delta_{ij}\,,
~~~u_{ab}=\delta_{ab}\,,
 \label{mkkw:flat metric:Eq}
 }
where $\delta_{ij}$, $\delta_{ab}$ are
the five-, three-dimensional Euclidean metrics,
respectively. 
The solution for $h$ and $h_k$ can be
obtained explicitly as
\Eqrsubl{mkkw:solutions1:Eq}{
h_w(t, y, z)&=&\bar{c}t+\tilde{c}
+\sum_{\ell}\frac{M_\ell}{\left[|\bm y-\bm y_\ell|^2
+4M|\bm z-\bm z_0|\right]^{7/2}},
 \label{mkkw:solution-r:Eq}\\
h_k(z)&=&\frac{M}{|\bm z-\bm z_0|},
 \label{mkkw:solution-s:Eq}
}
where $\bar{c}$, $\tilde{c}$, $M_\ell$ and $M$ are constant parameters,
and $\bm y_\ell$ and $\bm z_0$ are
constant vectors representing the positions of the branes.
Since the functions coincide, the locations of the 
branes will also coincide. 

For the static case, there is a classification of the multiple 
intersecting branes with the M-waves and/or KK-monopoles 
\cite{Bergshoeff:1996rn, Bergshoeff:1997tt}. The dynamical  
delocalized branes are also classified in \cite{Maeda:2009zi}. 
We show the intersection rule for the branes with M-wave and KK-monopoles,
which is summarized in Table \ref{table_1}. 
In the Table, circles indicate where the
brane world-volumes enter, $v$ represents the coordinate of the KK-monopole, 
and the time-dependent branes are indicated by $\surd$ for 
different solutions. 


\section{The intersection of dynamical branes in ten-dimensional theory}
  \label{sec:ten}
In this section, we apply the dynamical brane 
solutions to ten-dimensional string theory.
The ten-dimensional action
for the $p$-brane system in the Einstein frame 
can be written as
\Eqr{
S&=&\frac{1}{2\kappa^2}\int \left[R\ast{\bf 1}
 -\frac{1}{2}\ast d\phi \wedge d\phi
 -\frac{1}{2\cdot 3!}\e^{-\phi}\ast H_{(3)}\wedge H_{(3)}\right.
 \nn\\
 &&\left.
 -\sum_I\frac{1}{2\cdot (p_I+2)!}\e^{(3-p_I)\phi/2}
 \ast F_{(p_I+2)}\wedge F_{(p_I+2)}\right],
\label{Dp:action:Eq}
}
where $\kappa^2$ is the ten-dimensional gravitational constant, 
and $\phi$ is the dilaton, and 
$\ast$ is the Hodge operator in the ten-dimensional spacetime, 
and $H_{(3)}$, $F_{(p_I+2)}$ are 
3-, $(p_I+2)$-form field strength, respectively.  
We assume that the field strengths $H_{(3)}$, $F_{(p_I+2)}$ 
are given by following gauge potentials 
\Eq{
H_{(3)}=dB_{(2)}\,,~~~~
F_{(p_I+2)}=dC_{(p_I+1)}\,.
}

The field equations are given by
\Eqrsubl{Dp:field equations:Eq}{
&&\hspace{-1cm}R_{MN}=\frac{1}{2}\pd_M\phi \pd_N \phi
+\frac{1}{2\cdot 3!}\e^{-\phi}
\left[3H_{MAB} {H_N}^{AB}-\frac{1}{4}g_{MN} H^2_{(3)}\right]\nn\\
&&+\sum_I\frac{\e^{(3-p_I)\phi/2}}{2\cdot (p_I+2)!}
\left[(p_I+2)F_{MA_1\cdots A_{p_I+1}} {F_N}^{A_1\cdots A_{p_I+1}}
-\frac{1}{8}(p_I+1)g_{MN} F^2_{(p_I+2)}\right],
   \label{Dp:Einstein:Eq}\\
&&\hspace{-1cm}d\ast d\phi+\frac{1}{2\cdot 3!}
\e^{-\phi}\ast H_{(3)}\wedge H_{(3)}
-\sum_I\frac{(3-p_I)}{4\cdot (p_I+2)!}
\e^{(3-p_I)\phi/2}\ast F_{(p_I+2)}\wedge F_{(p_I+2)}=0,
   \label{Dp:scalar:Eq}\\
&&\hspace{-1cm}d\left[\e^{-\phi}\ast H_{(3)}\right]=0\,,
   \label{Dp:gauge-k:Eq}\\
&&\hspace{-1cm}d\left[\e^{(3-p_I)\phi/2}\ast F_{(p_I+2)}\right]=0\,.
   \label{Dp:gauge-r:Eq}
}
In what follows, we look for the possible configurations of intersecting
branes and present explicit solutions. The case with
waves or KK-monopoles will be also discussed. 

\subsection{The intersection involving two D$p$-brane}
Let us first discuss the dynamical solution of two D$p$-branes. 
The ten-dimensional metric thus takes the form
\Eqr{
ds^2&=&h^{(p-7)/8}(x, y, z)\bar{h}^{(p-7)/8}(z)\left[
q_{\mu\nu}(\Xsp)dx^{\mu}dx^{\nu}
+h(x, y, z)\gamma_{ij}(\Ysp_1)dy^idy^j\right.\nn\\
&&\left.+\bar{h}(z)w_{mn}(\Ysp_2)dv^mdv^n
+h(x, y, z)\bar{h}(z)u_{ab}(\Zsp)dz^adz^b\right], 
 \label{Dp:metric:Eq}
}
where $q_{\mu\nu}$ is the $(p-1)$-dimensional metric which
depends only on the $(p-1)$-dimensional coordinates $x^{\mu}$, 
$\gamma_{ij}$ is the two-dimensional metric which
depends only on the two-dimensional coordinates $y^i$, 
$w_{mn}$ is the two-dimensional metric which
depends only on the two-dimensional coordinates $w^m$, 
and finally $u_{ab}$ is the $(7-p)$-dimensional metric which
depends only on the $(7-p)$-dimensional coordinates $z^a$. 

We also assume that the scalar field $\phi$ and the gauge field 
strength $F_{\left(4\right)}$ are given by
\Eqrsubl{Dp:ansatz:Eq}{
\e^{\phi}&=&\left(h\bar{h}\right)^{(3-p)/4}\,,
  \label{Dp:ansatz for scalar:Eq}\\
F_{\left(p+2\right)}&=&d\left[h^{-1}(x, y, z)\right]\Omega(\Xsp)
\wedge\Omega(\Ysp_1)+d\left[\bar{h}^{-1}(z)\right]\Omega(\Xsp)
\wedge\Omega(\Ysp_2),
  \label{Dp:ansatz for gauge2:Eq}
}
where $\Omega(\Xsp)$, $\Omega(\Ysp_1)$, $\Omega(\Ysp_2)$ 
denote the volume $(p-1)$-, two-, two-form, respectively 
\Eqrsubl{Dp:volume:Eq}{
\Omega(\Xsp)&=&\sqrt{-q}\,dx^0\wedge dx^1 \wedge\cdots\wedge dx^{p-2}\,,
   \label{Dp:volume x:Eq}\\
\Omega(\Ysp_1)&=&\sqrt{\gamma}\,dy^1\wedge dy^2\,,
   \label{Dp:volume y1:Eq}\\
\Omega(\Ysp_2)&=&\sqrt{w}\,dv^1\wedge dv^2\,.
   \label{Dp:volume y2:Eq}
}
Here, $q$, $\gamma$, $w$ are the determinant of the metrics 
$q_{\mu\nu}$, $\gamma_{ij}$, $w_{mn}$.

In terms of ansatz for fields \eqref{Dp:metric:Eq} and 
\eqref{Dp:ansatz:Eq}, the field equations lead to
\Eqrsubl{Dp:solution1:Eq}{
&&R_{\mu\nu}(\Xsp)=0,~~~~R_{ij}(\Ysp_1)=0,~~~~
R_{mn}(\Ysp_2)=0,~~~~R_{ab}(\Zsp)=0,
   \label{Dp:Ricci:Eq}\\
&&h=h_0(x)+h_1(y, z)\,~~~
D_{\mu}D_{\nu}h_0=0,~~~~\bar{h}\lap_{\Ysp_1}h_1+\triangle_{\Zsp}h_1=0\,,
~~~\triangle_{\Zsp}\bar{h}=0\,,
   \label{Dp:warp:Eq}
 }
where $D_{\mu}$ is the covariant derivative with respect to
the metric $q_{\mu\nu}$, and $\triangle_{\Ysp_1}$, $\lap_{\Zsp}$ are
the Laplace operators on 
$\Ysp_1$, $\Zsp$ space, and $R_{\mu\nu}(\Xsp)$, 
$R_{ij}(\Ysp_1)$, $R_{mn}(\Ysp_2)$, $R_{ab}(\Zsp)$ are the Ricci tensors
 associated with the metrics $q_{\mu\nu}(\Xsp)$, $\gamma_{ij}(\Ysp_1)$, 
$w_{mn}(\Ysp_2)$, $u_{ab}(\Zsp)$, respectively. 

Let us consider the case
\Eq{
q_{\mu\nu}=\eta_{\mu\nu}\,,~~~
\gamma_{ij}=\delta_{ij}\,,~~~w_{mn}=\delta_{mn}\,,~~~
u_{ab}=\delta_{ab}\,,
 \label{Dp:flat metric:Eq}
 }
where $\eta_{\mu\nu}$ is the $(p-1)$-dimensional
Minkowski metric and $\delta_{ij}$, $\delta_{mn}$, $\delta_{ab}$ are
the two-, two-, $(7-p)$-dimensional Euclidean metrics, respectively. 
For $p\ne 3$ and $p\ne 5$, the solution for $h$ and $\bar{h}$ can be
obtained explicitly as
\Eqrsubl{Dp:solutions1:Eq}{
h(x, y, z)&=&c_{\mu}x^{\mu}+\tilde{c}
+\sum_{\ell}\frac{M_\ell}{\left[|\bm y-\bm y_\ell|^2
+\frac{4M}{(p-3)^2}|\bm z-\bm z_0|^{p-3}\right]^{\frac{2}{p-3}}},
 \label{Dp:solution-r:Eq}\\
\bar{h}(z)&=&\frac{M}{|\bm z-\bm z_0|^{5-p}},
 \label{Dp:solution-s:Eq}
}
where $c_{\mu}$, $\tilde{c}$, $M_\ell$ and $M$ are constant parameters,
and $\bm y_\ell$ and $\bm z_0$ are
constant vectors representing the positions of the branes.
Since the functions coincide, the locations of the 
branes will also coincide. 
In the case of $p=3$, the field equations give 
\Eqrsubl{Dp:solutions2:Eq}{
h_{\rm F}(x, y, z)&=&c_{\mu}x^{\mu}+\tilde{c}
+\sum_{\ell}M_\ell\left[|\bm y-\bm y_\ell|^2
-2M\ln|\bm z-\bm z_0|\right],
 \label{Dp:solution-r2:Eq}\\
h(z)&=&\frac{M}{|\bm z-\bm z_0|^2}\,.
 \label{Dp:solution-s2:Eq}
}
For $p=5$, the solution becomes
\Eqrsubl{Dp:solutions3:Eq}{
h(x, y, z)&=&c_{\mu}x^{\mu}+\tilde{c}
+\sum_{\ell}\frac{M_\ell}{\left[|\bm y-\bm y_\ell|^2
+M|\bm z-\bm z_0|^2\right]},
 \label{Dp:solution-r3:Eq}\\
\bar{h}(z)&=&M\ln|\bm z-\bm z_0|\,.
 \label{Dp:solution-s3:Eq}
}

\subsection{The intersection of D$p$-D$(p+2)$ branes}

Next we discuss the D$p$-branes ending on D$(p+2)$-branes. 
Let us consider the solution to be delocalized along 
the relative transverse direction of the D$p$-branes.
The ten-dimensional metric thus takes the form
\Eqr{
ds^2&=&h_p^{(p-7)/8}(x, y, z)h_{p+2}^{(p-5)/8}(z)\left[
q_{\mu\nu}(\Xsp)dx^{\mu}dx^{\nu}+h_p(x, y, z)\gamma_{ij}(\Ysp)dy^idy^j
\right.\nn\\
&&\left.+h_{p+2}(z)dv^2
+h_p(x, y, z)h_{p+2}(z)u_{ab}(\Zsp)dz^adz^b\right], 
 \label{p+2:metric:Eq}
}
where $q_{\mu\nu}$ is the $p$-dimensional metric which
depends only on the $p$-dimensional coordinates $x^{\mu}$, 
$\gamma_{ij}$ is the three-dimensional metric which
depends only on the three-dimensional coordinates $y^i$, 
and finally $u_{ab}$ is the $(6-p)$-dimensional metric which
depends only on the $(6-p)$-dimensional coordinates $z^a$. 

We also assume that the scalar field $\phi$ and the gauge field 
strengths $F_{\left(p+2\right)}$, $F_{\left(p+4\right)}$ are given by
\Eqrsubl{p+2:ansatz:Eq}{
\e^{\phi}&=&h_p^{(3-p)/4}h_{p+2}^{(1-p)/4}\,,
  \label{p+2:ansatz for scalar:Eq}\\
F_{\left(p+2\right)}&=&d\left[h_p^{-1}(x, y, z)\right]
\wedge\Omega(\Xsp)\wedge dv,
  \label{p+2:ansatz for gauge:Eq}\\
F_{\left(p+4\right)}&=&d\left[h_{p+2}^{-1}(z)\right]\wedge\Omega(\Xsp)
\wedge\Omega(\Ysp),
  \label{p+2:ansatz for gauge2:Eq}
}
where $\Omega(\Xsp)$ and $\Omega(\Ysp)$ 
denote the volume $p$-, 3-form, respectively 
\Eqrsubl{p+2:volume:Eq}{
\Omega(\Xsp)&=&\sqrt{-q}\,dx^0\wedge dx^1 \wedge\cdots\wedge dx^{p-1}\,,
   \label{p+2:volume x:Eq}\\
\Omega(\Ysp)&=&\sqrt{\gamma}\,dy^1\wedge dy^2\wedge dy^3\,.
   \label{p+2:volume y:Eq}
}
Here, $q$, $\gamma$ are the determinant of the metrics 
$q_{\mu\nu}$, $\gamma_{ij}$.

Under the assumptions \eqref{pf:metric:Eq} and 
\eqref{pf:ansatz:Eq}, the field equations lead to
\Eqrsubl{p+2:solution1:Eq}{
&&\hspace{-0.5cm}
R_{\mu\nu}(\Xsp)=0,~~~~R_{ij}(\Ysp)=0,~~~~R_{ab}(\Zsp)=0,
   \label{p+2:Ricci:Eq}\\
&&\hspace{-0.5cm}h_p=h_0(x)+h_1(y, z)\,,~~~
D_{\mu}D_{\nu}h_0=0, ~~~h_{p+2}\lap_{\Ysp}h_1+\triangle_{\Zsp}h_1=0\,,
~~~\triangle_{\Zsp}h_{p+2}=0\,,
   \label{p+2:warp:Eq}
 }
where  $D_{\mu}$ is the covariant derivative with respect to
the metric $q_{\mu\nu}$, and $\triangle_{\Ysp}$, $\lap_{\Zsp}$ are
the Laplace operators on 
$\Ysp$, $\Zsp$ space, and $R_{ij}(\Xsp)$, 
$R_{ij}(\Ysp)$, $R_{ab}(\Zsp)$ are the Ricci tensors 
associated with the metrics $q_{\mu\nu}(\Xsp)$, 
$\gamma_{ij}(\Ysp)$, $u_{ab}(\Zsp)$, respectively. 
Now we assume that the ten-dimensional metric is given by  
\Eq{
q_{\mu\nu}=\eta_{\mu\nu}\,,~~~
\gamma_{ij}=\delta_{ij}\,,~~~u_{ab}=\delta_{ab}\,,
 \label{p+2:flat metric:Eq}
 }
where $\eta_{\mu\nu}$ is the $p$-dimensional
Minkowski metric and $\delta_{ij}$, $\delta_{ab}$ are
the three-, $(6-p)$-dimensional Euclidean metrics,
respectively. 
For $p\ne 2$, $p\ne 4$, the solution for $h_p$ and $h_{p+2}$ can be
obtained explicitly as
\Eqrsubl{p+2:solutions1:Eq}{
h_p(x, y, z)&=&c_{\mu}x^{\mu}+\tilde{c}
+\sum_{\ell}\frac{M_\ell}{\left[|\bm y-\bm y_\ell|^2
+\frac{4M}{(p-2)^2}|\bm z-\bm z_0|^{p-2}\right]^{\frac{p+2}{2(p-2)}}},
 \label{p+2:solution-r:Eq}\\
h_{p+2}(z)&=&\frac{M}{|\bm z-\bm z_0|^{4-p}},
 \label{p+2:solution-s:Eq}
}
where $A$, $\tilde{c}$, $M_\ell$ and $M$ are constant parameters,
and $\bm y_\ell$ and $\bm z_0$ are
constant vectors representing the positions of the branes.

If we set $p=2$, the solution becomes 
\Eqrsubl{p+2:solutions2:Eq}{
h_2(x, y, z)&=&c_{\mu}x^{\mu}+\tilde{c}
+\sum_{\ell}M_\ell\left[|\bm y-\bm y_\ell|^2
-3M\ln|\bm z-\bm z_0|\right]\,,
 \label{p+2:solution-r2:Eq}\\
h_4(z)&=&\frac{M}{|\bm z-\bm z_0|^2}\,.
 \label{p+2:solution-s2:Eq}
}
Next we consider the case of $p=4$ . The solution is given by 
\Eqrsubl{p+2:solutions3:Eq}{
h_4(x, y, z)&=&c_{\mu}x^{\mu}+\tilde{c}
+\sum_{\ell}\frac{M_\ell}{\left[|\bm y-\bm y_\ell|^2
+M|\bm z-\bm z_0|^2\right]^{\frac{3}{2}}}\,,
 \label{p+2:solution-r3:Eq}\\
h_{6}(z)&=&M\ln |\bm z-\bm z_0|\,.
 \label{p+2:solution-s3:Eq}
}

\subsection{The D$p$-D$(p+4)$ brane system}
Now we consider the dynamical solution of D$p$-D$(p+4)$ brane system.  
Let us discuss the solution to be delocalized along 
the relative transverse direction of the D$p$-branes.
The ten-dimensional metric thus takes the form
\Eqr{
ds^2&=&h_{p}^{(p-7)/8}(x, y, z)h_{p+4}^{(p-3)/8}(z)\left[
q_{\mu\nu}(\Xsp)dx^{\mu}dx^{\nu}+h_{p}(x, y, z)\gamma_{ij}(\Ysp)dy^idy^j
\right.\nn\\
&&\left.+h_{p}(x, y, z)h_{p+4}(z)u_{ab}(\Zsp)dz^adz^b\right], 
 \label{p+4:metric:Eq}
}
where $q_{\mu\nu}$ is the $(p+1)$-dimensional metric which
depends only on the $(p+1)$-dimensional coordinates $x^{\mu}$, 
$\gamma_{ij}$ is the four-dimensional metric which
depends only on the four-dimensional coordinates $y^i$, 
and finally $u_{ab}$ is the $(5-p)$-dimensional metric which
depends only on the $(5-p)$-dimensional coordinates $z^a$. 

We also assume that the scalar field $\phi$ and the gauge field 
strengths $F_{\left(p+2\right)}$, $F_{\left(p+6\right)}$ are given by
\Eqrsubl{p+4:ansatz:Eq}{
\e^{\phi}&=&h_p^{(3-p)/4}h_{p+4}^{-(1+p)/4}\,,
  \label{p+4:ansatz for scalar:Eq}\\
F_{\left(p+2\right)}&=&d\left[h_p^{-1}(x, y, z)\right]
\wedge\Omega(\Xsp)\,,
  \label{p+4:ansatz for gauge:Eq}\\
F_{\left(p+6\right)}&=&d\left[h_{p+4}^{-1}(z)\right]
\wedge\Omega(\Xsp)\wedge\Omega(\Ysp),
  \label{p+4:ansatz for gauge2:Eq}
}
where $\Omega(\Xsp)$ and $\Omega(\Ysp)$ 
denote the volume $(p+1)$-, 4-form, respectively 
\Eqrsubl{p+4:volume:Eq}{
\Omega(\Xsp)&=&\sqrt{-q}\,dx^0\wedge dx^1 \wedge\cdots\wedge dx^{p}\,,
   \label{p+4:volume x:Eq}\\
\Omega(\Ysp)&=&\sqrt{\gamma}\,dy^1\wedge dy^2\wedge dy^3\wedge dy^4\,.
   \label{p+4:volume y:Eq}
}
Here, $q$ and $\gamma$ are the determinant of the metrics 
$q_{\mu\nu}$ and $\gamma_{ij}$.

In terms of ansatz for fields \eqref{p+4:metric:Eq} and 
\eqref{p+4:ansatz:Eq}, the field equations lead to
\Eqrsubl{p+4:solution1:Eq}{
&&\hspace{-0.5cm}
R_{\mu\nu}(\Xsp)=0,~~~~R_{ij}(\Ysp)=0,~~~~R_{ab}(\Zsp)=0,
   \label{p+4:Ricci:Eq}\\
&&\hspace{-0.5cm}h_p=h_0(x)+h_1(y, z),~~~
D_{\mu}D_{\nu}h_0=0, ~~~h_{p+4}\lap_{\Ysp}h_1+\triangle_{\Zsp}h_1=0\,,~~~
\triangle_{\Zsp}h_{p+4}=0\,,
   \label{p+4:warp:Eq}
 }
where  $D_{\mu}$ is the covariant derivative with respect to
the metric $q_{\mu\nu}$, and $\triangle_{\Ysp}$, $\lap_{\Zsp}$ are
the Laplace operators on $\Ysp$, $\Zsp$ space, and $R_{\mu\nu}(\Xsp)$, 
$R_{ij}(\Ysp)$, $R_{ab}(\Zsp)$ are the Ricci tensors 
associated with the metrics $q_{\mu\nu}(\Xsp)$, 
$\gamma_{ij}(\Ysp)$, $u_{ab}(\Zsp)$, respectively. 

Let us consider the case
\Eq{
q_{\mu\nu}=\eta_{\mu\nu}\,,~~~
\gamma_{ij}=\delta_{ij}\,,~~~u_{ab}=\delta_{ab}\,,
 \label{p+4:flat metric:Eq}
 }
where $\eta_{\mu\nu}$ is the $(p+1)$-dimensional
Minkowski metric and $\delta_{ij}$, $\delta_{ab}$ are
the four-, $(5-p)$-dimensional Euclidean metrics,
respectively. 
In the case of $p\ne 1$, $p\ne 3$, 
the solution for $h_p$ and $h_{p+4}$ can be expressed as
\Eqrsubl{p+4:solutions1:Eq}{
h_p(x, y, z)&=&c_{\mu}x^{\mu}+\tilde{c}
+\sum_{\ell}\frac{M_\ell}{\left[|\bm y-\bm y_\ell|^2
+\frac{4M}{(p-1)^2}|\bm z-\bm z_0|^{p-1}\right]^{\frac{p+1}{p-1}}},
 \label{p+4:solution-r:Eq}\\
h_{p+4}(z)&=&\frac{M}{|\bm z-\bm z_0|^{3-p}},
 \label{p+4:solution-s:Eq}
}
where $c_{\mu}$, $\tilde{c}$, $M_\ell$ and $M$ are constant parameters,
and $\bm y_\ell$ and $\bm z_0$ are
constant vectors representing the positions of the branes.

For $p=1$, we find 
\Eqrsubl{p+4:solutions2:Eq}{
h_1(x, y, z)&=&c_{\mu}x^{\mu}+\tilde{c}
+\sum_{\ell}M_\ell\left[|\bm y-\bm y_\ell|^2
-4M\ln|\bm z-\bm z_0|\right],
 \label{p+4:solution-r2:Eq}\\
h_{5}(z)&=&\frac{M}{|\bm z-\bm z_0|^{2}}\,.
 \label{p+4:solution-s2:Eq}
}
If we consider the D3-D7 brane system, the solution is given by 
\Eqrsubl{p+4:solutions3:Eq}{
h_3(x, y, z)&=&c_{\mu}x^{\mu}+\tilde{c}
+\sum_{\ell}\frac{M_\ell}{\left[|\bm y-\bm y_\ell|^2
+M|\bm z-\bm z_0|^2\right]^2},
 \label{p+4:solution-r3:Eq}\\
h_{7}(z)&=&M\ln|\bm z-\bm z_0|\,.
 \label{p+4:solution-s3:Eq}
}

\subsection{The intersection of D$p$-brane and KK-monopole}

Now we discuss the KK-monopole in the transverse space of D$p$-brane
with $p\le 4$. 
We assume that the ten-dimensional metric is given by
\Eqr{
ds^2&=&h^{(p-7)/8}(x, y, z)q_{\mu\nu}(\Xsp)dx^{\mu}dx^{\nu}
+h^{(p+1)/8}(x, y, z)\left[\gamma_{ij}(\Ysp)dy^idy^j\right.\nn\\
&&\left.+h_k(z)u_{ab}(\Zsp)dz^adz^b
+h_k^{-1}(z)\left(dv+A_adz^a\right)^2\right], 
 \label{pkk:metric:Eq}
}
where $q_{\mu\nu}$ is the $(p+1)$-dimensional metric which
depends only on the $(p+1)$-dimensional coordinates $x^{\mu}$, 
$\gamma_{ij}$ is the $(5-p)$-dimensional metric which
depends only on the $(5-p)$-dimensional coordinates $y^i$, 
and finally $u_{ab}$ is the three-dimensional metric which
depends only on the three-dimensional coordinates $z^a$. 

We also assume that the scalar field $\phi$ and the gauge field 
strength $F_{\left(p+2\right)}$ are given by
\Eqrsubl{pkk:ansatz:Eq}{
\e^{\phi}&=&h^{(3-p)/4}\,,
  \label{pkk:ansatz for scalar:Eq}\\
F_{\left(p+2\right)}&=&d\left[h^{-1}(x, y, z)\right]\wedge\Omega(\Xsp)\,,
  \label{pkk:ansatz for gauge:Eq}
}
where $\Omega(\Xsp)$ denotes the volume $(p+1)$-form 
\Eqr{
\Omega(\Xsp)=\sqrt{-q}\,dx^0\wedge dx^1\wedge \cdots\wedge dx^p\,.
   \label{pkk:volume:Eq}
}
Here, $q$ is the determinant of the metric $q_{\mu\nu}$.

Using the ansatz for fields \eqref{pkk:metric:Eq} and 
\eqref{pkk:ansatz:Eq}, the field equations lead to
\Eqrsubl{pkk:solution1:Eq}{
&&R_{\mu\nu}(\Xsp)=0,~~~~R_{ij}(\Ysp)=0,~~~~R_{ab}(\Zsp)=0,
   \label{pkk:Ricci:Eq}\\
&&h=h_0(x)+h_1(y, z),~~~~dh_k=\ast_{\Zsp}dA\,,
   \label{pkk:h:Eq}\\
&&D_{\mu}D_{\nu}h_0=0, ~~~~h_k\lap_{\Ysp}h_1+\triangle_{\Zsp}h_1=0\,,~~~
\triangle_{\Zsp}h_k=0\,,
   \label{pkk:warp:Eq}
 }
where $D_{\mu}$ is the covariant derivative with respect to
the metric $q_{\mu\nu}$, and 
$\ast_{\Zsp}$ denotes the Hodge operator on $\Zsp$, and 
$\triangle_{\Ysp}$, $\lap_{\Zsp}$ are
the Laplace operators on 
$\Ysp$, $\Zsp$ space, and $R_{\mu\nu}(\Xsp)$, 
$R_{ij}(\Ysp)$, $R_{ab}(\Zsp)$ are the Ricci tensors 
associated with the metrics $q_{\mu\nu}(\Xsp)$, 
$\gamma_{ij}(\Ysp)$, $u_{ab}(\Zsp)$, respectively. 

Now we set the metric : 
\Eq{
q_{\mu\nu}=\eta_{\mu\nu}\,,~~~\gamma_{ij}=\delta_{ij}\,,
~~~u_{ab}=\delta_{ab}\,,
 \label{pkk:flat metric:Eq}
 }
where $\eta_{\mu\nu}$ is the $(p+1)$-dimensional
Minkowski metric and $\delta_{ij}$, $\delta_{ab}$ are
the $(5-p)$-, three-dimensional Euclidean metrics,
respectively. 
The solution for $h$ and $h_k$ can be
obtained explicitly as
\Eqrsubl{pkk:solutions:Eq}{
h(x, y, z)&=&c_{\mu}x^{\mu}+\tilde{c}
+\sum_{\ell}\frac{M_\ell}{\left[|\bm y-\bm y_\ell|^2
+4M|\bm z-\bm z_0|\right]^{(7-p)/2}},
 \label{pkk:solution-r:Eq}\\
h_k(z)&=&\frac{M}{|\bm z-\bm z_0|},
 \label{pkk:solution-s:Eq}
}
where $c_{\mu}$, $\tilde{c}$, $M_\ell$ and $M$ are constant parameters,
and $\bm y_\ell$ and $\bm z_0$ are
constant vectors representing the positions of the branes.

\subsection{The intersection of D$p$-brane and plane wave}

We present the D$p$-brane with the plane wave propagating along
its longitudinal direction. 
We assume that the ten-dimensional metric takes the form
\Eqr{
ds^2&=&h^{(p-7)/8}(z)\left[-dt^2+dx^2
+\left\{h_w(t, y, z)-1\right\}\left(dt-dx\right)^2\right.\nn\\
&&\left.+\gamma_{ij}(\Ysp)dy^idy^j+h(z)u_{ab}(\Zsp)dz^adz^b\right], 
 \label{wp:metric:Eq}
}
where $\gamma_{ij}$ is the $(p-1)$-dimensional metric which
depends only on the $(p-1)$-dimensional coordinates $y^i$, 
and finally $u_{ab}$ is the $(9-p)$-dimensional metric which
depends only on the $(9-p)$-dimensional coordinates $z^a$. 

We also assume that the gauge field 
strength $F_{\left(p+2\right)}$ is given by
\Eqrsubl{wp:ansatz:Eq}{
\e^{\phi}&=&h^{(3-p)/4}\,,
  \label{wp:ansatz for scalar:Eq}\\
F_{\left(p+2\right)}&=&d\left[h^{-1}(z)\wedge dt
\wedge dx\wedge\Omega(\Ysp)\right],
  \label{wp:ansatz for gauge:Eq}
}
where $\Omega(\Ysp)$ denotes the volume $(p-1)$-form 
\Eqr{
\Omega(\Ysp)=\sqrt{\gamma}\,dy^1\wedge dy^2\cdots\wedge dy^{p-1}\,.
   \label{wp:volume:Eq}
}
Here, $\gamma$ is the determinant of the metric $\gamma_{ij}$.

In terms of ansatz for fields \eqref{wp:metric:Eq} and 
\eqref{wp:ansatz:Eq}, the field equations lead to
\Eqrsubl{wp:solution1:Eq}{
&&R_{ij}(\Ysp)=0,~~~~R_{ab}(\Zsp)=0,
   \label{wp:Ricci:Eq}\\
&&h_w=h_0(t)+h_1(y, z)\,,~~~
\pd_t^2h_0=0,~~~h\lap_{\Ysp}h_1+\triangle_{\Zsp}h_1=0\,,~~~
\triangle_{\Zsp}h=0\,,
   \label{wp:warp:Eq}
 }
where $\triangle_{\Ysp}$, $\lap_{\Zsp}$ are
the Laplace operators on 
$\Ysp$, $\Zsp$ space, and 
$R_{ij}(\Ysp)$, $R_{ab}(\Zsp)$ are the Ricci tensors
associated with the metrics $\gamma_{ij}(\Ysp)$, $u_{ab}(\Zsp)$, 
respectively. 
Now we set the ten-dimensional metric
\Eq{
\gamma_{ij}=\delta_{ij}\,,
~~~u_{ab}=\delta_{ab}\,,
 \label{wp:flat metric:Eq}
 }
where $\delta_{ij}$, $\delta_{ab}$ are
the $(p-1)$-, $(9-p)$-dimensional Euclidean metrics, respectively. 
Using \Eqref{wp:flat metric:Eq}, 
the solution for $p\ne 5$ and $p\ne 7$ can be written as
\Eqrsubl{wp:solutions1:Eq}{
h_w(t, y, z)&=&\bar{c}t+\tilde{c}
+\sum_{\ell}\frac{M_\ell}{\left[|\bm y-\bm y_\ell|^2
+\frac{4M}{(p-5)^2}|\bm z-\bm z_0|^{p-5}\right]^{\frac{p^2-8p+19}{2(p-5)}}}\,,
 \label{wp:solution-r:Eq}\\
h(z)&=&\frac{M}{|\bm z-\bm z_0|^{7-p}},
 \label{wp:solution-s:Eq}
}
where $\bar{c}$, $\tilde{c}$, $M_\ell$ and $M$ are constant parameters,
and $\bm y_\ell$ and $\bm z_0$ are
constant vectors representing the positions of the branes.
The solution \eqref{wp:solution-r:Eq} for $p=5$ becomes 
\Eqrsubl{wp:solutions2:Eq}{
h_w(t, y, z)&=&\bar{c}t+\tilde{c}
+\sum_{\ell}M_\ell\left[|\bm y-\bm y_\ell|^2
-(p-1)M\ln|\bm z-\bm z_0|\right]\,,
 \label{wp:solution-r2:Eq}\\
h(z)&=&\frac{M}{|\bm z-\bm z_0|^2}\,.
 \label{wp:solution-s2:Eq}
}
If we consider the case of $p=7$, 
the harmonic functions take the following form:
\Eqrsubl{wp:solutions3:Eq}{
h_w(t, y, z)&=&\bar{c}t+\tilde{c}
+\sum_{\ell}\frac{M_\ell}{\left[|\bm y-\bm y_\ell|^2
+M|\bm z-\bm z_0|^2\right]^3}\,,
 \label{wp:solution-r3:Eq}\\
h(z)&=&M\ln|\bm z-\bm z_0|\,.
 \label{wp:solution-s3:Eq}
}

\subsection{The pair intersection involving fundamental string}
Next we present intersecting fundamental string configurations 
of all the possible combinations.
The basic constituents of intersecting branes are D-branes, fundamental
string, solitonic NS5-brane, the KK-monopole and the plane wave. 

\subsubsection{The intersection of D$p$-brane and fundamental string}
First we discuss the D$p$-branes ending on fundamental string.  
Let us consider the solution to be delocalized along 
the relative transverse direction of the D$p$-branes.
The ten-dimensional metric thus takes the form
\Eqr{
ds^2&=&h_{\rm F}^{-3/4}(t, y, z)h^{(p-7)/8}(z)\left[-dt^2
+h_{\rm F}(t, y, z)\gamma_{ij}(\Ysp)dy^idy^j+h(z)dv^2\right.\nn\\
&&\left.+h_{\rm F}(t, y, z)h(z)u_{ab}(\Zsp)dz^adz^b\right], 
 \label{pf:metric:Eq}
}
where $\gamma_{ij}$ is the $p$-dimensional metric which
depends only on the $p$-dimensional coordinates $y^i$, 
and finally $u_{ab}$ is the $(8-p)$-dimensional metric which
depends only on the $(8-p)$-dimensional coordinates $z^a$. 

The scalar field $\phi$ and the gauge field 
strength $H_{\left(3\right)}$ are also assumed to be 
\Eqrsubl{pf:ansatz:Eq}{
\e^{\phi}&=&h_{\rm F}^{-1/2}h^{(3-p)/4}\,,
  \label{pf:ansatz for scalar:Eq}\\
H_{\left(3\right)}&=&d\left[h_{\rm F}^{-1}(t, y, z)\right]\wedge dt
\wedge dv,
  \label{pf:ansatz for gauge:Eq}\\
F_{\left(p+2\right)}&=&
d\left[h^{-1}(z)\right]\wedge dt \wedge\Omega(\Ysp),
  \label{pf:ansatz for gauge2:Eq}
}
where $\Omega(\Ysp)$ 
denotes the volume $p$-form 
\Eqr{
\Omega(\Ysp)=\sqrt{\gamma}\,dy^1\wedge dy^2 \wedge\cdots\wedge dy^p\,.
   \label{pf:volume:Eq}
}
Here, $\gamma$ is the determinant of the metric $\gamma_{ij}$.

If we use ansatz for fields \eqref{pf:metric:Eq} and 
\eqref{pf:ansatz:Eq}, the field equations give
\Eqrsubl{pf:solution1:Eq}{
&&R_{ij}(\Ysp)=0,~~~~R_{ab}(\Zsp)=0,
   \label{pf:Ricci:Eq}\\
&&h_{\rm F}=h_0(x)+h_1(y, z),~~~
\pd_t^2h_0=0\,, ~~~
h\lap_{\Ysp}h_1+\triangle_{\Zsp}h_1=0\,,~~~\triangle_{\Zsp}h=0\,,
   \label{pf:warp1-2:Eq}
 }
where $\triangle_{\Ysp}$, $\lap_{\Zsp}$ are
the Laplace operators on $\Ysp$, $\Zsp$ space, and
$R_{ij}(\Ysp)$, $R_{ab}(\Zsp)$ are the Ricci tensors 
associated with the metrics $\gamma_{ij}(\Ysp)$, $u_{ab}(\Zsp)$, 
respectively. 
We consider the case 
\Eq{
\gamma_{ij}=\delta_{ij}\,,~~~u_{ab}=\delta_{ab}\,,
 \label{pf:flat metric:Eq}
 }
where $\delta_{ij}$, $\delta_{ab}$ are
the $p$-, $(8-p)$-dimensional Euclidean metrics,
respectively. 
Under the metric \Eqref{pf:flat metric:Eq}, the solution of 
$h$ and $h_{\rm F}$ for $p\ne 4$ and $p\ne 6$ can be written by
\Eqrsubl{pf:solutions1:Eq}{
h_{\rm F}(t, y, z)&=&\bar{c}t+\tilde{c}
+\sum_{\ell}\frac{M_\ell}{\left[|\bm y-\bm y_\ell|^2
+\frac{4M}{(p-4)^2}|\bm z-\bm z_0|^{p-4}\right]^{\frac{p^2-6p+12}{2(p-4)}}},
 \label{pf:solution-r:Eq}\\
h(z)&=&\frac{M}{|\bm z-\bm z_0|^{6-p}},
 \label{pf:solution-s:Eq}
}
where $\bar{c}$, $\tilde{c}$, $M_\ell$ and $M$ are constant parameters,
and $\bm y_\ell$ and $\bm z_0$ are
constant vectors representing the positions of the branes.
For $p=4$, the solution becomes  
\Eqrsubl{pf:solutions2:Eq}{
h_{\rm F}(t, y, z)&=&\bar{c}t+\tilde{c}
+\sum_{\ell}M_\ell\left[|\bm y-\bm y_\ell|^2
-pM\ln|\bm z-\bm z_0|\right],
 \label{pf:solution-r2:Eq}\\
h(z)&=&\frac{M}{|\bm z-\bm z_0|^2}\,.
 \label{pf:solution-s2:Eq}
}
Next we consider the case of $p=6$. The solution is given by 
\Eqrsubl{pf:solutions3:Eq}{
h_{\rm F}(t, y, z)&=&\bar{c}t+\tilde{c}
+\sum_{\ell}\frac{M_\ell}{\left[|\bm y-\bm y_\ell|^2
+M|\bm z-\bm z_0|^2\right]^3},
 \label{pf:solution-r3:Eq}\\
h(z)&=&M\ln|\bm z-\bm z_0|\,.
 \label{pf:solution-s3:Eq}
}

\subsubsection{The intersection involving fundamental string and NS5-branes}
We discuss the solution of the fundamental strings with NS5-branes. 
We consider the case to be delocalized 
along one of the overall transverse directions. 
The ten-dimensional metric thus takes the form
\Eqr{
ds^2&=&h_{\rm F}^{-3/4}(x, y, z)h_{\rm NS}^{-1/4}(z)\left[
q_{\mu\nu}(\Xsp)dx^{\mu}dx^{\nu}
+h_{\rm F}(x, y, z)\gamma_{ij}(\Ysp)dy^idy^j\right.\nn\\
&&\left.+h_{\rm F}(x, y, z)h_{\rm NS}(z)u_{ab}(\Zsp)dz^adz^b\right], 
 \label{fns:metric:Eq}
}
where $q_{\mu\nu}$ is the two-dimensional metric which
depends only on the two-dimensional coordinates $x^{\mu}$, 
$\gamma_{ij}$ is the four-dimensional metric which
depends only on the four-dimensional coordinates $y^i$, 
and finally $u_{ab}$ is the four-dimensional metric which
depends only on the four-dimensional coordinates $z^a$. 

We also assume that the scalar field $\phi$ and the gauge field 
strength $H_{\left(3\right)}$ are given by
\Eqrsubl{fns:ansatz:Eq}{
\e^{\phi}&=&h_{\rm F}^{-1/2}h_{\rm NS}^{1/2}\,,
  \label{fns:ansatz for scalar:Eq}\\
H_{\left(3\right)}&=&d\left[h_{\rm F}^{-1}(x, y, z)\right]\wedge \Omega(\Xsp)
  +\e^{-\phi}\ast d\left[h_{\rm NS}^{-1}(z)\Omega(\Xsp)\wedge 
  \Omega(\Ysp)\right]\,,
  \label{fns:ansatz for gauge:Eq}
}
where $\Omega(\Xsp)$ and $\Omega(\Ysp)$ 
denote the volume two- and four-form 
\Eqrsubl{fns:volume:Eq}{
\Omega(\Xsp)&=&\sqrt{-q}\,dx^0\wedge dx^1\,,
   \label{fns:volume x:Eq}\\
\Omega(\Ysp)&=&\sqrt{\gamma}\,dy^1\wedge dy^2 \wedge dy^3\wedge dy^4\,.
   \label{fns:volume y:Eq}
}
Here, $q$ and $\gamma$ are the determinant of the metric $q_{\mu\nu}$ and 
$\gamma_{ij}$, respectively.

In terms of ansatz for fields \eqref{fns:metric:Eq} and 
\eqref{fns:ansatz:Eq}, the field equations lead to
\Eqrsubl{fns:solution1:Eq}{
&&\hspace{-0.7cm}R_{\mu\nu}(\Xsp)=0,~~~~R_{ij}(\Ysp)=0,~~~~R_{ab}(\Zsp)=0,
   \label{fns:Ricci:Eq}\\
&&\hspace{-0.7cm}h_{\rm F}=h_0(x)+h_1(y, z),~~~
D_{\mu}D_{\nu}h_0=0\,, ~~~
h_{\rm NS}\lap_{\Ysp}h_1+\triangle_{\Zsp}h_1=0\,,
~~~\triangle_{\Zsp}h_{\rm NS}=0\,,
   \label{fns:warp1-2:Eq}
 }
where $D_{\mu}$ is the covariant derivative with respect to
the metric $q_{\mu\nu}$, and 
$\triangle_{\Ysp}$, $\lap_{\Zsp}$ are
the Laplace operators on 
$\Ysp$, $\Zsp$ space, and $R_{\mu\nu}(\Xsp)$, 
$R_{ij}(\Ysp)$, $R_{ab}(\Zsp)$ are the Ricci tensors 
associated with the metrics $q_{\mu\nu}(\Xsp)$, 
$\gamma_{ij}(\Ysp)$, $u_{ab}(\Zsp)$, respectively. 
We assume that the ten-dimensional metric is given by 
\Eq{
q_{\mu\nu}=\eta_{\mu\nu}\,,~~~\gamma_{ij}=\delta_{ij}\,,~~~
u_{ab}=\delta_{ab}\,,
 \label{fns:flat metric:Eq}
 }
where $\eta_{\mu\nu}$ is the two-dimensional
Minkowski metric and $\delta_{ij}$, $\delta_{ab}$ are
the four-dimensional Euclidean metrics,
respectively. 
The solution for $h_{\rm F}$ and $h_{\rm NS}$ can be
obtained explicitly as
\Eqrsubl{fns:solutions1:Eq}{
h_{\rm F}(x, y, z)&=&c_{\mu}x^{\mu}+\tilde{c}
+\sum_{\ell}M_\ell\left[|\bm y-\bm y_\ell|^2
-4M\ln|\bm z-\bm z_0|\right]\,,
 \label{fns:solution-r:Eq}\\
h_{\rm NS}(z)&=&\frac{M}{|\bm z-\bm z_0|^2},
 \label{fns:solution-s:Eq}
}
where $c_{\mu}$, $\tilde{c}$, $M_\ell$ and $M$ are constant parameters,
and $\bm y_\ell$ and $\bm z_0$ are
constant vectors representing the positions of the branes.
If we delocalize the solution along one of the overall 
transverse directions, the solution can be written as 
\Eqrsubl{fns:solutions2:Eq}{
h_{\rm F}(x, y, z)&=&c_{\mu}x^{\mu}+\tilde{c}
+\sum_{\ell}\frac{M_\ell}{\left[|\bm y-\bm y_\ell|^2
+4M|\bm z-\bm z_0|\right]^3}\,,
 \label{fns:solution-r2:Eq}\\
h_{\rm NS}(z)&=&\frac{M}{|\bm z-\bm z_0|}\,.
 \label{fns:solution-s2:Eq}
}

\subsubsection{The pair involving fundamental string and one Kaluza-
Klein monopole}
Now we discuss the KK-monopole in the transverse space of the 
fundamental string.  
We set the ten-dimensional metric takes the form
\Eqr{
ds^2&=&h_{\rm F}^{-3/4}(x, y, z)q_{\mu\nu}(\Xsp)dx^{\mu}dx^{\nu}
+h_{\rm F}^{1/4}(x, y, z)\left[\gamma_{ij}(\Ysp)dy^idy^j\right.\nn\\
&&\left.+h_k(z)u_{ab}(\Zsp)dz^adz^b
+h_k^{-1}(z)\left(dv+A_adz^a\right)^2\right], 
 \label{kkf1:metric:Eq}
}
where $q_{\mu\nu}$ is the two-dimensional metric which
depends only on the two-dimensional coordinates $x^{\mu}$, 
$\gamma_{ij}$ is the four-dimensional metric which
depends only on the four-dimensional coordinates $y^i$, 
and finally $u_{ab}$ is the three-dimensional metric which
depends only on the three-dimensional coordinates $z^a$. 

We assume that the scalar field $\phi$ and the gauge field 
strength $H_{\left(3\right)}$ are given by
\Eqrsubl{kkf1:ansatz:Eq}{
\e^{\phi}&=&h_{\rm F}^{-1/2}\,,
  \label{kkf1:ansatz for scalar:Eq}\\
H_{\left(3\right)}&=&d\left[h_{\rm F}^{-1}(x, y, z)\right]\wedge\Omega(\Xsp),
  \label{kkf1:ansatz for gauge:Eq}
}
where $\Omega(\Xsp)$ 
denotes the volume two-form 
\Eqr{
\Omega(\Xsp)=\sqrt{-q}\,dx^0\wedge dx^1\,.
   \label{kkf1:volume:Eq}
}
Here, $q$ is the determinant of the metric $q_{\mu\nu}$.

In terms of ansatz for fields \eqref{kkf1:metric:Eq} and 
\eqref{kkf1:ansatz:Eq}, the field equations lead to
\Eqrsubl{kkf1:solution1:Eq}{
&&R_{\mu\nu}(\Xsp)=0,~~~~R_{ij}(\Ysp)=0,~~~~R_{ab}(\Zsp)=0,
   \label{kkf1:Ricci:Eq}\\
&&h_{\rm F}=h_0(x)+h_1(y, z),~~~
D_{\mu}D_{\nu}h_0=0\,, ~~~
h_k\lap_{\Ysp}h_1+\triangle_{\Zsp}h_1=0\,,~~~\triangle_{\Zsp}h_k=0\,,
   \label{kkf1:warp1-2:Eq}
 }
where $D_{\mu}$ is the covariant derivative with respect to
the metric $q_{\mu\nu}$, and 
$\triangle_{\Ysp}$, $\lap_{\Zsp}$ are
the Laplace operators on $\Ysp$, $\Zsp$ space, and $R_{\mu\nu}(\Xsp)$, 
$R_{ij}(\Ysp)$, $R_{ab}(\Zsp)$ are the Ricci tensors 
associated with the metrics $q_{\mu\nu}(\Xsp)$, 
$\gamma_{ij}(\Ysp)$, $u_{ab}(\Zsp)$, respectively. 

Now let us consider the case
\Eq{
q_{\mu\nu}=\eta_{\mu\nu}\,,~~~\gamma_{ij}=\delta_{ij}\,,
~~~u_{ab}=\delta_{ab}\,,
 \label{kkf1:flat metric:Eq}
 }
where $\eta_{\mu\nu}$ is the two-dimensional
Minkowski metric and $\delta_{ij}$, $\delta_{ab}$ are
the four-, three-dimensional Euclidean metrics,
respectively. 
The solution for $h_{\rm F}$ and $h_k$ can be
obtained explicitly as
\Eqrsubl{kkf1:solutions1:Eq}{
h_{\rm F}(x, y, z)&=&c_{\mu}x^{\mu}+\tilde{c}
+\sum_{\ell}\frac{M_\ell}{\left[|\bm y-\bm y_\ell|^2
+4M|\bm z-\bm z_0|\right]^3},
 \label{kkf1:solution-r:Eq}\\
h_k(z)&=&\frac{M}{|\bm z-\bm z_0|},
 \label{kkf1:solution-s:Eq}
}
where $c_{\mu}$, $\tilde{c}$, $M_\ell$ and $M$ are constant parameters,
and $\bm y_\ell$ and $\bm z_0$ are
constant vectors representing the positions of the branes.

\subsection{The intersection involving NS5-branes}
In this subsection, we discuss the intersecting brane  
involving NS5-branes. For the KK monopole, we cannot find 
partially delocalized solutions because 
the intersection does not have the relative transverse directions. 

\subsubsection{The intersection involving D$p$-brane and NS5-branes}

Let us first consider the D$p$-branes ending on NS5-branes. 
We discuss the solution to be delocalized along 
the relative transverse direction of the NS5-branes.
The ten-dimensional metric of D$p$-branes $(p\le 6)$ ending on 
NS5-branes thus takes the form
\Eqr{
ds^2&=&h^{(p-7)/8}(x, y, z)h_{\rm NS}^{-1/4}(z)\left[q_{\mu\nu}(\Xsp)
dx^{\mu}dx^{\nu}+h(x, y, z)\gamma_{ij}(\Ysp)dy^idy^j
+h_{\rm NS}(z)dv^2\right.\nn\\
&&\left.+h(x, y, z)h_{\rm NS}(z)u_{ab}(\Zsp)dz^adz^b\right], 
 \label{pns:metric:Eq}
}
where $q_{\mu\nu}$ is the $p$-dimensional metric which
depends only on the $p$-dimensional coordinates $x^{\mu}$, 
$\gamma_{ij}$ is the $(6-p)$-dimensional metric which
depends only on the $(6-p)$-dimensional coordinates $y^i$, 
and finally $u_{ab}$ is the three-dimensional metric which
depends only on the three-dimensional coordinates $z^a$. 

We assume that the scalar field $\phi$ and the gauge field 
strength $H_{\left(3\right)}$ are given by
\Eqrsubl{pns:ansatz:Eq}{
\e^{\phi}&=&h_{\rm NS}^{1/2}\,h^{(3-p)/4}\,,
  \label{pns:ansatz for scalar:Eq}\\
H_{\left(3\right)}&=&\e^{-\phi}\ast d\left[h_{\rm NS}^{-1}(z)
\,\Omega(\Xsp)\wedge \Omega(\Ysp)\right],
  \label{pns:ansatz for gauge:Eq}\\
F_{\left(p+2\right)}&=&d\left[h^{-1}(x, y, z)\right]\wedge\Omega(\Xsp)\,,
  \label{pns:ansatz for gauge2:Eq}
}
where $\Omega(\Xsp)$ and $\Omega(\Ysp)$ 
denote the volume $p$-form, $(6-p)$-form, respectively 
\Eqrsubl{pns:volume:Eq}{
\Omega(\Xsp)&=&\sqrt{-q}\,dx^0\wedge dx^1 \wedge\cdots\wedge dx^{p-1}\,,
   \label{pns:volume x:Eq}\\
\Omega(\Ysp)&=&\sqrt{\gamma}\,dy^1\wedge dy^2 \wedge\cdots\wedge dy^{6-p}\,.
   \label{pns:volume y:Eq}
}
Here, $q$ and $\gamma$ are the determinant of the metric $q_{\mu\nu}$, 
$\gamma_{ij}$, respectively.

Using the ansatz for fields \eqref{pns:metric:Eq} and 
\eqref{pns:ansatz:Eq}, the field equations lead to
\Eqrsubl{pns:solution1:Eq}{
&&R_{\mu\nu}(\Xsp)=0,~~~~R_{ij}(\Ysp)=0,~~~~R_{ab}(\Zsp)=0,
   \label{pns:Ricci:Eq}\\
&&h=h_0(x)+h_1(z),~~~
D_{\mu}D_{\nu}h_0=0\,,~~~
h_{\rm NS}\lap_{\Ysp}h_1+\triangle_{\Zsp}h_1=0\,,
~~~\triangle_{\Zsp}h_{\rm NS}=0\,,
   \label{pns:warp1-2:Eq}
 }
where $D_{\mu}$ is the covariant derivative constructed by the metric 
$q_{\mu\nu}$, and $\triangle_{\Ysp}$, $\lap_{\Zsp}$ are
the Laplace operators on $\Ysp$, $\Zsp$ space, and $R_{\mu\nu}(\Xsp)$, 
$R_{ij}(\Ysp)$, $R_{ab}(\Zsp)$ are the Ricci tensors
associated with the metrics $q_{\mu\nu}(\Xsp)$, 
$\gamma_{ij}(\Ysp)$, $u_{ab}(\Zsp)$, 
respectively. 
Let us consider the case
\Eq{
q_{\mu\nu}=\eta_{\mu\nu}\,,~~~
\gamma_{ij}=\delta_{ij}\,,~~~u_{ab}=\delta_{ab}\,,
 \label{pns:flat metric:Eq}
 }
where $\eta_{\mu\nu}$ is the $p$-dimensional
Minkowski metric and $\delta_{ij}$, $\delta_{ab}$ are
the $(6-p)$-, three-dimensional Euclidean metrics,
respectively. 
The solution for $h$ and $h_{\rm NS}$ can be
obtained explicitly as
\Eqrsubl{pns:solutions1:Eq}{
h(x, y, z)&=&c_{\mu}x^{\mu}+\tilde{c}
+\sum_{\ell}\frac{M_\ell}{\left[|\bm y-\bm y_\ell|^2
+4M|\bm z-\bm z_0|\right]^{\frac{(p-8)}{2}}},
 \label{pns:solution-r:Eq}\\
h_{\rm NS}(z)&=&\frac{M}{|\bm z-\bm z_0|},
 \label{pns:solution-s:Eq}
}
where $c_{\mu}$, $\tilde{c}$, $M_\ell$ and $M$ are constant parameters,
and $\bm y_\ell$ and $\bm z_0$ are
constant vectors representing the positions of the branes.

\subsubsection{The intersection of two NS5-branes}

Next we consider the solution of two NS5-brane. As we mentioned in 
Sec. \ref{sec:rs}, these intersect over 3 dimensions.  
We assume that the ten-dimensional metric is written by
\Eqr{
ds^2&=&h^{-1/4}_{\rm NS}(x, y, z)\bar{h}_{\rm NS}^{-1/4}(z)
\left[q_{\mu\nu}(\Xsp)
dx^{\mu}dx^{\nu}+h_{\rm NS}(x, y, z)\gamma_{ij}(\Ysp_1)dy^idy^j\right.\nn\\
&&\left.+\bar{h}_{\rm NS}(z)w_{mn}(\Ysp_2)dv^mdv^n
+h_{\rm NS}(x, y, z)\bar{h}_{\rm NS}(z)u_{ab}(\Zsp)dz^adz^b\right], 
 \label{ns:metric:Eq}
}
where $q_{\mu\nu}$ is the four-dimensional metric which
depends only on the four-dimensional coordinates $x^{\mu}$, 
$\gamma_{ij}$ is the two-dimensional metric which
depends only on the two-dimensional coordinates $y^i$, 
$w_{mn}$ is the two-dimensional metric which
depends only on the two-dimensional coordinates $v^m$, 
and finally $u_{ab}$ is the two-dimensional metric which
depends only on the two-dimensional coordinates $z^a$. 

We also assume that the scalar field $\phi$ and the gauge field 
strength $H_{\left(3\right)}$ are given by
\Eqrsubl{ns:ansatz:Eq}{
\e^{\phi}&=&\left(h_{\rm NS}\bar{h}_{\rm NS}\right)^{1/2}\,,
  \label{ns:ansatz for scalar:Eq}\\
H_{\left(3\right)}&=&\e^{-\phi}\ast d\left[h_{\rm NS}^{-1}(x, y, z)
\,\Omega(\Xsp)\wedge \Omega(\Ysp_2)+\bar{h}_{\rm NS}^{-1}(z)
\,\Omega(\Xsp)\wedge \Omega(\Ysp_1)\right],
  \label{ns:ansatz for gauge:Eq}
}
where $\Omega(\Xsp)$, $\Omega(\Ysp_1)$ and $\Omega(\Ysp_2)$ 
denote the volume four-form, two-form and two-form, respectively 
\Eqrsubl{ns:volume:Eq}{
\Omega(\Xsp)&=&\sqrt{-q}\,dx^0\wedge dx^1 \wedge dx^2\wedge dx^{3}\,,
   \label{ns:volume x:Eq}\\
\Omega(\Ysp_1)&=&\sqrt{\gamma}\,dy^1\wedge dy^2\,,
   \label{ns:volume y1:Eq}\\
\Omega(\Ysp_2)&=&\sqrt{w}\,dv^1\wedge dv^2\,.
   \label{ns:volume y2:Eq}
}
Here, $q$, $\gamma$ and $w$ are the determinants of the metrics $q_{\mu\nu}$, 
$\gamma_{ij}$, and $w_{mn}$, respectively.

In terms of ansatz for fields \eqref{ns:metric:Eq} and 
\eqref{ns:ansatz:Eq}, the field equations lead to
\Eqrsubl{ns:solution1:Eq}{
&&R_{\mu\nu}(\Xsp)=0,~~~~R_{ij}(\Ysp_1)=0,~~~~R_{mn}(\Ysp_2)=0,~~~~
R_{ab}(\Zsp)=0,
   \label{ns:Ricci:Eq}\\
&&h_{\rm NS}=h_0(x)+h_1(y, z),
   \label{ns:h:Eq}\\
&&D_{\mu}D_{\nu}h_0=0, 
~~~\bar{h}_{\rm NS}\lap_{\Ysp_1}h_1+\triangle_{\Zsp}h_1=0\,,
~~~\triangle_{\Zsp}\bar{h}_{\rm NS}=0\,,
   \label{ns:warp:Eq}
 }
where $D_{\mu}$ is the covariant derivative constructed by the metric 
$q_{\mu\nu}$, and $\triangle_{\Ysp_1}$, $\lap_{\Zsp}$ are
the Laplace operators on $\Ysp_1$, $\Zsp$ space, and $R_{\mu\nu}(\Xsp)$, 
$R_{ij}(\Ysp_1)$, $R_{mn}(\Ysp_2)$, $R_{ab}(\Zsp)$ are the Ricci tensors
associated with the metrics $q_{\mu\nu}(\Xsp)$, 
$\gamma_{ij}(\Ysp_1)$, $u_{ab}(\Zsp)$, 
respectively. 
As a special example, we consider the case
\Eq{
q_{\mu\nu}=\eta_{\mu\nu}\,,~~~\gamma_{ij}=\delta_{ij}\,,~~~
w_{mn}=\delta_{mn}\,,~~~u_{ab}=\delta_{ab}\,,
 \label{ns:flat metric:Eq}
 }
where $\eta_{\mu\nu}$ is the four-dimensional
Minkowski metric and $\delta_{ij}$, $\delta_{mn}$, $\delta_{ab}$ are
the two-, two-, two-dimensional Euclidean metrics,
respectively. 
The solution for $h_{\rm NS}$ and $\bar{h}_{\rm NS}$ can be
obtained explicitly as
\Eqrsubl{ns:solutions1:Eq}{
h_{\rm NS}(x, y, z)&=&c_{\mu}x^{\mu}+\tilde{c}
+\sum_{\ell}\frac{M_\ell}{|\bm y-\bm y_\ell|^2
+M|\bm z-\bm z_0|^2},
 \label{ns:solution-r:Eq}\\
\bar{h}_{\rm NS}(z)&=&\ln \left[M|\bm z-\bm z_0|\right],
 \label{ns:solution-s:Eq}
}
where $c_{\mu}$, $\tilde{c}$, $M_\ell$ and $M$ are constant parameters,
and $\bm y_\ell$ and $\bm z_0$ are
constant vectors representing the positions of the branes.

\subsubsection{The intersection involving plane wave and NS5-brane}

We present the NS5-brane with the plane wave propagating along
its longitudinal direction. 
We assume that the ten-dimensional metric takes the form
\Eqr{
ds^2&=&h_{\rm NS}^{-1/4}(z)\left[-dt^2+dx^2
+\left\{h_w(t, y, z)-1\right\}\left(dt-dx\right)^2\right.\nn\\
&&\left.+\gamma_{ij}(\Ysp)dy^idy^j+h_{\rm NS}(z)u_{ab}(\Zsp)dz^adz^b\right], 
 \label{wns:metric:Eq}
}
where $\gamma_{ij}$ is the four-dimensional metric which
depends only on the four-dimensional coordinates $y^i$, 
and finally $u_{ab}$ is the four-dimensional metric which
depends only on the four-dimensional coordinates $z^a$. 

We set the scalar field $\phi$ and the gauge field 
strength $H_{\left(3\right)}$ as follows: 
\Eqrsubl{wns:ansatz:Eq}{
\e^{\phi}&=&h_{\rm NS}^{1/2}\,,
  \label{wns:ansatz for scalar:Eq}\\
H_{\left(3\right)}&=&\e^{-\phi}\ast d\left[h_{\rm NS}^{-1}(z)\wedge dt
\wedge dx\wedge\Omega(\Ysp)\right],
  \label{wns:ansatz for gauge:Eq}
}
where $\Omega(\Ysp)$ 
denotes the volume four-form 
\Eqr{
\Omega(\Ysp)=\sqrt{\gamma}\,dy^1\wedge dy^2\wedge dy^3\wedge dy^4\,.
   \label{wns:volume:Eq}
}
Here, $\gamma$ is the determinant of the metric $\gamma_{ij}$.

In terms of ansatz for fields \eqref{wns:metric:Eq} and 
\eqref{wns:ansatz:Eq}, the field equations lead to
\Eqrsubl{wns:solution1:Eq}{
&&R_{ij}(\Ysp)=0,~~~~R_{ab}(\Zsp)=0,
   \label{wns:Ricci:Eq}\\
&&h_w=h_0(t)+h_1(y, z),~~~
\pd_t^2h_0=0\,,~~~
h_k\lap_{\Ysp}h_1+\triangle_{\Zsp}h_1=0\,,~~~\triangle_{\Zsp}h_{\rm NS}=0\,,
   \label{wns:warp1-2:Eq}
 }
where $\triangle_{\Ysp}$, $\lap_{\Zsp}$ are
the Laplace operators on $\Ysp$, $\Zsp$ space, and 
$R_{ij}(\Ysp)$, $R_{ab}(\Zsp)$ are the Ricci tensors
associated with the metrics $\gamma_{ij}(\Ysp)$, $u_{ab}(\Zsp)$, 
respectively. 
As a special example, we set the metric and the function $h_{\rm NS}$ 
\Eq{
\gamma_{ij}=\delta_{ij}\,,
~~~u_{ab}=\delta_{ab}\,,
 \label{wns:flat metric:Eq}
 }
where $\delta_{ij}$, $\delta_{ab}$ are
the four-dimensional Euclidean metrics, respectively. 
The solution for $h_{\rm NS}$ and $h_w$ can be
obtained explicitly as
\Eqrsubl{wns:solutions1:Eq}{
h_w(t, y, z)&=&\bar{c}t+\tilde{c}
+\sum_{\ell}M_\ell\left[|\bm y-\bm y_\ell|^2
-4M\ln|\bm z-\bm z_0|\right],
 \label{wns:solution-r:Eq}\\
h_{\rm NS}(z)&=&\frac{M}{|\bm z-\bm z_0|^2},
 \label{wns:solution-s:Eq}
}
where $\bar{c}$, $\tilde{c}$, $M_\ell$ and $M$ are constant parameters,
and $\bm y_\ell$ and $\bm z_0$ are
constant vectors representing the positions of the branes.

\subsection{The plane wave in the KK-monopole background}

We consider the plane wave propagating in the background of the 
KK-monopole. The solution of ten-dimensional metric is given by 
\Eqr{
ds^2&=&-dt^2+dx^2
+\left[h_w(t, y, z)-1\right]\left(dt-dx\right)^2
+\gamma_{ij}(\Ysp)dy^idy^j\nn\\
&&+h_k(z)u_{ab}(\Zsp)dz^adz^b
+h_k^{-1}(z)\left(dv+A_adz^a\right)^2, 
 \label{kkw:metric:Eq}
}
where $\gamma_{ij}$ is the four-dimensional metric which
depends only on the four-dimensional coordinates $y^i$, 
and finally $u_{ab}$ is the three-dimensional metric which
depends only on the three-dimensional coordinates $z^a$. 

The ten-dimensional metric and the function $h_k$ obey 
\Eqrsubl{kkw:solution1:Eq}{
&&R_{ij}(\Ysp)=0,~~~~R_{ab}(\Zsp)=0,
   \label{kkw:Ricci:Eq}\\
&&h_w=h_0(t)+h_1(y, z),~~~~\pd^2_th_0=0,~~~~
h_k\lap_{\Ysp}h_1+\triangle_{\Zsp}h_1=0\,,~~~\triangle_{\Zsp}h_k=0\,,
   \label{kkw:h:Eq}\\
&&dh_k=\ast_{\Zsp}dA\,,
   \label{kkw:warp:Eq}
 }
where $\ast_{\Zsp}$ is the Hodge operator in the Z space, and 
$\triangle_{\Ysp}$, $\lap_{\Zsp}$ are
the Laplace operators on $\Ysp$, $\Zsp$ space, and 
 $R_{ij}(\Ysp)$, $R_{ab}(\Zsp)$ are the Ricci tensors
associated with the metrics 
$\gamma_{ij}(\Ysp)$, $u_{ab}(\Zsp)$, 
respectively. 
Now we consider the case
\Eq{
\gamma_{ij}=\delta_{ij}\,,
~~~u_{ab}=\delta_{ab}\,,
 \label{kkw:flat metric:Eq}
 }
where $\delta_{ij}$, $\delta_{ab}$ are
the four-, three-dimensional Euclidean metrics,
respectively. 
The solution for $h$ and $h_k$ can be
obtained explicitly as
\Eqrsubl{kkw:solutions1:Eq}{
h(t, y, z)&=&\bar{c}t+\tilde{c}
+\sum_{\ell}\frac{M_\ell}{\left[|\bm y-\bm y_\ell|^2
+4M|\bm z-\bm z_0|\right]^3},
 \label{kkw:solution-r:Eq}\\
h_k(z)&=&\frac{M}{|\bm z-\bm z_0|},
 \label{kkw:solution-s:Eq}
}
where $\bar{c}$, $\tilde{c}$, $M_\ell$ and $M$ are constant parameters,
and $\bm y_\ell$ and $\bm z_0$ are
constant vectors representing the positions of the branes. 

There is a classification of the multiple intersecting branes solutions by  
\cite{Bergshoeff:1996rn, Bergshoeff:1997tt}. 
The dynamical delocalized branes in ten-dimensional theory  
are also classified in \cite{Minamitsuji:2010kb}. 
We again show the intersection rule for the branes with 
M-wave and KK-monopoles in Table \ref{table_1}. 
In the Table, circles indicate where the
brane world-volumes enter, $v$ represents the coordinate of the KK-monopole, 
and the time-dependent branes are indicated by $\surd$ for 
different solutions.

\section{Cosmology}
\label{sec:co}

In this section, we study the four-dimensional cosmology 
by using above solutions.
We assume an isotropic and homogeneous three-space
in the four-dimensional spacetime 
known as Friedmann-Robertson-Walker universe after compactification.
Now we set the $(p+1)$-dimensional Minkowski spacetime
with $q_{\mu\nu}(\Xsp)=\eta_{\mu\nu}(\Xsp)$. The functions $h_r$ and $h_s$
do not depend on the coordinate of $\Xsp$ space except for the time.
We discuss just the cases involving $p$-brane and KK-monopole because 
our Universe does not expand when the wave 
is time-dependent. Hence, we have no interesting case for wave solution. 

\subsection{The intersection of D$p_r$-D$p_s$ brane system}
Let us first discuss how these solutions are applied to our physical world  
in the case of D$p_r$-D$p_s$ brane system. Suppose that
our Universe is a part of branes. 
Since our Universe is isotropic and homogeneous, same branes must 
contain this whole three dimensions.
The $D$-dimensional metric (\ref{rs:metric:Eq}) can be expressed as
\Eq{
ds^2=-hdt^2+ds^2(\tilde{\Xsp})+ds^2(\Ysp_1)+ds^2(\Ysp_2)+ds^2(\Zsp),
   \label{rsc:metric:Eq}
}
where line elements $ds^2(\tilde{\Xsp})$, $ds^2(\Ysp_1)$, $ds^2(\Ysp_2)$,
 and $ds^2(\Zsp)$ are defined by 
\Eqrsubl{rsc:cmetric:Eq}{
ds^2(\tilde{\Xsp})&\equiv&h\delta_{PQ}(\tilde{\Xsp})d\theta^{P}d\theta^{Q},\\
ds^2(\Ysp_1)&\equiv&h^{b_r}_r(t, y, z)
h_s^{a_s}(t, v, z)\gamma_{ij}(\Ysp_1)dy^idy^j,\\
ds^2(\Ysp_2)&\equiv&h^{a_r}_r(t, y, z)
h_s^{b_s}(t, v, z)w_{mn}(\Ysp_2)dv^{m}dv^{n},\\
ds^2(\Zsp)&\equiv&h^{b_r}_r(t, y, z)h_s^{b_s}(t, v, z)u_{ab}(\Zsp)dz^adz^b,\\
h&\equiv&h^{a_r}_r(t, y, z)h_s^{a_s}(t, v, z).
 }
Here, $\delta_{PQ}(\tilde{\Xsp})$ denotes the $p$-dimensional Euclidean
metric, and $\theta^P$ is the coordinate of the $p$-dimensional Euclid
space $\tilde{\Xsp}$.

Now we set $h_s=h_s(z)$ and
$h_r=At+h_1(y, z)$. Then, the $D$-dimensional metric
(\ref{rsc:cmetric:Eq}) is given by
\Eqr{
ds^2&=&h_s^{a_s}\left[1+\left(\frac{\tau}{\tau_0}\right)^{-2/(a_r+2)}
h_1\right]^{a_r}
\left[-d\tau^2+\left(\frac{\tau}{\tau_0}\right)^{2a_r/(a_r+2)}
\delta_{PQ}(\tilde{\Xsp})d\theta^Pd\theta^Q\right.\nn\\
&&+\left\{1+\left(\frac{\tau}{\tau_0}\right)^{-2/(a_r+2)}h_1\right\}
\left(\frac{\tau}{\tau_0}\right)^{2b_r/(a_r+2)}
\gamma_{ij}(\Ysp_1)dy^idy^j\nn\\
&&+h_s\left(\frac{\tau}{\tau_0}\right)^{2a_r/(a_r+2)}
w_{mn}(\Ysp_2)dv^mdv^n\nn\\
&&\left.+h_s\left\{1+\left(\frac{\tau}{\tau_0}\right)^{-2/(a_r+2)}
h_1\right\}\left(\frac{\tau}{\tau_0}\right)^{2b_r/(a_r+2)}
u_{ab}(\Zsp)dz^adz^b\right],
   \label{rsc:metric1:Eq}
}
where the cosmic time $\tau$ is defined by
\Eq{
\frac{\tau}{\tau_0}=\left(At\right)^{(a_r+2)/2},~~~~\tau_0=
\frac{2}{\left(a_r+2\right)A}.
}
For $h_r=h_r(z)$ and $h_s=At+k_1(v, z)$, the metric
(\ref{rsc:cmetric:Eq}) is written by replacing $a_r$ and $h_1(y, z)$ 
with $a_s$ and $k_1(v, z)$.

In the following, 
we apply the solution to lower-dimensional effective theory.
We compactify $d(\equiv d_1+d_2+d_3+d_4)$ dimensions to fit our Universe,
where $d_1$, $d_2$, $d_3$, and $d_4$ are the compactified dimensions with
respect to the $\tilde{\Xsp}$, $\Ysp_1$, $\Ysp_2$, and $\Zsp$ spaces.
The $D$-dimensional metric (\ref{rsc:metric:Eq}) is then expressed as
\Eq{
ds^2=ds^2(\Msp)+ds^2(\Nsp)\,,
   \label{rsc:metric2:Eq}
}
where $ds^2(\Msp)$ denotes the $(D-d)$-dimensional metric and
$ds^2(\Nsp)$ is the line element of compactified dimensions.
We can rewrite the $(D-d)$-dimensional metric in the Einstein frame 
by the conformal transformation
\Eq{
ds^2(\Msp)=h_r^Bh_s^Cds^2(\bar{\Msp})\,,
}
where $B$ and $C$ are given by 
\Eq{
B=\frac{-(a_r+1)d+d_1+d_3}{D-d-2},~~~~~~C=\frac{-(a_s+1)d+d_2+d_4}{D-d-2}.
}
The $(D-d)$-dimensional metric in the Einstein frame is written by 
\Eqr{
ds^2(\bar{\Msp})&=&h_r^{B'}h_s^{C'}\left[-dt^2+\delta_{P'Q'}
(\tilde{\Xsp}')d\theta^{P'}d\theta^{Q'}
+h_r\gamma_{k'l'}({\Ysp_1}')dy^{k'}dy^{l'}\right.\nn\\
&&\left.+h_sw_{m'n'}({\Ysp_2}')dv^{m'}dv^{n'}
+h_rh_su_{a'b'}({\Zsp}')dz^{a'}dz^{b'}\right],
  \label{rsc:metric-m:Eq}
}
where $B'$ and $C'$ are defined by $B'=-B+a_r$ and $C'=-C+a_s$,
and $\tilde{\Xsp}'$, ${\Ysp_1}'$, ${\Ysp_2}'$, and ${\Zsp}'$ denote
the $(p-d_1)$-, $(p_s-p-d_2)$-, $(p_r-p-d_3)$-, and
$(D+p-p_r-p_s-d_4)$-dimensional spaces, respectively.

If we set $h_r=At+h_1$, 
the metric (\ref{rsc:metric-m:Eq}) is thus rewritten as
\Eqr{
ds^2(\bar{\Msp})&=&h_s^{C'}\left[1+\left(\frac{\tau}{\tau_0}\right)
^{-2/(B'+2)}h_1\right]^{B'}\left[-d\tau^2+
\left(\frac{\tau}{\tau_0}\right)^{2B'/(B'+2)}
\delta_{P'Q'}(\tilde{\Xsp}')d\theta^{P'}d\theta^{Q'}\right.\nn\\
&&+\left\{1+\left(\frac{\tau}{\tau_0}\right)^{-2/(B'+2)}h_1\right\}
\left(\frac{\tau}{\tau_0}\right)^{2(B'+1)/(B'+2)}
\gamma_{k'l'}({\Ysp_1}')dy^{k'}dy^{l'}\nn\\
&&+h_s\left(\frac{\tau}{\tau_0}\right)^{2B'/(B'+2)}
w_{m'n'}({\Ysp_2}')dv^{m'}dv^{n'}\nn\\
&&\left.+h_s\left\{1+\left(\frac{\tau}{\tau_0}\right)^{-2/(B'+2)}
h_1\right\}\left(\frac{\tau}{\tau_0}\right)^{2(B'+1)/(B'+2)}
u_{a'b'}({\Zsp}')dz^{a'}dz^{b'}\right],
  \label{rsc:mr:Eq}
}
where we have introduced the cosmic time $\tau$ defined by
\Eq{
\frac{\tau}{\tau_0}=\left(At\right)^{(B'+2)/2},~~~~\tau_0=
\frac{2}{\left(B'+2\right)A}.
}
For $h_s=At+k_1$ and $\pd_{\mu}h_r=0$,
one can find results similar to
(\ref{rsc:metric1:Eq}) and (\ref{rsc:mr:Eq}). 
Hence, we cannot find the solution which
gives an accelerating expansion of our Universe. 

The power exponents of the scale factor of possible four-dimensional 
Universe are given by $a(\tilde{\Msp})\propto \tau^{\lambda(\tilde{\Msp})}$,
where $\tilde{\Msp}$ denotes the spatial part of the spacetime $\Msp$, 
$\tau$ denotes the cosmic time, and $a(\tilde{\Msp})$ and
$a_{\rm E}(\tilde{\Msp})$ are the scale factors of the space
$\tilde{\Msp}$ in Jordan and Einstein frames with the exponents carrying
the same suffices, respectively. 

The time dependence in the metric comes from only one
brane in the intersections. Then, the obtained expansion law is not 
complicated. In our solutions, one may have to
compactify the vacuum bulk space as well as the brane world volume 
in order to find an expanding universe. 
Since the fastest expanding case in the Jordan frame
has the power $\lambda(\tilde{\Msp})<1/2$, the power is so small that 
solutions cannot give a realistic expansion law like that in 
the matter-dominated era ($a\propto \tau^{2/3}$)
or that in the radiation-dominated era ($a\propto \tau^{1/2}$).

When we discuss the dynamics on the four-dimensional Einstein frame 
after compactifying the extra dimensions, 
the power exponents are differently depending on 
how we compactify the extra dimensions even within one solution.
For M-brane in the eleven-dimensional theory, 
we give the power exponent of the fastest expansion of 
our four-dimensional Universe in the Einstein frame.  
Since the solution again imply that the expansion is too small, 
we have to conclude that in order to find a realistic expansion
of the Universe in this type of models, one has to add matter 
fields on the brane.

These are the same results as the case of the delocalized intersecting 
brane solutions. 
For the solutions (\ref{rsc:mr:Eq}) involving two intersecting brane 
in the ten- or eleven-dimensional theories which are related to 
the supergravity, we can see 
the power exponents of the scale factor of possible four-dimensional 
Friedmann-Robertson-Walker cosmological models in 
the Tables in \cite{Minamitsuji:2010kb}.

%
\subsection{The intersection of brane and KK-monopole system}
Next we study the dynamical intersecting brane solutions including 
KK-monopoles. 
We should look for whether there is a solution 
with an isotropic and homogeneous three space. 
The $D$-dimensional metric (\ref{k:metric:Eq}) can be expressed as
\Eq{
ds^2=-h^a(t, y, z)dt^2+ds^2(\tilde{\Xsp})+ds^2(\Ysp)+ds^2(\Zsp)
+h^b(t, y, z)h_k^{-1}(z)\left(dv+A_adz^a\right)^2,
   \label{kc:metric:Eq}
}
where we have defined
\Eqrsubl{kc:metric1:Eq}{
ds^2(\tilde{\Xsp})&\equiv&h^a(t, y, z)
\delta_{PQ}(\tilde{\Xsp})d\theta^{P}d\theta^{Q},\\
ds^2(\Ysp)&\equiv&h^b(t, y, z)\gamma_{ij}(\Ysp)dy^idy^j,\\
ds^2(\Zsp)&\equiv&h^b(t, y, z)h_k(z)u_{ab}(\Zsp)dz^adz^b\,.
 }
Here, $a$, $b$ are given by \eqref{k:paremeter:Eq}\,, and  
$\delta_{PQ}(\tilde{\Xsp})$ denotes the $p$-dimensional Euclidean
metric, and $\theta^P$ is the coordinate of the $p$-dimensional Euclid
space $\tilde{\Xsp}$. 

Now we set $h=At+h_1(y, z)$. The $D$-dimensional metric
(\ref{kc:metric:Eq}) can be written as
\Eqr{
ds^2&=&\left[1+\left(\frac{\tau}{\tau_0}\right)^{-2/(a+2)}
h_1\right]^{a}
\left[-d\tau^2+\left(\frac{\tau}{\tau_0}\right)^{2a/(a+2)}
\delta_{PQ}(\tilde{\Xsp})d\theta^Pd\theta^Q\right.\nn\\
&&+\left\{1+\left(\frac{\tau}{\tau_0}\right)^{-2/(a+2)}h_1\right\}
\left(\frac{\tau}{\tau_0}\right)^{2b/(a+2)}
\left\{\gamma_{ij}(\Ysp)dy^idy^j+h_ku_{ab}(\Zsp)dz^adz^b\right.\nn\\
&&\left.\left. +h_k^{-1}\left(dv+A_adz^a\right)^2\right\}\right],
   \label{kc:metric-a:Eq}
}
where the cosmic time $\tau$ is defined by
\Eq{
\frac{\tau}{\tau_0}=\left(At\right)^{(a+2)/2},~~~~\tau_0=
\frac{2}{\left(a+2\right)A}.
}

In the following, 
we again apply the solution to lower-dimensional effective theory.
We compactify $d(\equiv d_1+d_2+d_3)$ dimensions to fit our Universe,
where $d_1$, $d_2$, and $d_3$ denote the compactified dimensions with
respect to the $\tilde{\Xsp}$, $\Ysp$, and $\Zsp$ spaces.
The $D$-dimensional metric (\ref{kc:metric:Eq}) is then written as
\Eq{
ds^2=ds^2(\Msp)+ds^2(\Nsp),
   \label{kc:metric2:Eq}
}
where $ds^2(\Msp)$ denotes the $(D-d)$-dimensional metric and
$ds^2(\Nsp)$ is the line element of compactified dimensions.

Now we can rewrite the $(D-d)$-dimensional metric in the Einstein frame 
in terms of the conformal transformation
\Eq{
ds^2(\Msp)=h^Bh_k^Cds^2(\bar{\Msp})\,,
}
where constants $B$ and $C$ are given by 
\Eq{
B=-\frac{ad_1+b(d_2+d_3)}{D-d-2},~~~~~~C=-\frac{d_3}{D-d-2}.
}
Then, the $(D-d)$-dimensional metric in the Einstein frame can be 
expressed as
\Eqr{
ds^2(\bar{\Msp})&=&h^{B'}h_k^{-C}\left[-dt^2+\delta_{P'Q'}
(\tilde{\Xsp}')d\theta^{P'}d\theta^{Q'}
+h^{4/N}\left\{\gamma_{k'l'}({\Ysp}')dy^{k'}dy^{l'}\right.\right.\nn\\
&&\left.\left.+h_ku_{a'b'}({\Zsp}')dz^{a'}dz^{b'}
+h_k^{-1}\left(dv+A_{a'}dz^{a'}\right)^2\right\}
\right],
  \label{kc:metric-m:Eq}
}
where $B'$ is defined by $B'=-B+a$,
and $\tilde{\Xsp}'$, ${\Ysp}'$, and ${\Zsp}'$ denote
the $(p-d_1)$-, $(D-5-p-d_2)$-, and $(3-d_3)$-dimensional spaces, 
respectively.

If we set $h=At+h_1(y, z)$, 
the metric (\ref{kc:metric-m:Eq}) is thus rewritten by
\Eqr{
ds^2(\bar{\Msp})&=&h_s^{C'}\left[1+\left(\frac{\tau}{\tau_0}\right)
^{-2/(B'+2)}h_1\right]^{B'}\left[-d\tau^2+
\left(\frac{\tau}{\tau_0}\right)^{2B'/(B'+2)}
\delta_{P'Q'}(\tilde{\Xsp}')d\theta^{P'}d\theta^{Q'}\right.\nn\\
&&+\left\{1+\left(\frac{\tau}{\tau_0}\right)^{-2/(B'+2)}h_1\right\}
\left(\frac{\tau}{\tau_0}\right)^{2(B'+\frac{4}{N})/(B'+2)}
\left\{\gamma_{k'l'}({\Ysp}')dy^{k'}dy^{l'}\right.\nn\\
&&\left.\left.+h_ku_{a'b'}({\Zsp}')dz^{a'}dz^{b'}
+h_k^{-1}\left(dv+A_{a'}dz^{a'}\right)^2\right\}\right],
  \label{kc:metric-m2:Eq}
}
where we have introduced the cosmic time $\tau$ defined by
\Eq{
\frac{\tau}{\tau_0}=\left(At\right)^{(B'+2)/2},~~~~\tau_0=
\frac{2}{\left(B'+2\right)A}.
}

If we choose $N=4$, 
the $(D-d)$-dimensional metric \eqref{kc:metric-m2:Eq} shows that 
there is no solution which exhibits an accelerating expansion 
of our Universe. 

The Friedmann-Robertson-Walker 
cosmological solutions with an isotropic 
and homogeneous three-space for the solutions 
(\ref{kc:metric-m2:Eq}) are listed in Table~\ref{table_1} 
for M-branes, Table~\ref{table_2} for D-branes, 
and Table~\ref{table_3} for F1 and NS5-branes.
We find the power exponents of the scale factor of four-dimensional 
cosmological models as 
$a(\tilde{\Msp})\propto \tau^{\lambda(\tilde{\Msp})}$,
where $\tilde{\Msp}$ is the spatial part of the spacetime $\Msp$, 
$a(\tilde{\Msp})$ and $a_{\rm E}(\tilde{\Msp})$ is the scale factors 
of our Universe $\tilde{\Msp}$ in Jordan and Einstein frames with the 
exponents carrying the same suffices, respectively.

In the KK-monopole solution, the expansion law is easily obtained
because the time dependence in the metric comes from only 
$p$-brane in the intersections. We can find an expanding universe even 
if one may compactify the vacuum bulk space as well as the brane world 
volume. However, it is impossible to obtain the cosmological model whose 
scale factor has the power $\lambda(\tilde{\Msp})>1/2$ in the Jordan frame. 
Then they cannot give a realistic expansion law like that in 
the matter-dominated era ($a\propto \tau^{2/3}$)
or that in the radiation-dominated era ($a\propto \tau^{1/2}$).

The power exponents in the four-dimensional Einstein frame after 
compactifing the extra dimensions are different depending on 
how we compactify the extra dimensions even within one solution.
For M-brane and D-brane in the ten- or eleven-dimensional theory, 
we give the power exponent of the fastest expansion of 
our four-dimensional Universe in the Einstein frame in 
Table \ref{table_4}.
However, the expansion is too small to find a realistic expansion
of the Universe in the KK-monopole background. 
Then it is necessary to add some corrections in the background to 
obtain a realistic cosmological solution.


\section{Discussions}
  \label{sec:discussions}
In this paper, we have studied 
dynamical solutions of $p$-brane. 
In the case of partially localized static intersecting brane solutions, 
even the metric ansatz in terms of harmonic functions which 
would generalize such restricted metric ansatz for the delocalized 
brane case is known, we could mention the explicit expressions for 
harmonic functions. 
We have applied simple coordinate transformations to the differential 
equations satisfied by the harmonic functions in order to bring 
them to the forms of partial differential equations which have known
explicit solutions. The Einstein equations give the intersection rules 
which the dynamical brane have to obey. 
Because of the simplicity of the intersection rule, it is easy to work 
out obtaining the explicit analytical form of the solution for the
field equations in the $D$-dimensional projective.  
The intersection rules have led to the results that 
harmonic functions
can be written by linear combinations of terms depending on both
coordinates of worldvolume and transverse space. 
Moreover, in the case of ten- or eleven-dimensional theories, 
the form of the harmonic function implies that the dynamical solution becomes 
the supersymmetric one if we drop the time dependence. We also find 
that the field strengths vanish if we take a limit where the coordinate 
dependence with respect to transverse space becomes much smaller. 
This turned out to be vacuum solutions if the scalar field is trivial  
because in this limit the scalar and gauge fields do not contribute the 
energy momentum tensor, which presumably does not affect the model. 
We can understand this as a Kasner-type metric.
This feature is expected to be 
seen in a wide class of supersymmetric 
solutions beyond the examples considered in the present paper. 

The dynamical solutions include the dilaton coupling parameter $N$ in 
which appears the exact forms of the field strengths. 
We observed that obtaining the explicit analytical form of dynamical 
solutions is almost impossible with $N\ne 4$, since harmonic 
functions that specify branes now satisfy coupled partial differential 
equations. 
If we set $N=4$, these are apparently related to the classical 
solutions of string theory and certainly have a lot of attractive 
properties. Firstly, these solutions were obtained by replacing
the time-independent warp factor of the static solution with the 
time-dependent function. 
Secondly, our solutions can contain only one function depending on both time 
as well as overall or relative transverse space coordinates because 
the Einstein equations tell us that either (i) 
two branes depend only on the coordinates along the relative and 
overall transverse directions or (ii) while one brane is completely 
dynamical the other brane has to depend only on the 
coordinates along the relative and overall transverse directions. 
A new class of solutions where all harmonic functions depend on time
is more challenging.

We have shown that 
each solution gives a FLRW universe if we regard 
the homogeneous and isotropic part of the brane world-volume
or transverse space as our 
spacetime. However, the power of the scale factor is so small that 
the solutions of field equations cannot give a realistic expansion law. 
This means that we have to consider additional matter on the brane 
in order to get a realistic expanding universe. The solutions implies 
that as the number $p$ increases, the power of the scale factor
becomes large. We have found that the intersection with D6-brane 
in ten-dimensional theory gives the fastest expansion of our Universe 
because the three-dimensional spatial space of our Universe stays in 
the transverse space to the D6-brane. Though the power of the scale factor
for the transverse space in solutions with the D7- or D8-branes is larger 
than those with the D6-brane, the number of the transverse space to 
these branes is less than three. Hence, these solutions cannot provide 
the isotropic universe if we assume that the transverse space
to the brane is the part of our Universe. These are same results as 
the case of delocalized brane solutions \cite{Minamitsuji:2010kb, 
Minamitsuji:2010uz}.

The method described here can of course be applied in this model to all
other intersecting brane systems involving more than three branes. 
The same equations, given by \eqref{rs:warp1-1:Eq} will lead to 
the coordinate dependence of the metric because of the ansatz for the fields. 
Hence, only one of branes 
will appear exhibit time dependence, 
as we have already discussed in Secs.~\ref{sec:eleven} and 
\ref{sec:ten} through \ref{sec:two}.  
A serious problem is the difficulty of constructing realistic 
cosmological models such as inflationary universe scenario of 
the early universe or the accelerating expansion of the present universe. 
This problem is of course avoided in delocalized brane solutions 
with particular coupling 
between scalar field and gauge field strength, which does not apparently 
relate to the classical solutions of string theory. 
Then, beyond the model discussed here, obtaining the accelerating expansion 
of our Universe in a string theory, there is more realistic 
problem of setting the field ansatz, coupling constant and internal 
space in a theory of coupled scalar, gauge and gravitational fields. 
This is more complicated, because even in the case of using the same field 
ansatz as the supersymmetric solutions there are coupled
partial differential equations which in general do not have simple form. 
Hence, that alone should not prevent the method described here from being 
applicable to realistic theories, at least for D$p_r$-D$p_s$ branes, 
since a lot of terms in the field equations cannot 
be eliminated by including enough fields. 
If the coupling constant that involve fields have a single parameter $N$ 
attached to matter fields, then we can introduce a parameter for the 
fields by coupling scalar field with the harmonic functions to
the gauge field strengths. But it is not clear how to deal with 
the relation between the string theory and containing parameters 
to which are attached two or more field strengths. 
This raises the question whether the dynamical brane solution is really 
related to the supersymmetric solutions because the value of the 
coupling constant in these solutions are in general severely restricted. 
It is difficult to obtain the de Sitter compactification model which is 
consistent with the string theory \cite{Maeda:2010aj, Minamitsuji:2011gn, 
Minamitsuji:2011gp}.
There is something mysterious about this. The actual calculations in
this paper were done for a fixed field ansatz. Dynamical delocalized brane 
solutions would have been done in the same way, which had never give 
an accelerating expansion of universe in the string theory. 

The lower-dimensional effective theory for the intersection of two branes
and branes on KK-monopole or wave could almost have been discussed
with same calculation as in the case of dynamical delocalized branes. 
The moduli potential in the lower-dimensional effective theories 
after compactifications gives the flat direction. Hence, the 
solutions we have obtained may give some moduli instabilities. 
It would be necessary to introduce some corrections in the effective theory 
to fix the volume and shape moduli of the internal space. Otherwise, 
the moduli instabilities will grow unless the global or local minimum of
the potential. 
Such an effective theory was briefly mentioned in \cite{Minamitsuji:2010kb}, 
and proposed and developed in some detail by 
\cite{Kodama:2005cz, Maeda:2009zi}.

\section*{Acknowledgments}
The work of M.M. was supported by Yukawa fellowship.
K.U. would like to thank H. Kodama, M. Sasaki, N. Ohta and T. Okamura
for continuing encouragement.
K.U. is supported by Grant-in-Aid for Young Scientists (B) of JSPS Research,
under Contract No. 20740147.





\begin{table}[p]
\caption{\baselineskip 14pt
Pair intersections between M-brane and KK-monopole  
in $D=11$ with dependence on overall transverse
coordinates.. }
\label{twoM}
{\scriptsize
\begin{center}
\begin{tabular}{|c||c|c|c|c|c|c|c|c|c|c|c|c||c||c|c|c|}
\hline
Case&&0&1&2&3&4&5&6&7&8&9&10& & $\tilde{\Msp}$ & $\lambda(\tilde{\Msp})$
& $\lambda_{\rm E}(\tilde{\Msp})$
\\
\hline
&M2 & $\circ$ & $\circ$ & $\circ$ &   &   &   &&&&&&
 $\surd$ & & $\lambda(\Ysp)=1/4$ &
$\lambda_{\rm E}(\Ysp)=\frac{-3+d_1}{-12+2d_1+d_2+d_3}$
\\
\cline{3-13}
M2-KK&KK & $\circ$ &$\circ$ &  $\circ$ & $\circ$ &$\circ$ & $\circ$& $\circ$ 
&  &$A_1$& $A_2$ & $A_3$ &
 &  $\Ysp$ \& $v$ \& $\Zsp$ & $\lambda(v)$=1/4 &
$\lambda_{\rm E}(v)=\frac{-3+d_1}{-12+2d_1+d_2+d_3}$ 
\\
\cline{3-13}
&$x^N$ & $t$ & $x^1$ & $x^2$ & $y^1$ & $y^2$ & $y^3$ & $y^4$ & $v$
& $z^1$ & $z^2$ & $z^3$ &
& & $\lambda(\Zsp)$=1/4 &
$\lambda_{\rm E}(\Zsp)=\frac{-3+d_1}{-12+2d_1+d_2+d_3}$ 
\\
\hline
\hline
&M5 & $\circ$ & $\circ$ & $\circ$ & $\circ$ & $\circ$ & $\circ$ &&&&& 
& $\surd$ &  $\tilde{\Xsp}$ & $\lambda(\tilde{\Xsp})=-1/5$ &
$\lambda_{\rm E}(\tilde{\Xsp})=\frac{3-d_2-d_3}{-15+2d_1+d_2+d_3}$
\\
\cline{3-13}
M5-KK&KK & $\circ$ &$\circ$ &  $\circ$ & $\circ$ &$\circ$ & $\circ$& $\circ$ 
&  &$A_1$& $A_2$ & $A_3$ &
&$\Ysp$ \& $v$ \& $\Zsp$ & $\lambda(\Ysp)$=2/5 &
$\lambda_{\rm E}(\Ysp)=\frac{6-d_1}{15-2d_1-d_2-d_3}$ 
\\
\cline{3-13}
&$x^N$ & $t$ & $x^1$ & $x^2$ & $x^3$ & $x^4$ & $x^5$ & $y$ & $v$
& $z^1$ & $z^2$ & $z^3$ &
& & $\lambda(v)=\lambda(\Zsp)$=2/5 &
$\lambda_{\rm E}(v)=\lambda_{\rm E}(\Zsp)=\frac{6-d_1}{15-2d_1-d_2-d_3}$ 
\\
\hline
\end{tabular}
\end{center}
}
\label{table_1}
\end{table}

\begin{table}[p]
\caption{\baselineskip 14pt
Pair intersections between D$p~(p\le 4)$-brane and KK-monopole  
in $D=10$ with dependence on overall transverse
coordinates. 
}
\label{twoDI-1}
{\scriptsize
\begin{center}
\begin{tabular}{|c||c|c|c|c|c|c|c|c|c|c|c||c||c|c|c|}
\hline
Branes&&0&1&2&3&4&5&6&7&8&9& & $\tilde{\Msp}$ & $\lambda(\tilde{\Msp})$
& $\lambda_{\rm E}(\tilde{\Msp})$
\\
\hline
&D0 &  $\circ$ &&&&&&&&&&$\surd$
&  & $\lambda(\Ysp)$=1/9 & $\lambda_{\rm E}(\Ysp_1)=
\frac{1}{9-d_2-d_3}$
\\
\cline{3-12}
D0-KK&KK & $\circ$ & $\circ$ & $\circ$  & $\circ$ & $\circ$
& $\circ$ & &$A_1$&$A_2$&$A_3$&
&$\Ysp$ \& $v$ \& $\Zsp$ &$\lambda(v)$=1/9  & $\lambda_{\rm E}(v)=
\frac{1}{9-d_2-d_3}$
\\
\cline{3-12}
&$x^N$ & $t$ & $y^1$ & $y^2$ & $y^3$ & $y^4$ & $y^5$ & $v$
& $z^1$ & $z^2$ & $z^3$ &
&& $\lambda(\Zsp)$=1/9  &$\lambda_{\rm E}(\Zsp)=
\frac{1}{9-d_2-d_3}$
\\
\hline
\hline
&D1 & $\circ$ &  $\circ$  &&   &&&&&&& $\surd$
& & $\lambda(\Ysp)$=1/5&
$\lambda_{\rm E}(\Ysp)=\frac{2-d_1}{10-2d_1-d_2-d_3}$
\\
\cline{3-12}
D1-KK&KK & $\circ$ &$\circ$& $\circ$ & $\circ$ & $\circ$ & $\circ$ &  &
 $A_1$  & $A_2$  & $A_3$ && $\Ysp$ \& $v$ \& $\Zsp$ & $\lambda(v)$=1/5 &
 $\lambda_{\rm E}(v)=\frac{2-d_1}{10-2d_1-d_2-d_3}$
\\
\cline{3-12}
&$x^N$ & $t$ & $x$ & $y^1$ & $y^2$ & $y^3$ & $y^4$ & $v$
& $z^1$ & $z^2$ & $z^3$ &
&& $\lambda(\Zsp)$=1/5 & 
$\lambda_{\rm E}(\Zsp)=\frac{2-d_1}{10-2d_1-d_2-d_3}$
\\
\hline
\hline
&D2 & $\circ$ & $\circ$ & $\circ$ &&&&&&&& $\surd$
& & $\lambda(\Ysp)$=3/11 &
$\lambda_{\rm E}(\Ysp)=\frac{3-d_1}{11-2d_1-d_2-d_3}$
\\
\cline{3-12}
D2-KK&KK & $\circ$ & $\circ$ & $\circ$ & $\circ$ & $\circ$ & $\circ$ & 
 & $A_1$ & $A_2$ & $A_3$ & & $\Ysp$ \& $v$ \& $\Zsp$ & $\lambda(v)$=3/11 &
 $\lambda_{\rm E}(v)=\frac{3-d_1}{11-2d_1-d_2-d_3}$
\\
\cline{3-12}
&$x^N$ & $t$ & $x^1$ & $x^2$ & $y^1$ & $y^2$ & $y^3$ & $v$
& $z^1$ & $z^2$ & $z^3$ &
&&  $\lambda(\Zsp)$=3/11 & 
$\lambda_{\rm E}(\Zsp)=\frac{3-d_1}{11-2d_1-d_2-d_3}$
\\
\hline
\hline
&D3 & $\circ$ & $\circ$ & $\circ$ & $\circ$ &&&&&&& $\surd$
& $\tilde{\Xsp}$ & $\lambda(\tilde{\Xsp})=-1/3$ &
$\lambda_{\rm E}(\tilde{\Xsp})=\frac{d_2+d_3-4}{12-2d_1-d_2-d_3}$
\\
\cline{3-12}
D3-KK&KK & $\circ$ & $\circ$ & $\circ$ & $\circ$ & $\circ$ & $\circ$ & 
 & $A_1$ & $A_2$ & $A_3$ & & $\Ysp$ \& $v$ \& $\Zsp$ & $\lambda(\Ysp)$=1/3 &
 $\lambda_{\rm E}(\Ysp)=\frac{4-d_1}{12-2d_1-d_2-d_3}$
\\
\cline{3-12}
&$x^N$ & $t$ & $x^1$ & $x^2$ & $x^3$ & $y^1$ & $y^2$ & $v$
& $z^1$ & $z^2$ & $z^3$ &
&&  $\lambda(v)=\lambda(\Zsp)$=1/3 & 
$\lambda_{\rm E}(v)=\lambda_{\rm E}(\Zsp)=\frac{4-d_1}{12-2d_1-d_2-d_3}$
\\
\hline
\hline
&D4 & $\circ$ & $\circ$ & $\circ$ & $\circ$ & $\circ$ &&&&&& $\surd$
& & $\lambda(\tilde{\Xsp})=-3/13$ &
$\lambda_{\rm E}(\tilde{\Xsp})=\frac{d_2+d_3-3}{13-2d_1-d_2-d_3}$
\\
\cline{3-12}
D4-KK&KK & $\circ$ & $\circ$ & $\circ$ & $\circ$ & $\circ$ & $\circ$ & 
 & $A_1$ & $A_2$ & $A_3$ & & $\Ysp$ \& $v$ \& $\Zsp$ & $\lambda(\Ysp)$=5/13 &
 $\lambda_{\rm E}(\Ysp)=\frac{5-d_1}{13-2d_1-d_2-d_3}$
\\
\cline{3-12}
&$x^N$ & $t$ & $x^1$ & $x^2$ & $x^3$ & $x^4$ & $y$ & $v$
& $z^1$ & $z^2$ & $z^3$ &
&&  $\lambda(v)=\lambda(\Zsp)$=5/13 & 
$\lambda_{\rm E}(v)=\lambda_{\rm E}(\Zsp)=\frac{5-d_1}{13-2d_1-d_2-d_3}$
\\
\hline
\end{tabular}
\end{center}
}
\label{table_2}
\end{table}

\begin{table}[p]
\caption{\baselineskip 14pt
Pair intersections between fundamental string and KK-monopole  
in $D=10$ with dependence on overall transverse coordinates. 
}
\label{two f-1}
{\scriptsize
\begin{center}
\begin{tabular}{|c||c|c|c|c|c|c|c|c|c|c|c||c||c|c|c|}
\hline
Branes&&0&1&2&3&4&5&6&7&8&9& & $\tilde{\Msp}$ & $\lambda(\tilde{\Msp})$
& $\lambda_{\rm E}(\tilde{\Msp})$
\\
\hline
&F1 & $\circ$ &  $\circ$  &&   &&&&&&& $\surd$
& & $\lambda(\Ysp)$=1/5&
$\lambda_{\rm E}(\Ysp)=\frac{2-d_1}{10-2d_1-d_2-d_3}$
\\
\cline{3-12}
F1-KK&KK & $\circ$ &$\circ$& $\circ$ & $\circ$ & $\circ$ & $\circ$ &  &
 $A_1$  & $A_2$  & $A_3$ && $\Ysp$ \& $v$ \& $\Zsp$ & $\lambda(v)$=1/5 &
 $\lambda_{\rm E}(v)=\frac{2-d_1}{10-2d_1-d_2-d_3}$
\\
\cline{3-12}
&$x^N$ & $t$ & $x$ & $y^1$ & $y^2$ & $y^3$ & $y^4$ & $v$
& $z^1$ & $z^2$ & $z^3$ &
&& $\lambda(\Zsp)$=1/5 & 
$\lambda_{\rm E}(\Zsp)=\frac{2-d_1}{10-2d_1-d_2-d_3}$
\\
\hline
\end{tabular}
\end{center}
}
\label{table_3}
\end{table}

\begin{table}[p]
\caption{\baselineskip 14pt
The power exponent of the fastest expansion in the Einstein frame
for M-brane, D-brane, fundamental string in KK-monopole background. 
``TD" in the table shows which brane is time dependent.
}
\begin{center}
{\scriptsize
\begin{tabular}{|c|c|c||c|c||c|}
\hline
Branes& TD &dim$(\Msp)$  &$\bar{\Msp}$
&$(d_1, d_2, d_3)$&
$\lambda_{\rm E}(\bar{\Msp})$
\\
\hline
\hline
M2-KK&M2& 8& $\Ysp$ \& $v$ \& Z & (0, 1, 2) &1/3
\\
\cline{2-6}
&M2&8& $\Ysp$ \& Z & (0, 3, 0) & 1/3 
\\
\hline
M5-KK&M5&7& $\tilde{\Xsp}$ \& $\Ysp$\& $v$ \& Z & (2, 0, 2)  & $-1/9$ 
\\
\cline{2-6}
&M5&9& $\tilde{\Xsp}$ \& $\Ysp$\& $v$ \& Z & (0, 0, 2) &6/13 
\\
\hline
D0-KK&D0&6& $\Ysp$\& $v$ \& Z & (0, 2, 2) &1/5 
\\
\cline{2-6}
&D0&5& $\Ysp$\& $v$ \& Z & (0, 3, 2) &1/4 
\\
\hline
D1-KK&D1&7& $\Ysp$\& $v$ \& Z & (0, 1, 2) &2/7 
\\
\cline{2-6}
&D1&5& $\Ysp$\& $v$ \&  Z & (0, 3, 2) &2/5 
\\
\hline
D2-KK&D2&8& $\Ysp$\& $v$ \& Z & (0, 0, 2)& 1/3 
\\
\cline{2-6}
&D2&6& $\Ysp$\& $v$ \& Z & (0, 2, 2) & 3/7 
\\
\hline
D3-KK&D3&7&$\tilde{\Xsp}$ \& $\Ysp$ \& $v$ \& Z
&(0, 1, 2) & $-1/9$ 
\\
\cline{2-6}
&D3&7&$\tilde{\Xsp}$ \& $\Ysp$\& $v$ \& $\Zsp$ & (0, 1, 2) & 4/9 
\\
\hline
D4-KK&D4&8&$\tilde{\Xsp}$ \& $\Ysp$ \& $v$ \& Z
&(0, 0, 2) & $-1/11$ 
\\
\cline{2-6}
&D4&8&$\tilde{\Xsp}$ \& $\Ysp$\& $v$ \& $\Zsp$ & (0, 0, 2) & 5/11 
\\
\hline
F1-KK&F1&7& $\Ysp$\& $v$ \& Z & (0, 1, 2) &2/7 
\\
\cline{2-6}
&F1&5& $\Ysp$\& $v$ \&  Z & (0, 3, 2) &2/5 
\\
\hline
\end{tabular}
}
\label{table_4}
\end{center}
\end{table}


\begin{thebibliography}{99}

\bibitem{Binetruy:2007tu}
  P.~Binetruy, M.~Sasaki and K.~Uzawa,
  ``Dynamical D4-D8 and D3-D7 branes in supergravity,''
  Phys.\ Rev.\  D {\bf 80} (2009) 026001
  [arXiv:0712.3615 [hep-th]].

\bibitem{Maeda:2009zi}
  K.~i.~Maeda, N.~Ohta and K.~Uzawa,
  ``Dynamics of intersecting brane systems -- Classification and their
  applications --,''
  JHEP {\bf 0906} (2009) 051
  [arXiv:0903.5483 [hep-th]].

\bibitem{Gibbons:2009dr}
  G.~W.~Gibbons and K.~i.~Maeda,
  ``Black Holes in an Expanding Universe,''
  Phys.\ Rev.\ Lett.\  {\bf 104} (2010) 131101
  [arXiv:0912.2809 [gr-qc]].

\bibitem{Minamitsuji:2010kb}
  M.~Minamitsuji, N.~Ohta and K.~Uzawa,
  ``Cosmological intersecting brane solutions,''
  Phys.\ Rev.\  D {\bf 82} (2010) 086002
  [arXiv:1007.1762 [hep-th]].

\bibitem{Minamitsuji:2010uz}
  M.~Minamitsuji and K.~Uzawa,
  ``Cosmology in $p$-brane systems,''
  Phys.\ Rev.\  D {\bf 83} (2011) 086002
  [arXiv:1011.2376 [hep-th]].

\bibitem{Gibbons:2005rt}
  G.~W.~Gibbons, H.~Lu and C.~N.~Pope,
  ``Brane worlds in collision,''
  Phys.\ Rev.\ Lett.\  {\bf 94} (2005) 131602
  [arXiv:hep-th/0501117].

\bibitem{Maeda:2009tq}
  K.~i.~Maeda, N.~Ohta, M.~Tanabe and R.~Wakebe,
  ``Supersymmetric Intersecting Branes in Time-dependent Backgrounds,''
  JHEP {\bf 0906} (2009) 036
  [arXiv:0903.3298 [hep-th]].

\bibitem{Maeda:2009ds}
  K.~i.~Maeda and M.~Nozawa,
  ``Black Hole in the Expanding Universe from Intersecting Branes,''
  Phys.\ Rev.\  D {\bf 81} (2010) 044017
  [arXiv:0912.2811 [hep-th]].
  
\bibitem{Maeda:2010yk}
  K.~i.~Maeda, N.~Ohta, M.~Tanabe and R.~Wakebe,
  ``Supersymmetric Intersecting Branes on the Waves,''
  JHEP {\bf 1004} (2010) 013
  [arXiv:1001.2640 [hep-th]].

\bibitem{Minamitsuji:2010fp}
  M.~Minamitsuji, N.~Ohta and K.~Uzawa,
  ``Dynamical solutions in the 3-Form Field Background in the
  Nishino-Salam-Sezgin Model,''
  Phys.\ Rev.\  D {\bf 81} (2010) 126005
  [arXiv:1003.5967 [hep-th]].

\bibitem{Maeda:2010ja}
  K.~i.~Maeda and M.~Nozawa,
  ``Black Hole in the Expanding Universe with Arbitrary Power-Law Expansion,''
  Phys.\ Rev.\  D {\bf 81} (2010) 124038
  [arXiv:1003.2849 [gr-qc]].

\bibitem{Nozawa:2010zg}
  M.~Nozawa and K.~i.~Maeda,
  ``Cosmological rotating black holes in five-dimensional fake supergravity,''
  Phys.\ Rev.\  D {\bf 83} (2011) 024018
  [arXiv:1009.3688 [hep-th]].

\bibitem{Maeda:2011sh}
  K.~i.~Maeda and M.~Nozawa,
  ``Black hole solutions in string theory,''
  Prog.\ Theor.\ Phys.\ Suppl.\  {\bf 189} (2011) 310
  [arXiv:1104.1849 [hep-th]].

\bibitem{Yang:1999ze}
  H.~s.~Yang,
  ``Localized intersecting brane solutions of $D=11$ supergravity,''
  arXiv:hep-th/9902128.

\bibitem{Tseytlin:1996as}
  A.~A.~Tseytlin,
  ``Extreme dyonic black holes in string theory,''
  Mod.\ Phys.\ Lett.\  A {\bf 11} (1996) 689
  [arXiv:hep-th/9601177].

\bibitem{Tseytlin:1997cs}
  A.~A.~Tseytlin,
  ``Composite BPS configurations of $p$-branes in ten-dimensions and
  eleven-dimensions,''
  Class.\ Quant.\ Grav.\  {\bf 14} (1997) 2085
  [arXiv:hep-th/9702163].

\bibitem{Youm:1997hw}
  D.~Youm,
  ``Black holes and solitons in string theory,''
  Phys.\ Rept.\  {\bf 316} (1999) 1
  [arXiv:hep-th/9710046].

\bibitem{Youm:1999zs}
  D.~Youm,
  ``Partially localized intersecting BPS branes,''
  Nucl.\ Phys.\  B {\bf 556} (1999) 222
  [arXiv:hep-th/9902208].

\bibitem{Tseytlin:1996bh}
  A.~A.~Tseytlin,
  ``Harmonic superpositions of M-branes,''
  Nucl.\ Phys.\  B {\bf 475} (1996) 149
  [arXiv:hep-th/9604035].

\bibitem{Argurio:1997gt}
  R.~Argurio, F.~Englert and L.~Houart,
  ``Intersection rules for $p$-branes,''
  Phys.\ Lett.\  B {\bf 398} (1997) 61
  [arXiv:hep-th/9701042].

\bibitem{Argurio:1998cp}
  R.~Argurio,
  ``Brane physics in M-theory,''
  arXiv:hep-th/9807171.

\bibitem{Ohta:1997gw}
  N.~Ohta,
  ``Intersection rules for non-extreme $p$-branes,''
  Phys.\ Lett.\  B {\bf 403} (1997) 218
  [arXiv:hep-th/9702164].

\bibitem{Kodama:2005fz}
  H.~Kodama and K.~Uzawa,
  ``Moduli instability in warped compactifications of the type IIB
  supergravity,''
  JHEP {\bf 0507} (2005) 061
  [arXiv:hep-th/0504193].

\bibitem{Kodama:2005cz}
  H.~Kodama and K.~Uzawa,
  ``Comments on the four-dimensional effective theory for warped
  compactification,''
  JHEP {\bf 0603} (2006) 053
  [arXiv:hep-th/0512104].

\bibitem{FigueroaO'Farrill:1999tx}
  J.~M.~Figueroa-O'Farrill,
  ``Breaking the M waves,''
  Class.\ Quant.\ Grav.\  {\bf 17} (2000) 2925
  [arXiv:hep-th/9904124].

\bibitem{FigueroaO'Farrill:2002tc}
  J.~M.~Figueroa-O'Farrill and J.~Simon,
  ``Supersymmetric Kaluza-Klein reductions of M waves and MKK monopoles,''
  Class.\ Quant.\ Grav.\  {\bf 19} (2002) 6147
  [arXiv:hep-th/0208108].

\bibitem{Bergshoeff:1997tt}
  E.~Bergshoeff, M.~de Roo, E.~Eyras, B.~Janssen and J.~P.~van der Schaar,
  ``Intersections involving monopoles and waves in eleven-dimensions,''
  Class.\ Quant.\ Grav.\  {\bf 14} (1997) 2757
  [arXiv:hep-th/9704120].

\bibitem{Bergshoeff:1996rn}
  E.~Bergshoeff, M.~de Roo, E.~Eyras, B.~Janssen and J.~P.~van der Schaar,
  ``Multiple intersections of D-branes and M-branes,''
  Nucl.\ Phys.\  B {\bf 494} (1997) 119
  [arXiv:hep-th/9612095].

\bibitem{Maeda:2010aj}
  K.~i.~Maeda, M.~Minamitsuji, N.~Ohta and K.~Uzawa,
  ``Dynamical $p$-branes with a cosmological constant,''
  Phys.\ Rev.\  D {\bf 82} (2010) 046007
  [arXiv:1006.2306 [hep-th]].

\bibitem{Minamitsuji:2011gn}
  M.~Minamitsuji and K.~Uzawa,
  ``Spectrum from the warped compactifications with the de Sitter universe,''
  JHEP {\bf 1207} (2012) 154
  [arXiv:1103.5325 [hep-th]].
  
\bibitem{Minamitsuji:2011gp}
  M.~Minamitsuji and K.~Uzawa,
  ``Warped de Sitter compactifications,''
  JHEP {\bf 1201} (2012) 142
  [arXiv:1103.5326 [hep-th]].

\end{thebibliography}
\end{document}